\documentclass[12pt]{article}

\usepackage{amsmath}
\usepackage{amssymb}
\usepackage{epsfig}
\usepackage{graphicx,color}
\usepackage{cite}
\usepackage{multirow}
\usepackage{longtable}
\usepackage{lscape}
\usepackage{bbm}
\usepackage{mathrsfs}

\allowdisplaybreaks[4] 

\newcommand{\capdef}{}
\newcommand{\mycaption}[2][\capdef]{\renewcommand{\capdef}{#2}%
        \caption[#1]{{\footnotesize #2}}}
\makeatletter
\renewcommand{\fnum@table}{\textbf{\tablename~\thetable}}
\renewcommand{\fnum@figure}{\textbf{\figurename~\thefigure}}
\makeatother

\newcounter{myenumi}

\renewcommand{\themyenumi}{\roman{myenumi}}
{\end{list}}

\newlength{\myem}
\newcounter{mysubequation}[equation]

\makeatletter
\renewcommand{\section}{\@startsection{section}{1}{0em}{-\baselineskip}%
{\baselineskip}{\normalfont\large\bfseries}}
\renewcommand{\subsection}%
{\@startsection{subsection}{2}{0em}{-0.7\baselineskip}%
{0.7\baselineskip}{\normalfont\bfseries}}
\makeatother

\newcommand{\bi}{\begin{itemize}}
\newcommand{\ei}{\end{itemize}}

\newcommand{\be}{\begin{equation}}
\newcommand{\ee}{\end{equation}}
\newcommand{\bea}{\begin{eqnarray}}
\newcommand{\eea}{\end{eqnarray}}

\newcommand{\ldm}{\Delta m_{31}^2}
\newcommand{\sdm}{\Delta m_{21}^2}
\newcommand{\deltacp}{\delta_{\mathrm{CP}}}
\newcommand{\stheta}{\sin^2 2 \theta_{13}}

\newcommand{\CP}{{CP}}
\newcommand{\eps}{\varepsilon}
\newcommand{\ie}{{\it i.e.}}

\newcommand{\eg}{{\it e.g.}}

\newcommand{\cf}{{\it cf.}}

\newcommand{\fig}{Fig.}
\newcommand{\Fig}{Fig.}

\newcommand{\Ref}{Ref.}

\newcommand{\Tab}{Table}

\newcommand{\figu}[1]{\fig~\ref{fig:#1}}

\newcommand{\ma}{\Delta m^2_{31}}
\newcommand{\chr}{\mbox{$\breve{\rm C}$erenkov~}}

\newcommand{\stch}{\sin^2 2\theta_{13}}

\newcommand{\bb}{$\beta$-Beam~}

\newcommand{\br}{$^8$B~}
\newcommand{\li}{$^8$Li~}
\newcommand{\he}{$^6$He~}
\newcommand{\neon}{$^{18}$Ne~}

\newcommand{\bra}[1]{\ensuremath{\langle #1 |}}   
\newcommand{\ket}[1]{\ensuremath{| #1 \rangle}}   

\DeclareMathOperator{\Real}{Re}

\title{EURONU WP6 2009 yearly report: Update of the physics potential of Nufact, superbeams and betabeams}

\author{Contributors:
J.~Bernabeu$^{\rm a}$, M. Blennow$^{\rm b}$, P.~Coloma$^{\rm c}$, A. ~Donini$^{\rm a,c}$, \\C.~Espinoza$^{\rm a}$,
 E.~Fern\'andez-Martinez$^{\rm b}$, P. Hern\'andez$^{\rm a}$, P.~Huber$^{\rm d}$,  \\
 J.~Kopp$^{\rm e}$,  A.~Longhin$^{\rm f}$,  J.~L\'opez-Pav\'on$^{\rm c}$, M.~Mezzetto$^{\rm g}$, T.~Ota$^{\rm h}$,  \\ T.~Schwetz$^{\rm i}$ and  W.~Winter$^{\rm h}$
 \\
\\
{\it $^{\rm a}$ Dep. de F\'{\i}sica Te\'orica and IFIC, Universidad de Valencia and CSIC,}
\\
{\it  Apart. 22085, 46071 Valencia}
 \\
 {\it $^{\rm b}$ Max-Planck-Insitute F\"ur Physik, F\"ohringer Ring, }
 \\
 {\it 80805 M\"unchen, Germany}
  \\
{\it $^{\rm c}$ Dep. de F\'{\i}sica Te\'orica and Instituto de F\'{\i}sica Te\'orica, UAM-CSIC,}
 \\
 {\it Cantoblanco, 28049 Madrid, Spain.}
 \\
 {\it  $^{\rm d}$ Department of Physics, IPNAS, Virginia Tech, Blacksburg,} 
 \\
 {\it Virginia, VA 24061, USA.}
\\
 {\it $^{\rm e}$ Theoretical Physics Department, Fermilab, P. O. Box 500, }
 \\
 {\it Batavia, IL60510 USA.}
\\
  {\it $^{\rm f}$ Commissariat \`a l'Energie Atomique (CEA), Dir. des Sciences de la Mati\`ere, }
  \\
  {\it 25 Rue Leblanc, Paris 75015, France} 
\\
  {\it $^{\rm g}$ INFN, Sezione di Padova, Via Marzolo 8, }
  \\
  {\it 35131 Padova, Italy.}
  \\
     {\it $^{\rm h}$ Institute F\"ur Theoretische Physik und Astrophysik, Universit\"at W\"urzburg, }\\
    {\it  97074  W\"urzburg, Germany.}
          \\
   {\it $^{\rm i}$ Max-Planck-Institute F\"ur Physik, P.O. Box 103980, }\\
   {\it 69029 Heidelberg, Germany}
  }

\begin{document}

\maketitle

\begin{flushright}
{EURONU-WP6-10-19}
\end{flushright}

\newpage

\section{Neutrino masses and new physics}

The main motivation of a future neutrino physics programme is to understand what the new physics associated to neutrino masses is. We know for sure that new degrees of freedom must be added to the Standard Model (e.g. right-handed neutrinos) at some energy scale $\Lambda$. If $\Lambda$ is much larger than the electroweak scale, there is a natural explanation of why neutrinos are so light. Indeed the effects of {\it any} such new physics must be generically  well described at low energies by an effective Lagrangian which contains the Standard Model, plus a tower of higher dimensional operators constructed with the SM fields and satisfying all the gauge symmetries:
 \begin{eqnarray}
{\mathcal L} = {\mathcal L}_{SM} + \sum_i \frac{{\alpha}_i}{\Lambda} {\mathcal O}_i^{d=5} + \sum_i \frac{\beta_i}{\Lambda^2} {\mathcal O}_i^{d=6} + ...
\label{eq:eft}
\end{eqnarray} 
The effective operators, ${\mathcal O}_i$,  are ordered by their mass dimension, since the higher the dimension, the higher the power of $\Lambda$ that suppresses them.  The dominant operator is therefore the lowest dimensional one, with $d=5$, which is precisely the Weinberg's operator:
\begin{eqnarray}
{\mathcal O}^{d=5} = {\bar L}^c \Phi \Phi L ,
\label{eq:weinberg}
\end{eqnarray}
which, as is well known, induces three new ingredients to the minimal SM:
\begin{itemize}
\item Neutrino masses
\item Lepton mixing
\item Lepton number violation
\end{itemize}
In this context, neutrino masses are very small, because they 
come from an effective operator which is  suppressed by a high energy scale. 
If we go to operators of $d=6$, that are suppressed by two powers of $\Lambda$, these will generically induce new physics in dipole moments, rare decays, etc. Beyond $d=6$ we would find operators inducing non-standard neutrino interactions (NSI).  

It is also possible that the scale  $\Lambda$  is at or below the electroweak scale, or in other words that neutrino masses are linked to light degrees of freedom, {\it i.e.} a {\it hidden} sector which we have not detected yet, because it is weakly interacting. Such scenarios do not offer an explanation of why neutrinos are light, but neutrinos are the natural messengers with such hidden sectors, since they are the only particles in the SM carrying no conserved charge. Such 
new physics could be related to other fundamental problems in particle physics such as the origin of dark matter and dark energy.

Even though it is not guaranteed that we can fully understand the new physics associated to neutrino masses by measuring them, it is quite clear that we have a good chance to learn something more about it by testing the Standard scenario of 3$\nu$ mixing with future and more precise neutrino experiments. In particular we should be able to measure all the fundamental parameters: three mass eigenstates ($m_1^2, m_2^2, m_3^2$), three angles  ($\theta_{12}, \theta_{13}, \theta_{23}$) and one or three CP-violating phases  ($\delta, \alpha_1, \alpha_2$). But, also, it will be very important  to search for new physics 
beyond neutrino masses and mixings, in particular for those effects that are generic in many models of neutrino masses, such as violations of unitarity, non-standard interactions or the presence of light sterile species. To some extent these searches can also be improved in future facilities and this should be evaluated. Typically such analyses imply dealing with a much larger parameter space, which calls for new tools to perform the fits, in particular Montecarlo methods. 

\begin{figure}[htbp]
\begin{center}
\includegraphics[width=8cm]{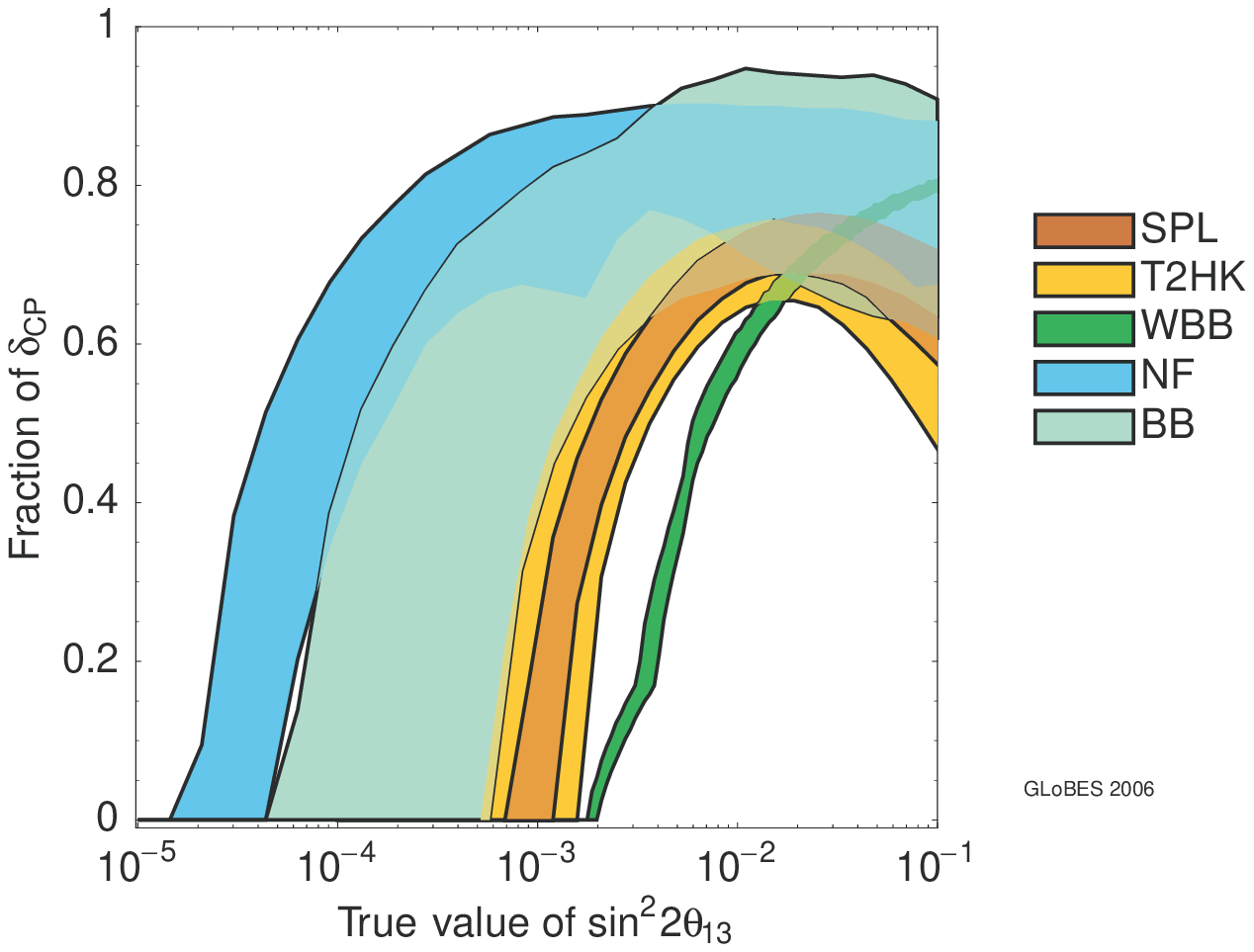}\includegraphics[width=8cm]{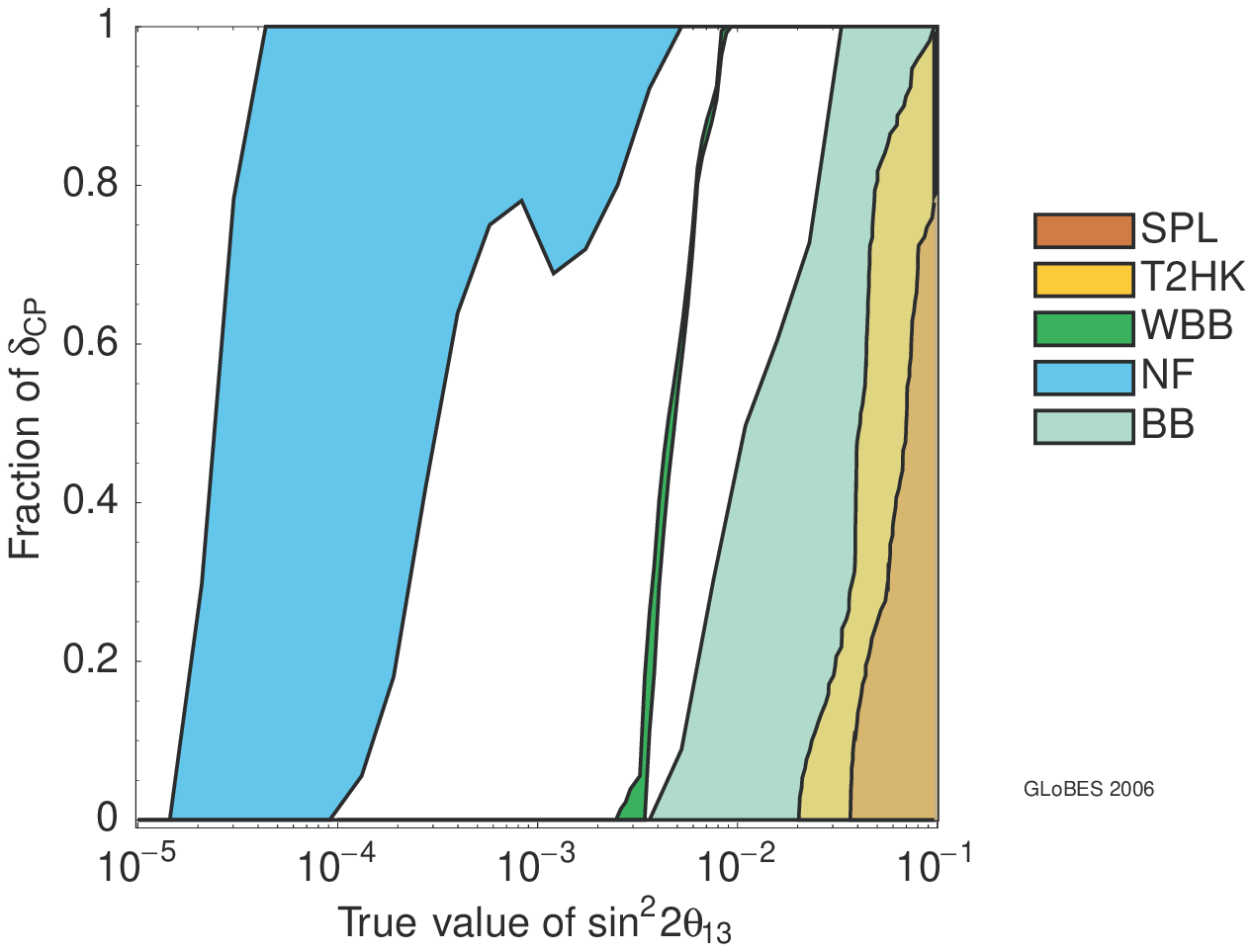}
\caption{Comparison of the physics reach of different future facilities in leptonic CP violation (left) and the neutrino mass hierarchy (right). Taken from \cite{Bandyopadhyay:2007kx}.}
\label{fig:iss}
\end{center}
\end{figure}

Many studies in the last ten years have shown that we can measure the unknown angle $\theta_{13}$, discover leptonic CP violation and determine the neutrino hierarchy in more precise neutrino oscillation experiments, searching for the subleading channel $\nu_e \leftrightarrow \nu_\mu$ in the atmospheric range. A starting point for the Physics Work Package (WP6) of the EURONU project are the results of the International Scoping Study (ISS)  summarized in Figs.~\ref{fig:iss} \cite{Bandyopadhyay:2007kx}. 

In this first report of WP6 activities the following new results are presented:
\begin{itemize}
\item Sect.~\ref{sec:upcoming}: Re-evaluation of the physics reach of the upcoming generation of experiments to measure $\theta_{13}$ and $\delta$ (see Ref.~Ê\cite{Huber:2009cw}).
\item Sect.~\ref{sec:mcubes}: New tools to explore a larger parameter space as needed beyond the standard scenario (see Ref.~\cite{Blennow:2009pk}).
\item Sect.~\ref{sec:nufact}: Neutrino Factory
\begin{enumerate}
\item Sect.~\ref{sec:sterile}: evaluation of the physics reach of a Nufact regards sterile neutrinos (see Ref.~\cite{Donini:2008wz}).
\item Sect.~\ref{sec:nsi}: evaluation of the physics reach of a Nufact as regards non-standard interactions (see Ref.~\cite{Kopp:2008ds}).
\item Sect.~\ref{sec:nonunit}: evaluation of the physics reach of a Nufact as regards violation of unitarity (see Ref.~\cite{Antusch:2009pm}).
\item Sect.~\ref{sec:taus}: critical assessment on long baseline $\tau$-detection at Nufact \cite{DoniniKopp}.
\item Sect.~\ref{sec:near}: new physics searches at a near detector in a Nufact (see Ref.~\cite{Tang:2009na}).
\end{enumerate}
\item Sect.~\ref{sec:bbeam}: Beta-beams
\begin{enumerate}
\item Sect.~\ref{sec:enrique}: choice of ions and location for a $\gamma = 100$ CERN-based $\beta$-beam (from \cite{FernandezMartinez:2009hb}).
\item Sect.~\ref{sec:atmo}: re-evaluation of atmospheric neutrino background for the $\gamma = 100$ $\beta$-beam scenario (from \cite{FernandezMartinez:2009hb}).
\item Sect.~\ref{sec:twobaseline}: study of a two baseline $\beta$-beam (see Ref.~\cite{Choubey:2009ks}).
\item Sect.~\ref{sec:mossbauer}: measuring absolute neutrino mass with Beta Beams (see Ref.~\cite{Lindroos:2009mx}).
\item Sect.~\ref{sec:mono}: progress on monochromatic $\beta$-beams (see Ref.~\cite{Bernabeu:2009np}).
\end{enumerate}
\item Sect.~\ref{sec:mezzetto}: Update of the physics potential of the SPL super-beam (from \cite{Longhin}).
\end{itemize}

\section{Up-coming oscillation experiments}
\label{sec:upcoming}

The main purpose of the upcoming generation of experiments will be to
discover sub-leading effects in neutrino oscillations. This includes
the determination of the small lepton mixing angle $\theta_{13}$,
establishing CP violation (CPV) in neutrino oscillations for a value
of the Dirac CP phase $\deltacp \neq 0, \pi$, and identification of
the type of the neutrino mass hierarchy (MH), which can be normal
($\ldm > 0$) or inverted ($\ldm < 0$).
There are several neutrino oscillation experiments currently under
construction, which are expected to start data taking soon. These are
the reactor neutrino experiments Double Chooz~\cite{Ardellier:2004ui},
Daya Bay~\cite{Guo:2007ug}, RENO~\cite{Kim:2008zzb} and the
accelerator experiments T2K~\cite{Itow:2001ee} and
NO$\nu$A~\cite{Ambats:2004js}. In Ref.~\cite{Huber:2009cw} the potential of
this next generation of experiments towards the three tasks mentioned
above has been evaluated. While the primary goal for all of these
experiments is the discovery of the yet unknown mixing angle
$\theta_{13}$, it might also be interesting to ask the question,
whether there is some chance to address also CPV and MH, in case
$\theta_{13}$ is relatively large.  The analysis of Ref.~\cite{Huber:2009cw}
updates previous works~\cite{Minakata:2002jv, Huber:2002rs,
Huber:2003pm, Huber:2004ug, McConnel:2004bd} with respect to the now
settled parameters for the considered experiments.

\Tab~\ref{tab:nom} summarizes the key parameters of the considered
experiments, for further details see~\cite{Huber:2009cw}. The analysis is
performed by using the GLoBES
software~\cite{Huber:2004ka,Huber:2007ji}; the corresponding {\tt
glb}-files are available at the GLoBES web-page~\cite{Huber:2004ka}
including detailed technical information on the simulation.  In all
cases the strategy is to follow as close as possible the original
Letters of Intent (LOIs) or Technical Design Reports (TDRs). We have
made sure that our sensitivities agree with the ``official'' curves
from the corresponding collaborations under the same assumptions.
For the sensitivity analyses we use the oscillation parameter values
from \Ref~\cite{Schwetz:2008er}: $\sdm=7.65 \times 10^{-5} \,
\mathrm{eV}^2$, $|\ldm|=2.40 \times 10^{-5} \, \mathrm{eV}^2$, $\sin^2
\theta_{12}=0.304$, and $\sin^2 \theta_{23}=0.500$, unless stated
otherwise. We impose external $1\sigma$ errors on $\sdm$ (3\%) and
$\theta_{12}$ (4\%) as conservative estimates for the current
measurement errors, as well as $\ldm$ (5\%) for reactor experiments if
analyzed without beam experiments. In addition, we include a 2\%
matter density uncertainty.

\begin{table}[b]
\begin{tabular}{lccrrlr}
\hline
Setup & $t_\nu$ [yr] & $t_{\bar{\nu}}$ [yr] & $P_{\mathrm{Th}}$ or
$P_{\mathrm{Target}}$ & $L$ [km] & Detector technology & $m_{\mathrm{Det}}$ \\
\hline
 Double Chooz & - & 3 &  8.6 GW & 1.05 & Liquid scintillator &  8.3 t\\
 Daya Bay     & - & 3 & 17.4 GW & 1.7  & Liquid scintillator &  80 t \\
 RENO         & - & 3 & 16.4 GW & 1.4  & Liquid scintillator & 15.4 t \\
 T2K          & 5 & - & 0.75~MW & 295  & Water Cerenkov      & 22.5 kt \\
 NO$\nu$A     & 3 & 3 & 0.7~MW  & 810  & TASD                &   15 kt \\
\hline
\end{tabular}
\mycaption{\label{tab:nom} Summary of the standard setups at their nominal luminosities.}
\end{table}

In Ref.~\cite{Huber:2009cw} various performance indicators have been
considered for the nominal configurations of the experiments, such as
sensitivity to $\theta_{13}$, potential for large $\theta_{13}$,
accuracy to the atmospheric parameters $\theta_{23}$, $|\Delta
m^2_{31}|$, CP-violation, and mass hierarchy. In the following we show
as an important result the prospective time evolution of the
sensitivity to $\theta_{13}$. These calculations are based as much as
possible on official statements of the collaborations. Although the
assumed schedules and proton beam plans may turn out to be not
realistic in some cases, our toy scenario will be illustrative to show
the key issues for the individual experiments within the global
neutrino oscillation context.  The sensitivities are shown as a
function of time assuming that data are continously analyzed and
results are available immediately.

The key assumptions for our toy scenario are as follows.  Double Chooz
starts late 2009 and runs 1.5 years with far detector only, then with
far and near detector.  RENO and Daya Bay start mid 2010 and mid 2011,
respectively, with all detectors on-line.  T2K starts late 2009 with
virtually 0~MW beam power, which increases linearly to 0.75~MW reached
in 12/2012. From then we assume the full target power of 0.75~MW. The
beam runs only with neutrinos.  NO$\nu$A starts mid 2012 with full
beam (0.7~MW), but 2.5~kt detector mass only.  Then the detector mass
increases linearly to 15~kt in 01/2014.  From then we assume the full
detector mass of 15~kt. The beam runs with neutrinos first, until the
equivalent of three years operation at nominal luminosity (\cf,
\Tab~\ref{tab:nom}) is reached, \ie, 03/2016. Then it switches
(possibly) to antineutrinos and runs at least until 2019.

\begin{figure}[t!]
\begin{center}
\includegraphics[width=0.48\textwidth]{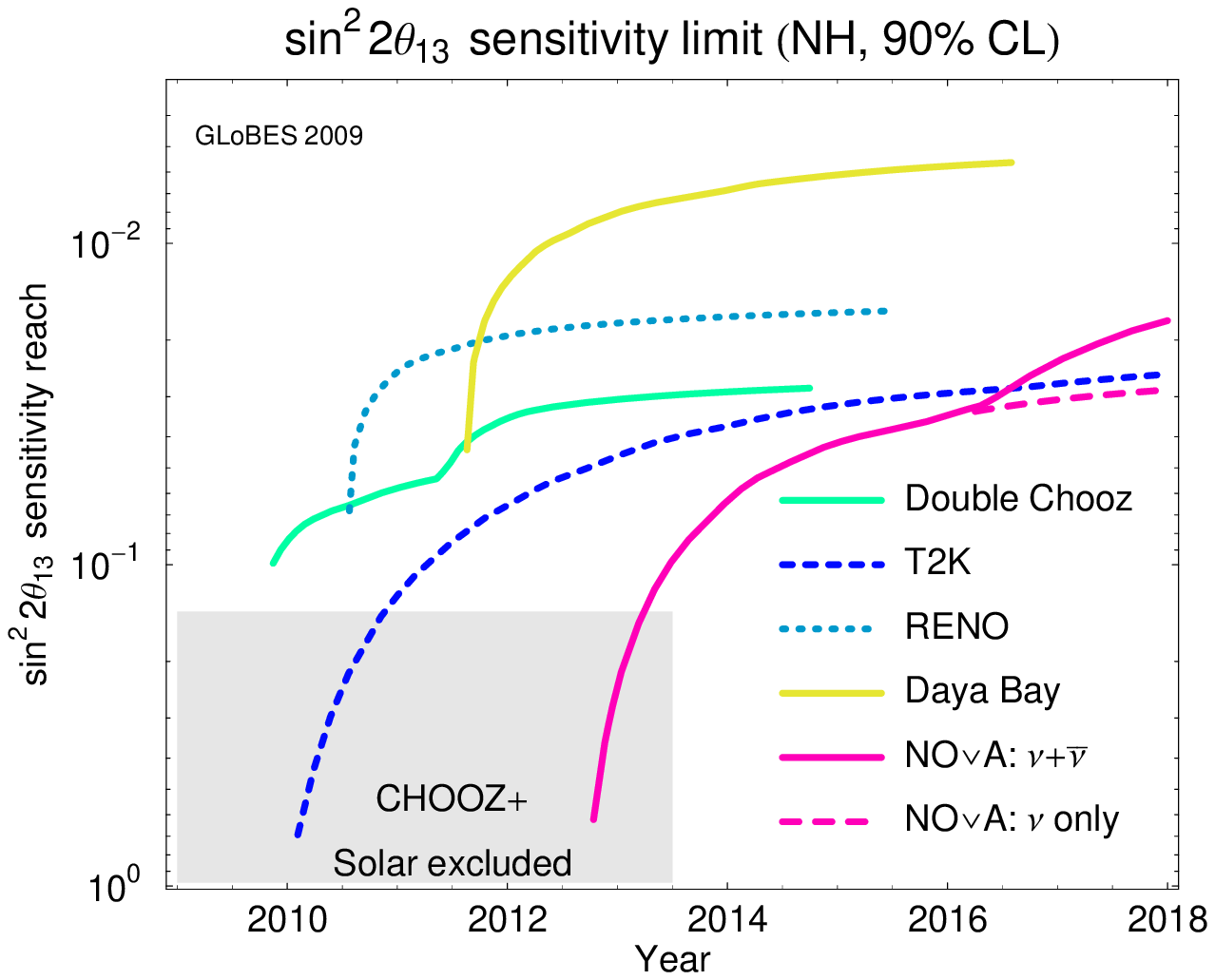}\hspace*{0.01\textwidth}
\includegraphics[width=0.48\textwidth]{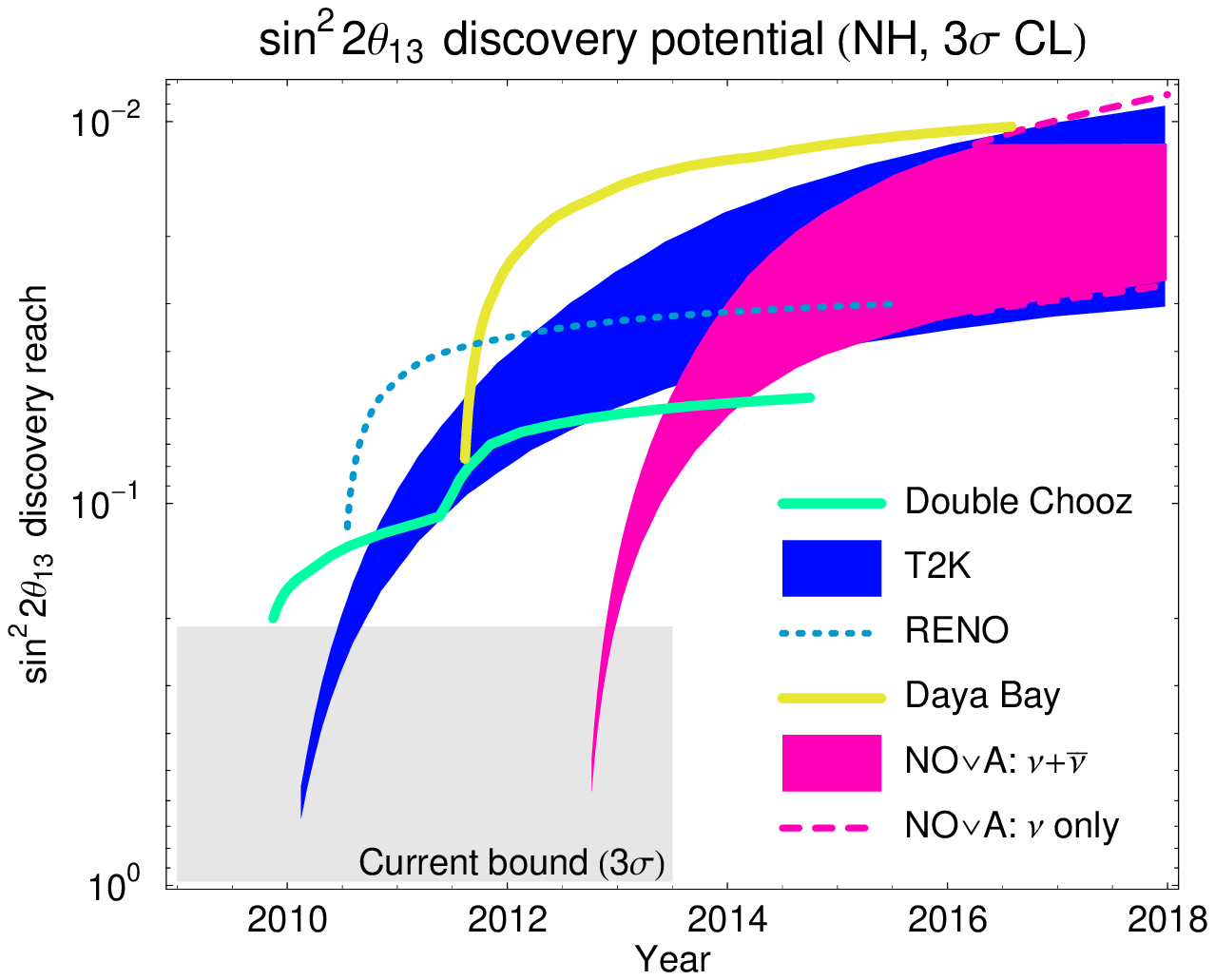} 
\end{center}
\caption{\label{fig:th13evol} Left: Evolution of the $\theta_{13}$
sensitivity limit as a function of time (90\% CL), \ie, the 90\%~CL
limit which will be obtained if the true $\theta_{13}$ is zero. Right:
Evolution of the $\theta_{13}$ discovery potential as a function of
time ($3\sigma$~CL), \ie, the smallest value of $\theta_{13}$ which
can be distinguished from zero at $3\sigma$.  The bands reflect the
(unknown) true value of $\deltacp$.  In both panels we assume normal
hierarchy. Taken from Ref.~\cite{Huber:2009cw}.}
\end{figure}

We show the $\theta_{13}$ sensitivity limit (bound on $\theta_{13}$ in
case of no signal) as a function of time in \figu{th13evol} (left). We
observe that the global sensitivity limit will be dominated by reactor
experiments. As soon as operational, Daya Bay will dominate the global
limit. For Daya Bay, time is not critical, but matching the
systematics or statistics goals is.\footnote{The Daya Bay assumptions
of a systematical error of 0.18\%, fully uncorrelated among all
detectors is more aggressive than for other reactor experiments. For
example, if the systematic error is at the level of 0.6\% (such as
assumed in Double Chooz) and uncorrelated among modules, the Daya Bay
sensitivity of $\sin^22\theta_{13}= 0.0066$ deteriorates to
$\sin^22\theta_{13}\simeq 0.01$. If on the other hand the systematic
error is $0.38\%$ (the Daya Bay ``baseline'' value) and assumed to be
fully correlated among modules at one site the limit would correspond
roughly to the one obtained for an uncorrelated error of $0.38\%
\times \sqrt{N} \simeq 0.76\%$ for $N=4$ modules at the far site. This
will lead to a limit of
$\sin^22\theta_{13}\simeq0.012$~\cite{McFarlane:2009xx}.}
If the assumed schedules of both, Double Chooz and Daya Bay are
matched, Double Chooz will dominate the $\theta_{13}$ sensitivity for
about two years in the absence of RENO. If available, RENO, on the
other hand, will dominate the $\theta_{13}$ sensitivity if it is
operational significantly before the end of 2011. As a peculiarity,
the $\theta_{13}$ sensitivity of NO$\nu$A is improved by switching to
antineutrinos. However, the global limit will at that time be
dominated by the reactor experiments.

The $\theta_{13}$ discovery potential (smallest $\theta_{13}$ which
can be distinguished from zero) is shown in \figu{th13evol} (right) as
a function of time. For the beam experiments, the dependence on the
true value of $\deltacp$ is shown as shaded region, whereas the
reactor experiments are not affected by the true $\deltacp$. There is
a small dependence on the true mass hierarchy for the beam
experiments, here we choose a true normal hierarchy.
The comparison of the left and right panels in \figu{th13evol} shows
that suitable values of $\deltacp$ may significantly improve the
discovery potential of beams compared to their sensitivity
limit. Indeed, for favorable values of $\deltacp$ the discovery reach
of beams can be similar to the one of Daya Bay, whereas the
sensitivity limit is more like the one from Double Chooz.

\bigskip

If $\theta_{13}$ is close to its present bound, and hence will be
discovered rather soon, it might be interesting to investigate,
whether ``modest upgrades'' to the proposed setups of T2K and NO$\nu$A
might allow to address the issues of CP violation and mass hierarchy
determination. With ``modest upgrades'' we mean modifications of
existing equipment and infrastructure. This includes a longer running
time and an upgraded beam power for both accelerator experiments and
the addition of antineutrino running in T2K.  To be specific, we
assume that a proton driver is installed for T2K, which increases the
beam power from 0.75 to 1.66~MW, linearly from 2015 to
2016~\cite{T2KUPGRADE}, and for NO$\nu$A we assume a linear increase
from 0.7 to 2.3~MW from March 2018 to March 2019 according to
``Project~X''~\cite{PROJECTX}.
We consider these upgraded beams for T2K and NO$\nu$A combined with
reactor data, and we have performed a global optimization for
the switching between neutrinos and antineutrinos in both beams.

\begin{figure}[tp]
\begin{center}
\includegraphics[width=\textwidth]{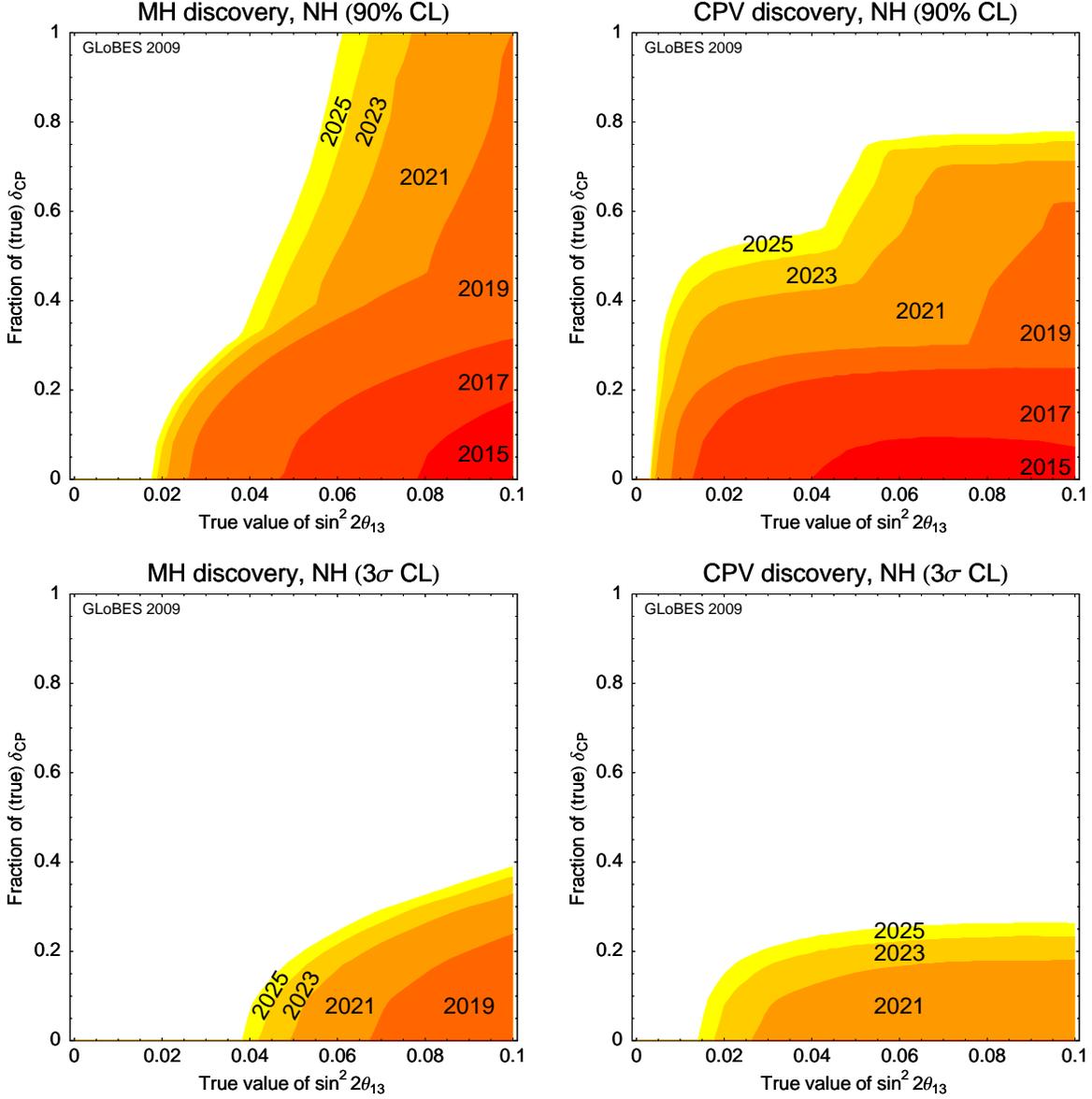}
\end{center}
\caption{\label{fig:tslUPGR} Mass hierarchy (left panels) and CP
  violation (right panels) discovery potentials as a function of true
  $\stheta$ and fraction of true $\deltacp$ for T2K+NO$\nu$A
  (including beam upgrades and global $\nu/\bar\nu$-optimization) and
  reactor experiments. The upper panels are for 90\% CL, the lower
  panels for $3 \sigma$ CL.  The different shadings corresponds to
  different points of time. Taken from Ref.~\cite{Huber:2009cw}.}
\end{figure}

\figu{tslUPGR} shows the discovery potential as a function of true
$\stheta$ and fraction of true $\deltacp$ for times from 2015 to 2025.
From the upper row of this figure we conclude that at the 90\%
confidence level, there will be hints for the MH and CPV for $\stheta
\gtrsim 0.05$ for most values of $\deltacp$ around 2025.
However, certainly a 90\%~CL is not sufficient to make any meaningful
statement about a discovery. Therefore, we show in the lower row of
\figu{tslUPGR} the corresponding results at $3\sigma$~CL. Obviously
the sensitivity regions reduce drastically, however, we see from the
figure that assuming both beams upgraded, a fully optimized
neutrino/antineutrino run plan, and data from reactors a
non-negligible discovery potential at $3\sigma$ will be reached in
2025. The mass hierarchy can be identified for $\stheta \gtrsim 0.05$
for about 20\% to 40\% of $\deltacp$ values, whereas CPV can be
discovered for $\stheta \gtrsim 0.02$ for 25\% of $\deltacp$
values. In both cases, MH and CPV, there is sensitivity for values of
$\deltacp$ around $3\pi/2$ ($\pi/2$) if the true hierarchy is normal
(inverted). This is related to the sign of the matter effect, see,
\eg, \Ref~\cite{Winter:2003ye} for a discussion.

Although ``minor upgrades'' of existing facilities may provide a
non-negligible sensitivity to CP violation and the mass hierarchy,
there is high risk associated with this strategy, since for about 75\%
of all possible values for $\deltacp$ no discovery will be possible at
the 3$\sigma$ level. Therefore, we conclude that the upcoming
generation of oscillation experiments may lead to interesting
indications for the mass hierarchy and CP violation, but it is very
likely that an experiment beyond the upcoming superbeams (including
reasonable upgrades) will be required to confirm these hints.

\section{New tools: MonteCUBES}
\label{sec:mcubes}




Many models of new physics that have been discussed in the context of neutrino physics, and in particular long-baseline neutrino oscillation experiments, expand the dimensionality of the neutrino oscillation parameter space. Since the most general standard neutrino oscillation parameter space already includes six free parameters, it is essentially impossible to perform a deterministic analysis of the full parameter space when also including new physics. The common practice has been to fix all of the new physics parameters except for one or two as well as several of the standard oscillation parameters. While this can give a hint at the dependence on the new physics, specific features requiring the interplay of several parameters may be lost. Examples of the new physics are non-standard neutrino interactions (NSI) and non-unitarity of the lepton mixing matrix.

The Monte Carlo Utility Based Experiment Simulator (MonteCUBES) software \cite{Blennow:2009pk,mcubeshome} was introduced to address this issue. The software is a plugin for the General Long Baseline Experiment Simulator (GLoBES) \cite{Huber:2004ka,Huber:2007ji} and compiles and installs on computers where GLoBES has already been installed.

\subsection{The MonteCUBES plugin}

MonteCUBES has been designed to sample the neutrino oscillation parameter space using Markov Chain Monte Carlo (MCMC) mehods, which makes it ideal for studying high-dimensional parameter spaces. We will here give a short introduction to MCMCs before discussing the features included in MonteCUBES.

\subsubsection{Markov Chain Monte Carlos}

A MCMC is designed to create a representative sample of a probability distribution $P(x)$, where $x$ represents a set of parameters, using stochastic methods. The most common implementation of this is the Metropolis--Hastings algorithm, which is defined as follows:
\begin{enumerate}
\item Pick a starting point $\bar x$ in parameter space.
\item Using a transition probability function $W(\bar x \to x')$, pick a test point $x'$.
\item Compute the acceptance probability $\alpha$ as
 \[
  \alpha = \min\left(1, \frac{P(x')}{P(\bar x)}\frac{W(x'\to \bar x)}{W(\bar x \to x')}\right).
 \]
\item Pick a random number $r$ in $[0,1]$.
\item If $r < \alpha$, then put $\bar x = x'$.
\item Add $\bar x$ to the list of samples.
\item Repeat steps 2-6 until a sufficiently large number of samples has been produced.
\end{enumerate}
The particular choice of $\alpha$ implies that $P(x)$ is the equilibrium distribution.


The advantage of using a MCMC instead of using deterministic algorithms for exploring large parameter spaces is that, when analyzing an experiment with a high-dimensional parameter space with a classic method, a lot of computer power will be spent on computing $\chi^2$ functions or likelihoods in regions which are not of physical interest.
Instead, the MH algorithm is designed to mainly explore the physically interesting region also in high-dimensional parameter spaces (the breaking point where the MH algorithm becomes more suitable is generally around $n = 4$).

\begin{figure}
\begin{center}
\includegraphics[width=0.48\textwidth]{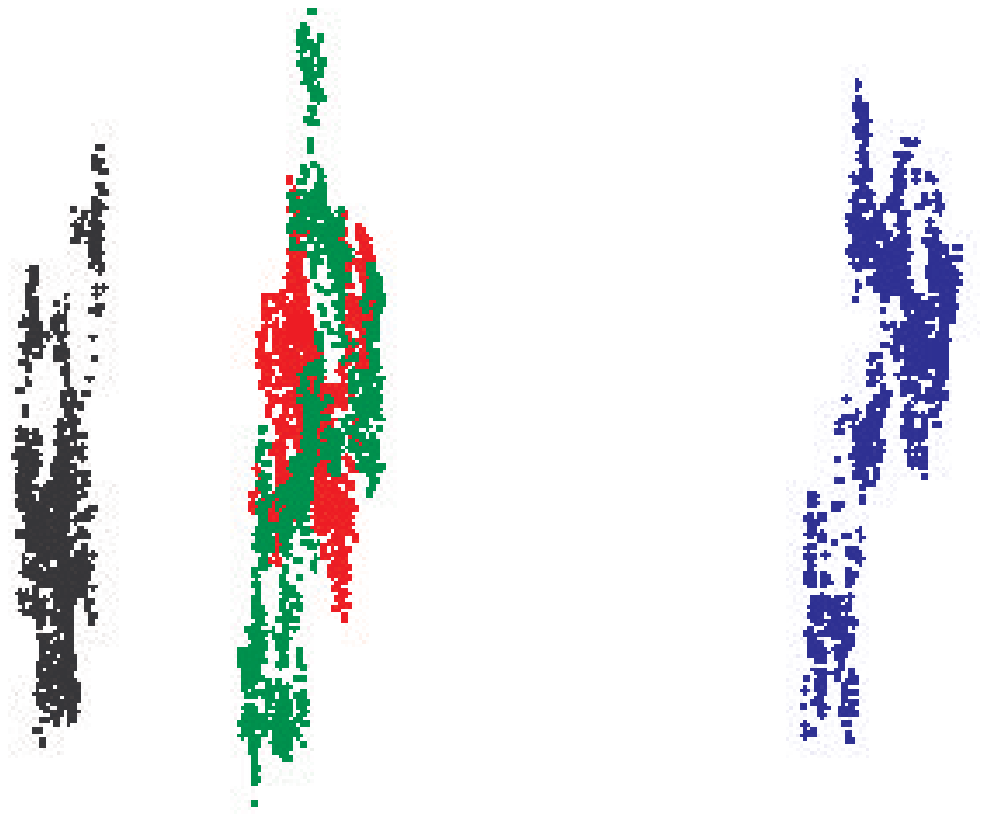} 
\includegraphics[width=0.48\textwidth]{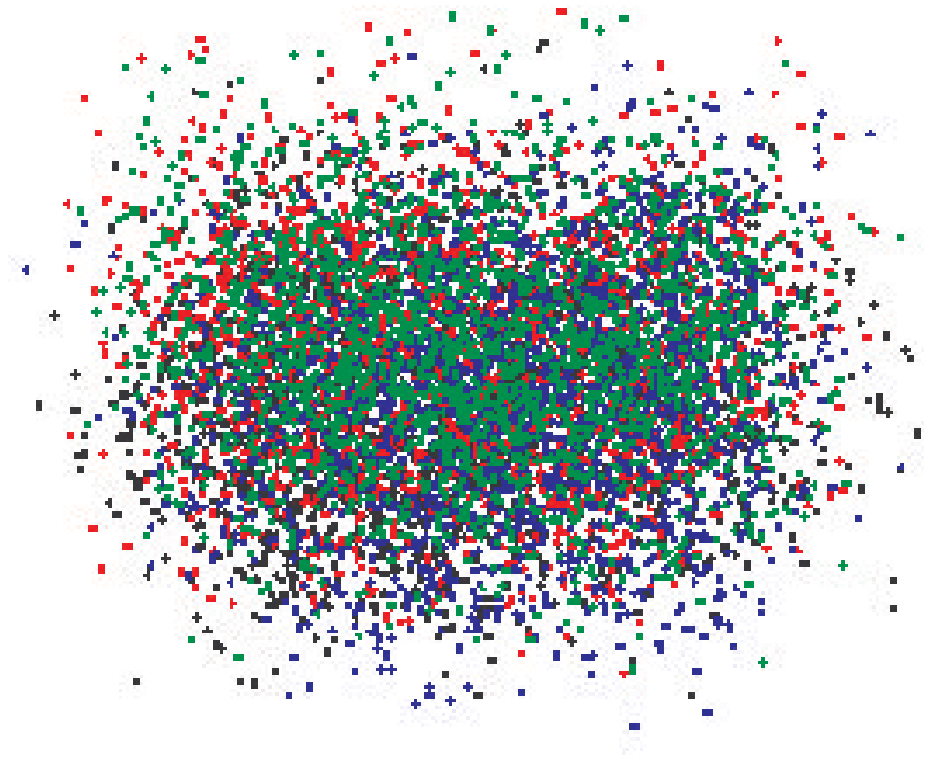}
\caption{\label{fig:convergence} Samples scattered in parameter space for chains with bad (left) and good (right) convergence, respectively. 
See Ref.~\cite{mcubeshome} for details.}
\end{center}
\end{figure}

In step 7 of the description of the MH algorithm, we have stated that the algorithm should be repeated until a sufficient number of samples has been reached. It is therefore natural to define what is ment by this. Of course, the number of samples must be high enough so that a fairly good approximation of the probability distribution can be reconstructed. However, even though $P(x)$ is the equilibrium distribution, it can take some time to reach this equilibrium. As an example, in \Fig~\ref{fig:convergence}, we show four chains that have sampled the same distribution with both bad and good convergence.
In the case of bad convergence, the chains are separate, while in the case of good convergence, they all have more or less the same distribution. In MonteCUBES, there are built-in convergence checks, based on running several different MCMCs in parallel and comparing the variance within one chain with the variance in the combined sample (see \Ref~\cite{Gelman:1992zz}). This is done in run-time and the simulations can be run for as long as it takes to converge, or until a maximum number of samples has been reached.


The most straightforward use of MCMCs for experimental analysis is through the use of Bayesian statistics. We use the MCMC to sample the posterior probability distribution $P(\theta|d)$, where $\theta$ is a parameter and $d$ a data set, which according to Bayes' theorem can be written as
\begin{equation}
 P(\theta|d) = \frac{P(d|\theta)P(\theta)}{P(d)} \equiv \frac{L_d(\theta)\pi(\theta)}{M},
\end{equation}
where $L_d(\theta)$ is the likelihood of getting the data $d$ given the oscillation parameters $\theta$, $\pi(\theta)$ is a prior which summarizes our previous knowledge (or assumptions) on the distribution of $\theta$, and $M$ is a normalization constant.

Once the posterior probability distribution has been sampled, it can be projected onto any subspace of the parameters by simply ignoring the values of other parameters. The Bayesian regions of the $p$ most likely parameter values are then defined as the smallest regions containing the probability $p$. Note that this differs from the frequentist confidence regions at level $p$, which are the regions of parameter space such that the actual experimental outcome is among the $p$ less extreme outcomes\footnote{Here, ``less extreme'' refers to a predefined ordering of the outcomes, such as double-sided intervals, upper/lower bounds, or Feldman-Cousins ordering \cite{Feldman:1997qc}.}.

\subsubsection{The C-library features}

The C-library part of MonteCUBES provides all the methods necessary in order to perform MCMC simulations for neutrino oscillation experiments using the same experimental definition files as GLoBES. MonteCUBES uses a significant amount of the GLoBES functionality, which means that MonteCUBES can be used with any previously defined experimental definitions.

From a technical perspective, the basic features of MonteCUBES are very easy to use and provides a simple Metropolis sampling with a gaussian step proposal function $W(x\to y)$. It also provides several options for the checking of convergence and burn-in\footnote{The ``burn'' of a MCMC is a number of samples in the beginning of the simulation that are thrown away. This ensures that the actual chains will start closer to equilibrium, which will result in faster convergence.}.

\subsubsection{The graphical interface}

In addition to the C-library features, MonteCUBES is distributed along with a graphical user interface (GUI) for Matlab. The GUI provides methods for converting the C-library output into plots for diagnosis of the chains and for displaying the actual physical results. A perl script for users who do not have access to Matlab is under development.

\begin{figure}\begin{center}
 \includegraphics[width=0.48\textwidth]{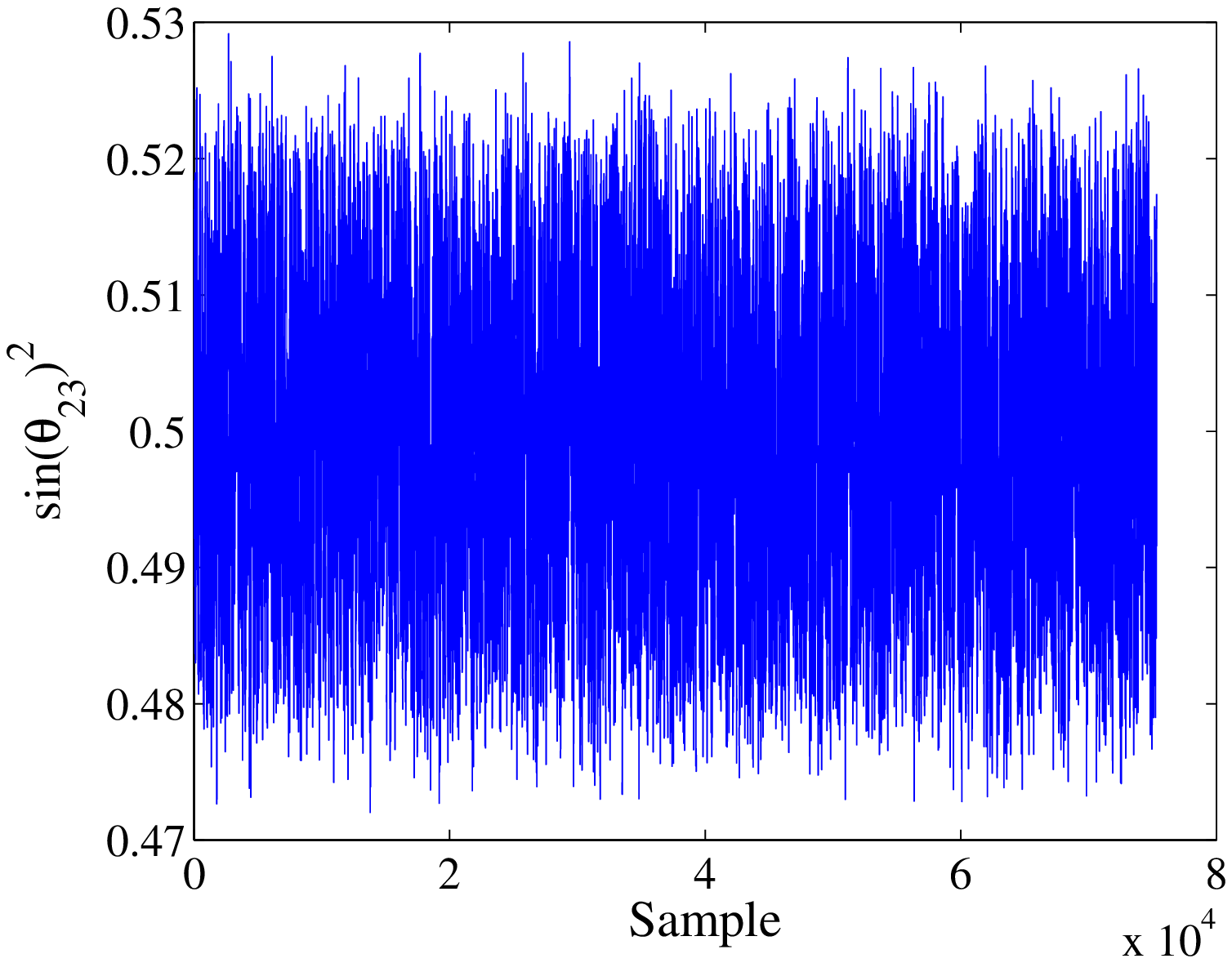} \includegraphics[width=0.48\textwidth]{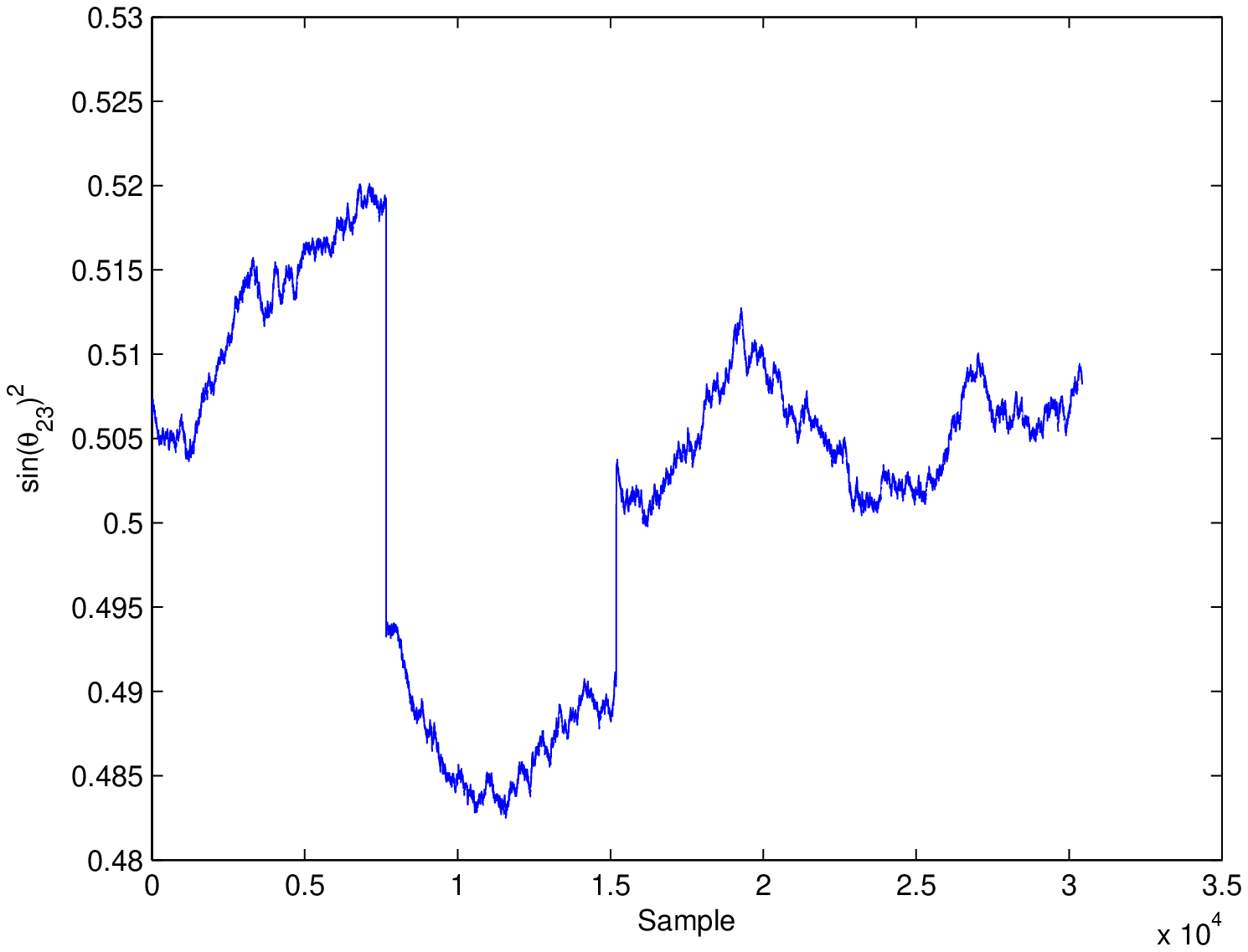} 
 \caption{\label{fig:chprog} Examples of one-dimensional chain progression plots for chains with good (left) as well as bad (right) convergence. See Ref.~\cite{mcubeshome} for details.}
 \end{center}
\end{figure}

Although MonteCUBES has the possibility to check convergence of the chains in run-time, it can be of interest to also check them manually using visual aids. For this purpose, the GUI provides two plotting options, the one-dimensional chain progression (see \Fig~\ref{fig:chprog}) and the two-dimensional parameter scatter (similar to \Fig~\ref{fig:convergence}).
This type of plots can also be useful to tune the typical step size needed in the step proposal function $W(x\to y)$ in order to give fast convergence. Typically, the step size should be set to about the size of the region that is being explored by the chains.


The GUI also provides several possibilities for displaying the results of simulations. The basic type of plot is that of the most likely parts of the posterior probability distribution in one, two, or three dimensions.
Apart from simply plotting the results against the parameters used in the simulation (such as $\theta_{23})$, the GUI provides the possibility to plot the results against an arbitrary function of the parameters (such as $\sin^2(2\theta_{23})$). The GUI also includes the possibility of applying post-simulation priors through a weight function.

\subsection{Advanced features}

In addition to the basic functionality provided by MonteCUBES, it also contains several advanced features. First, the MCMC itself is highly customizable, allowing an advanced user to use essentially any step proposal function. As an intermediate step to this is the use of degeneracy steps, which provides the user with a powerful tool for exploring degenerate solutions in the same simulation (using only gaussian steps, this would require very long simulations - especially if the degeneracies are well separated).

MonteCUBES is also distributed along with two additional features not related to the MCMC simulations. First, it includes the GLoBES implementation of two scenarios of new physics, NSI and non-unitarity. The second feature is the possibility to set the experimental event rates explicitly, which can be useful for analyzing actual data.


We have summarized the main features of MonteCUBES, a plugin for the simulation software GLoBES that provides the possibility to analyze neutrino oscillation experiments using MCMC methods. While doing so, we have discussed the advantages of using stochastic methods for exploring high-dimensional parameter spaces and why this is particlularly applicable to new physics in neutrino oscillations. Since MonteCUBES is a plugin for GLoBES, it can use any previously written experiment definition files and while the plugin itself is very simple to use, users already familiar with GLoBES will find it particularly easy to get started.

\section{Review of Nufact Physics Potential}
\label{sec:nufact}

\subsection{Sterile Neutrinos at a Nufact}
\label{sec:sterile}

In the last ten years,  extensive work by many different collaborations has been devoted to the study of the potential of a Neutrino Factory facility to measure the parameters of the standard three-family PMNS mixing matrix that are still unknown
(e.g., the mixing angle $\theta_{13}$; the leptonic CP-violating phase $\delta$; 
and  the hierarchy between neutrino mass, i.e. the sign of the atmospheric mass difference). 
Since some consensus on the principal features of the Neutrino Factory design to be fullfilled to achieve these goals (see the Report of the {\em International Scoping Study for a Neutrino Factory and a Superbeam Facility}, 
\cite{Bandyopadhyay:2007kx}, and 
the reports of the {\em International Design Study of a Neutrino Factory} for details),
it is important to assess the potential of such a facility to study new physics beyond the standard three-family oscillations. 

Within the possible straightforward extensions of the three-family oscillation model, we have studied in 
Ref.~\cite{Donini:2008wz}
the inclusion of one new light singlet fermion and the corresponding sensitivity of a Neutrino Factory facility
to the four-family PMNS mixing matrix between three active neutrinos and the {\em sterile} one.
The possible existence of a fourth light neutrino state was advanced to explain the LSND experiment data, 
which were consistent with $\nu_\mu \to \nu_e$ oscillations driven by a $\Delta m^2 \sim 1$ eV$^2$ mass difference, 
with an effective mixing angle $\sin^2 2 \theta_{LSND} \sim 10^{-2}$ \cite{LSND}. 
However, the MiniBooNE experiment (that was designed to check the LSND data) has found a negative
result in the region of interest \cite{AguilarArevalo:2007it}. 
If we discard the possibility that O(1) eV singlet fermions could exist with an effective mixing angle 
$\theta_{LSND} \sim 3^\circ$, it is still interesting to ask which is the ultimate sensitivity of existing, planned or 
future facilities to the parameter space of models with extra light singlet fermions. 

Preliminar studies of four-neutrino models were performed at early stages of the Neutrino Factory optimization, 
\cite{Donini:1999jc,Donini:2001xy,Donini:2001xp}, albeit using an ideal detector. 
A detailed study of the potential to study the four-neutrino model at the CNGS beam using the OPERA detector
has been presented in Ref.~\cite{Donini:2007yf}.

\subsubsection{The setups}

Notice that this study can be performed either using the ISS/IDS design (optimized to look for three-family oscillation
signals) or to look for modification of the baseline design that could improve the potential of the facility to new 
physics while keeping most of its potential to look for standard observables such as $\theta_{13}$ and $\delta$, 
that would represent an unavoidable background to the search for deviations from the three-family PMNS model. 

For this reason, we have compared two Neutrino Factory setups: 
\begin{enumerate}
\item ISS/IDS-inspired Neutrino Factory

It consists of a Neutrino Factory with 20 GeV muons stored into two storage rings aiming at two 
different baselines, with 50 Kton magnetized iron detectors of the MIND-type \cite{Abe:2007bi} 
located at $L = 4000$ Km and $L = 7500$ Km, respectively. The muons that decay in the straight section 
of the storage rings aiming at the detectors (useful muons) are
$5 \times 10^{20}$ per year per baseline (i.e., a total of $1 \times 10^{21}$ useful muon decays per year).
\item High-energy Neutrino Factory

In this case, we consider 50 GeV muons, again  stored into two storage rings aiming at two 
different baselines, with 50 Kton magnetized iron detectors of the Hybrid-MIND-type \cite{Abe:2007bi}
located at $L = 3000$ Km and $L = 7500$ Km, respectively. The Hybrid-MIND detector consists of 
a 4 Kton magnetized Emulsion Cloud Chamber (MECC) with iron plates intertwined with emulsion layers
next to a 50 Kton MIND-type detector. The MECC section of the detector is needed to detect $\tau$ produced
in $\nu_e \to \nu_\tau$ ("silver channel") and $\nu_\mu \to \nu_\tau$ ("discovery channel"). 
The muons that decay in the straight section of the storage rings aiming at the detectors (useful muons) are
$2 \times 10^{20}$ per year per baseline (i.e., a total of $4 \times 10^{20}$ useful muon decays per year).
\end{enumerate}

Realistic estimates of the efficiencies, of the backgrounds and of the systematic errors have been taken into account
(see Ref. ~\cite{Donini:2008wz} for details).

\subsubsection{Results}

\begin{figure}[ht]
\begin{tabular}{cc}
\hspace{-0.55cm}
\includegraphics[width=0.45\textwidth]{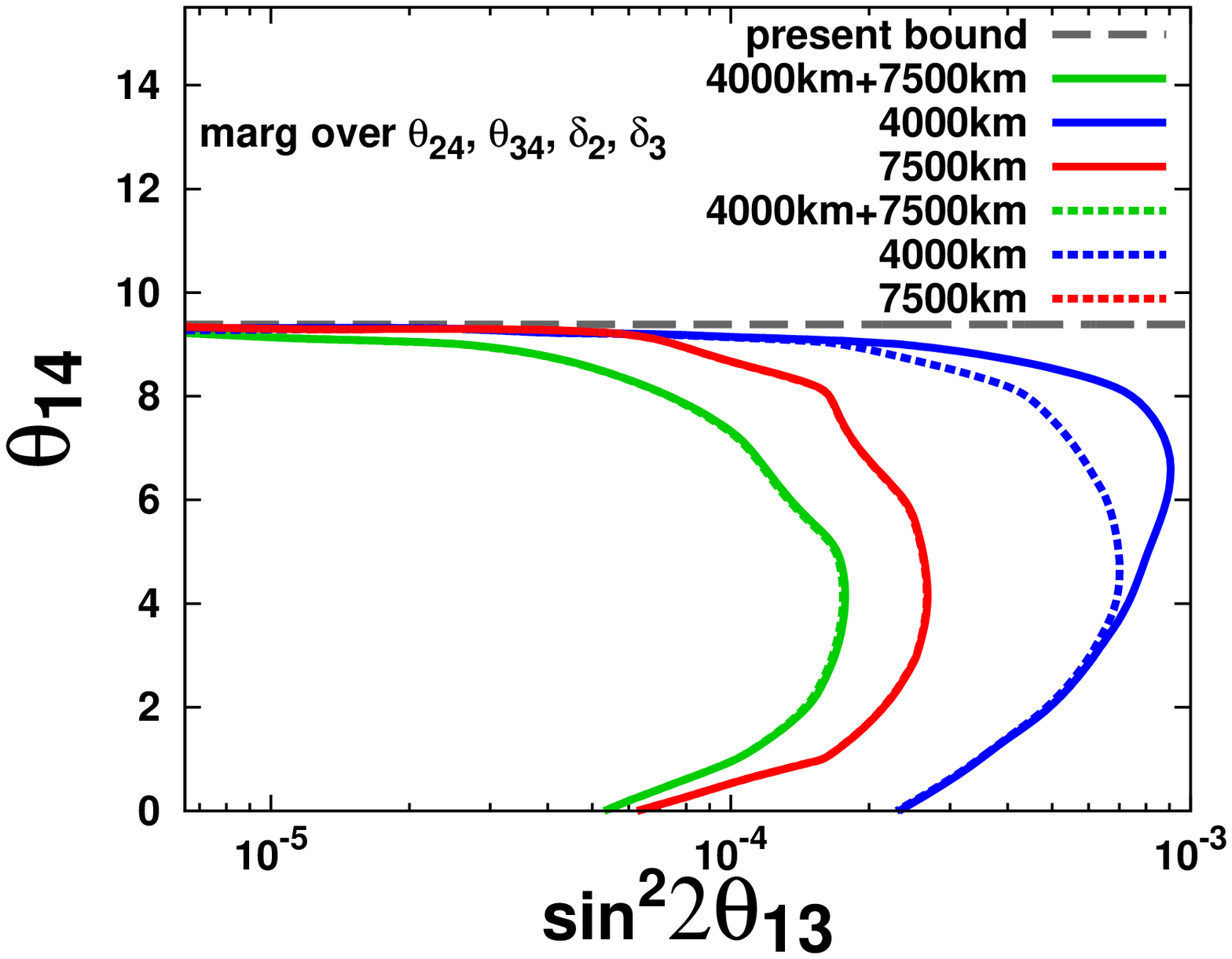} &
\includegraphics[width=0.45\textwidth]{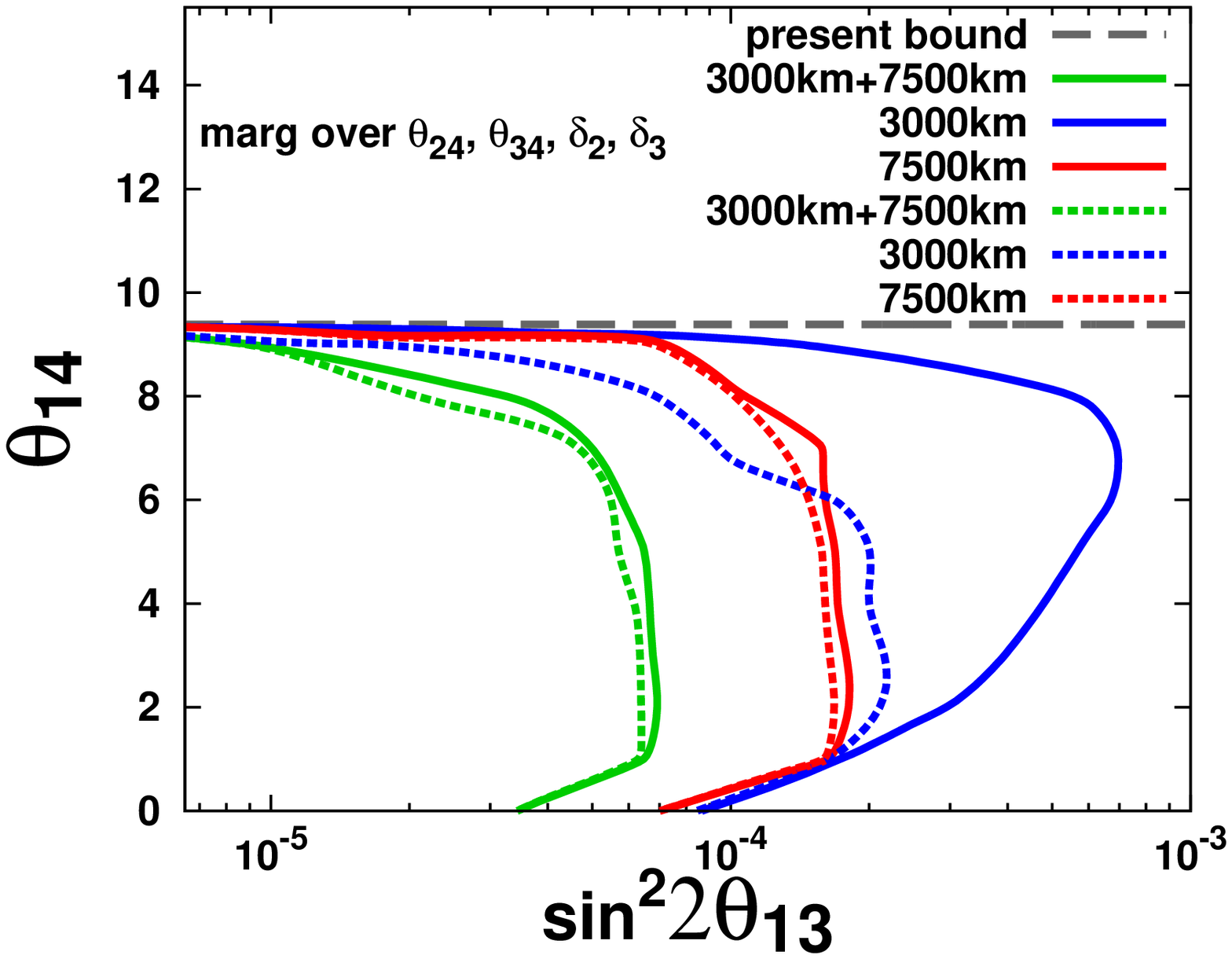} \\
\includegraphics[width=0.42\textwidth]{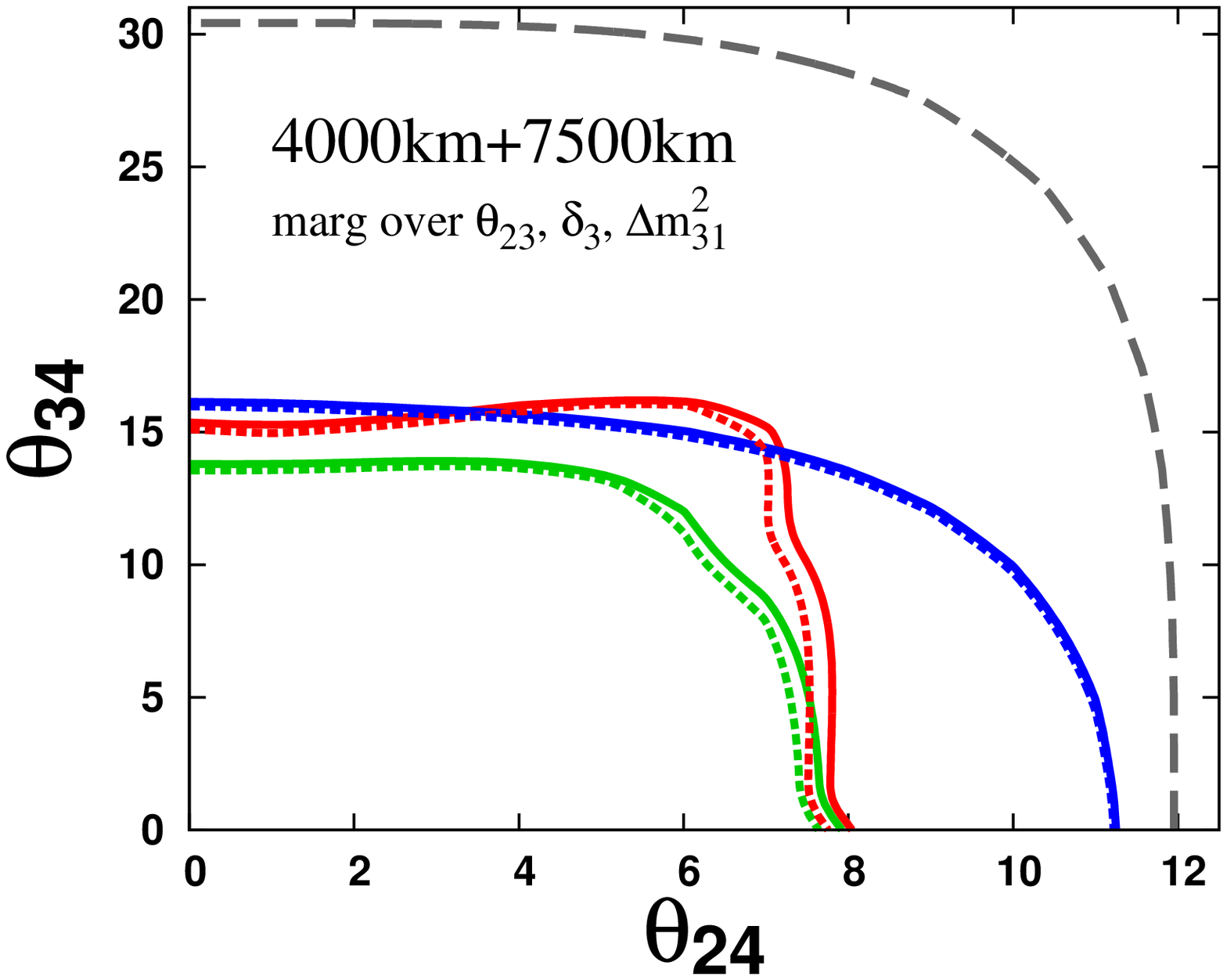} &
\includegraphics[width=0.42\textwidth]{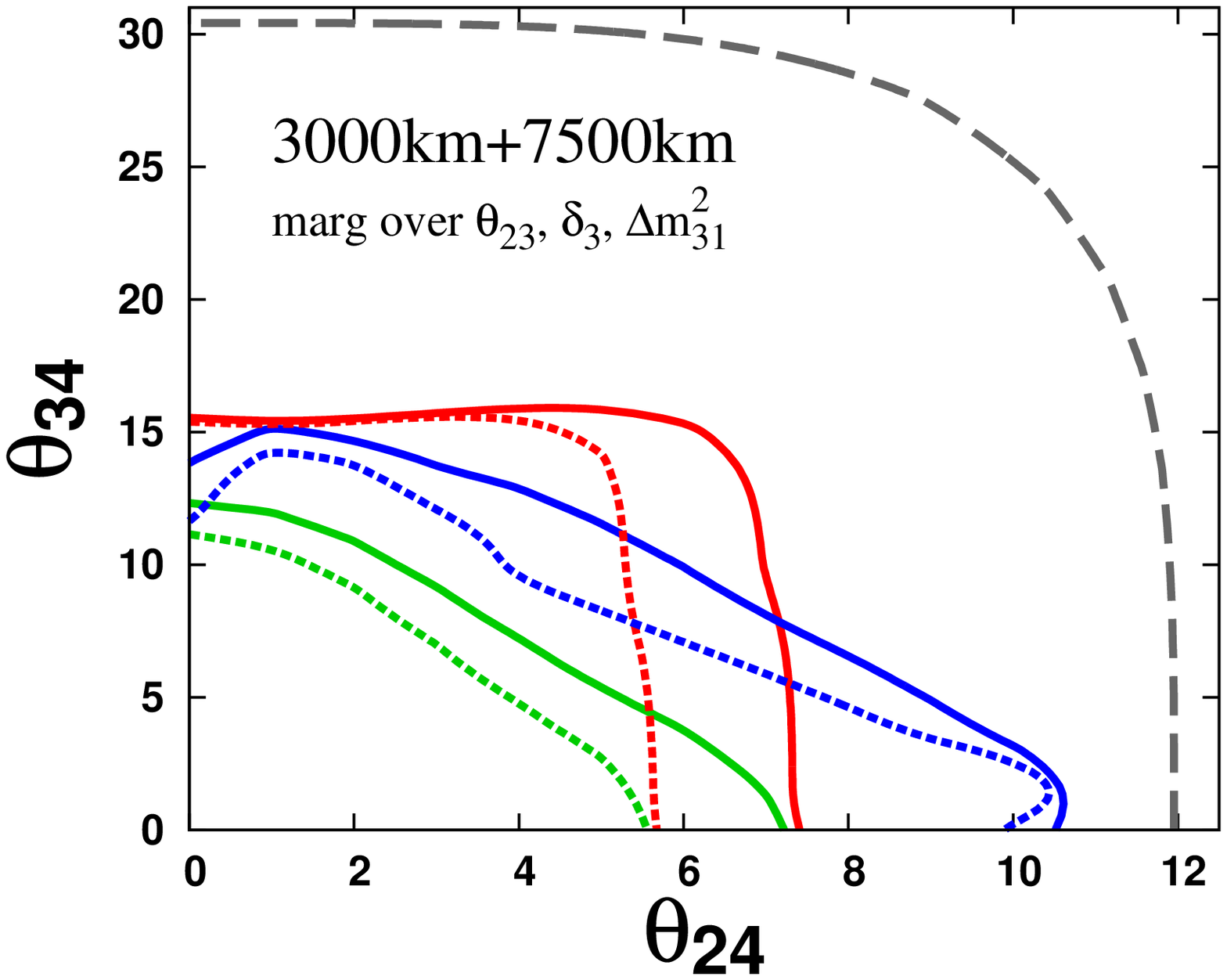} 
\end{tabular}
\begin{tiny}
\caption{\label{fig:results}%
     Sensitivity limit at 90\% CL in the ($\sin^2 2\theta_{13}$, $\theta_{14}$) plane (above) 
     and in the ($\theta_{24},\theta_{34}$) plane (below), marginalizing over the parameters not shown.
     The dashed grey line represent  the present bounds on these parameters.
     The solid lines refer to MIND data ($\nu_e \to \nu_\mu, \nu_\mu \to \nu_\mu$), only.
     Dashed lines stand for Hybrid-MIND data (combination of the previous channels with $\nu_e \to \nu_\tau$ and 
     $\nu_\mu \to \nu_\tau$ data). The colors are: blue for the shortest baseline; red for 
      longest baseline; green for the combination of the two baselines.
      Left panels: 20 GeV NF; Right panel: 50 GeV NF. Taken from Ref.~\cite{Donini:2008wz}.
            }
\end{tiny}
\end{figure}

We have studied in Ref.~\cite{Donini:2008wz} the extension of the standard three-family oscillation model with
one extra singlet fermions, much heavier than the three active ones. This model is called the "3+1" model, 
and we parametrize the four-family mixing matrix as follows: 
\begin{equation}
U_{PMNS} = U_{34} (\theta_{34}) U_{24} (\theta_{24}) U_{23} (\theta_{23}, \delta_3) 
                        U_{14} (\theta_{14}) U_{13} (\theta_{13},\delta_2) U_{12} (\theta_{12},\delta_1)
\end{equation}

The active-sterile mixing angles $\theta_{14}$ and $\theta_{24}$ are strongly bounded by electron disappearance
experiments, and $s_{14} \sim s_{24} \sim s_{13}$. The angle $\theta_{34}$ can, on the other hand, reach values
as large as $\theta_{34} \sim 30^\circ$.  It is therefore useful to expand in the following small parameters:
\begin{eqnarray}
\epsilon \equiv \theta_{34}\sim \sqrt{ \theta_{13}} \sim \sqrt{ \theta_{14}}\sim\;
\sqrt{ \theta_{24}}\sim\sqrt{\delta \theta_{23}}\;\lesssim 4\times 10^{-1} \, ,  \nonumber
\end{eqnarray}
with $\delta \theta_{23} = \theta_{23} - \pi/4$. At third order in $\epsilon$, the oscillation probabilities in matter are: 
\begin{eqnarray}   
P_{ee}&\sim& 1+O\left( \epsilon^4 \right)\,;\;\;\;\; P_{e\mu}\sim P_{e\tau}\;\sim\; P_{e s}\; \sim\;O\left(\epsilon^4 \right), 
\nonumber \\
\label{eq:ps} 
P_{\mu\mu}&=&1- \sin^2\frac{\Delta_{31} L}{2}
-2\left( A_nL\right)s_{24}\,s_{34}\cos\delta_3\sin \Delta_{31} L + O\left( \epsilon^4 \right) \; ,\\
P_{\mu\tau}&=&\left( 1-s_{34}^2 \right) \sin^2\frac{\Delta_{31} L}{2}
+\left\lbrace s_{24}\,s_{34}\sin\delta_3 +  
.2\left( A_nL\right)s_{24}\,s_{34}\cos\delta_3\right\rbrace \sin\Delta_{31}L+ O\left( \epsilon^4 \right) ,\nonumber
\end{eqnarray}
where $\Delta_{31}=\Delta m_{31}^2/2E$, and $|\Delta m^2_{31}|$ is $\Delta m^2_{atm}$, which is determined by
the two-flavor analysis of the atmospheric neutrino data.
The matter density parameter $A_n$ is $A_n = \sqrt{2} G_F n_n/2$, with $n_n$ the number density of neutrons \cite{Dziewonski:1981xy,prem}.
We can see from these oscillation probabilities that $P_{\mu \tau}$ has an $O(\epsilon^2)$ sensitivity to sterile neutrinos through the $s_{34}^2$ terms, whereas both $P_{\mu\tau}$ and $P_{\mu\mu}$ have sensitivity to the non-standard
CP-violating phase $\delta_3$ through the $O(\epsilon^3)$ term in $s_{24} s_{34}$. The potential relevance 
of the $\nu_\mu \to \nu_\tau$ channel to search for non-standard physics motivates the high-energy Neutrino Factory
setup, designed to improve the sensitivity to $\nu_e \to \nu_\tau$ and $\nu_\mu \to \nu_\tau$.
Since the $\nu_\tau N$ CC cross-section is strongly suppressed with respect to the $\nu_e N,\nu_\mu N$ CC ones at low neutrino energies due to the $\tau$ mass, a beam with high-energy neutrinos (with respect to the ISS/IDS setup) is desirable to increase the statistics in these channels. 

In Fig.~\ref{fig:results} we present  the sensitivity at 90\% CL to $\theta_{13}$ and to the active-sterile mixing angles
$\theta_{14}, \theta_{24}$ and $\theta_{34}$ in case of a null result. In the top panels, we can see that the 50 GeV
Neutrino Factory setup (right) performs a bit better than the 20 GeV one (left), and that the inclusion of silver channel data
($\nu_e \to \nu_\mu$) plays a marginal role once two baselines are considered.
In the bottom panels, on the other hand, we see that the 50 GeV facility performs much better than the 20 GeV
one (when the two baselines are combined) and that the inclusion of $\nu_\mu \to \nu_\tau$ data has a non-negligible 
impact on the sensitivities. 

\begin{figure}[t]
\begin{tabular}{cc}
\hspace{-0.55cm}
\includegraphics[width=0.45\textwidth]{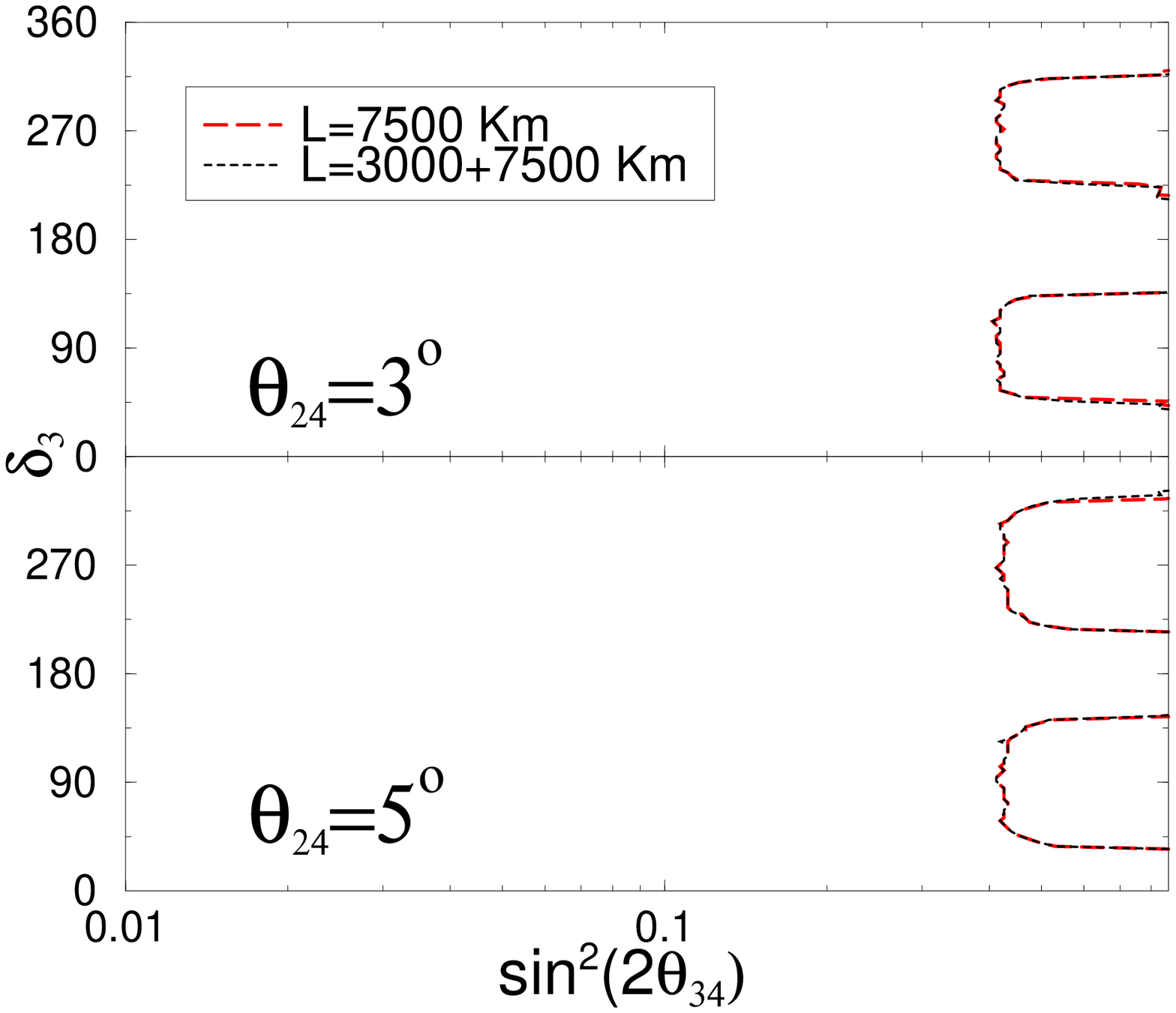} &
\includegraphics[width=0.45\textwidth]{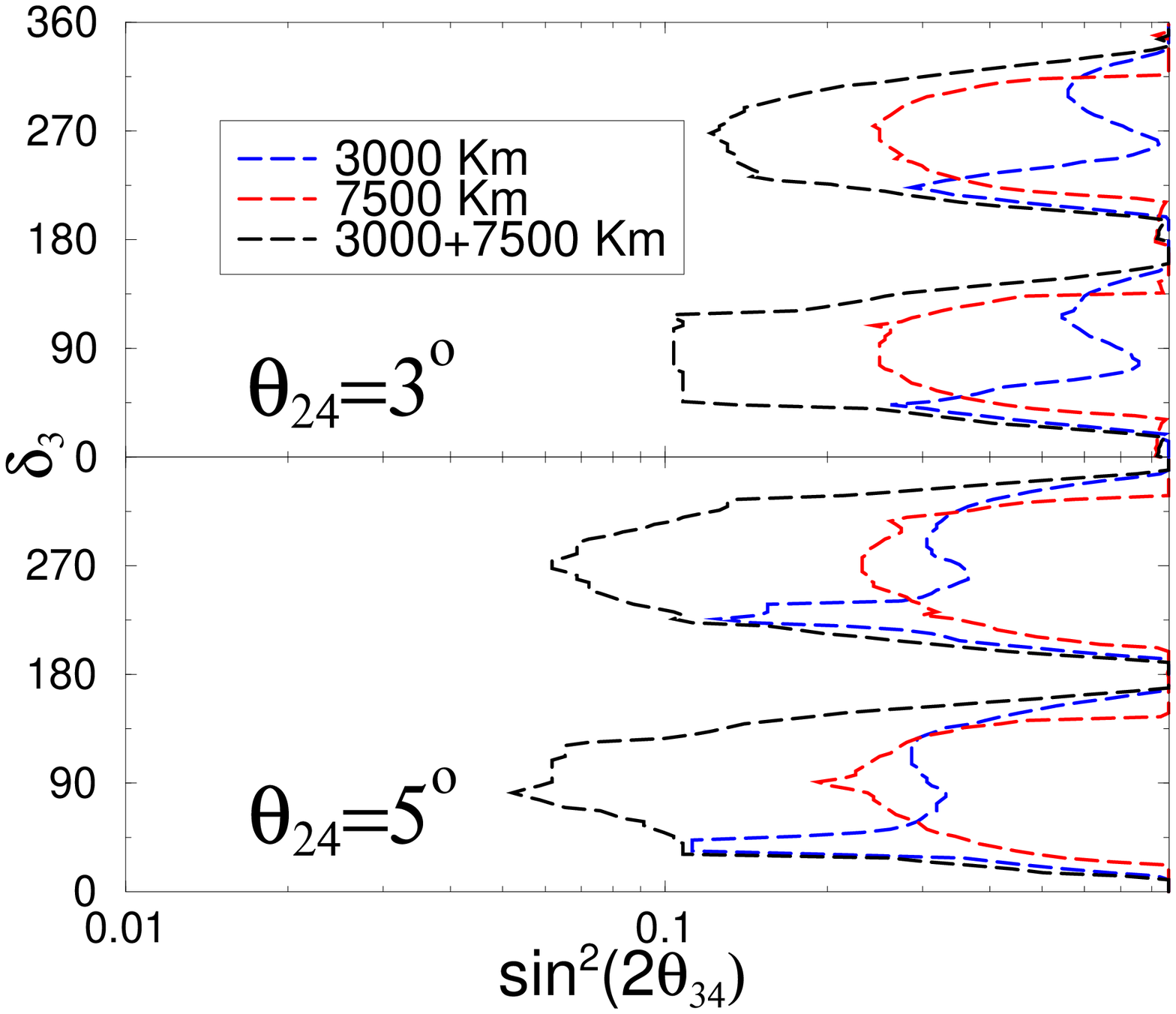} \\
\end{tabular}
\begin{tiny}
\caption{\label{fig:cp}%
     The 99\% CL $\delta_3$-discovery potential in the ($\theta_{34},\delta_3$)-plane at the 50 GeV Neutrino Factory, 
     for two values of $\theta_{24}$, $\theta_{24} = 3^\circ$ (above) and $5^\circ$ (below). 
     Left: $\nu_\mu \to \nu_\mu$ data, only; 
     Right: combination of $\nu_\mu \to \nu_\mu$ and $\nu_\mu \to \nu_\tau$ data.
     Blue lines: 3000 Km data; red lines: 7500 Km data; black lines: combination of the two baselines. Taken from Ref.~\cite{Donini:2008wz}.
         }
\end{tiny}
\end{figure}

In Fig.~\ref{fig:cp} we can see the $\delta_3$-discovery potential of the 50 GeV Neutrino Factory in case of a positive
result as a function of $\theta_{34}$ for $\theta_{24} = 3^\circ, 5^\circ$ (since the sensitivity of $P_{\mu\mu},P_{\mu\tau}$
to $\delta_3$ depends on the term proportional to $s_{24} s_{34}$, for $\theta_{24} = 0$ the sensitivity is lost).
In this case, the inclusion of the $\nu_\mu \to \nu_\tau$ data is very important: the discovery potential 
goes from $\sin^2 \theta_{34} \sim 4 \times 10^{-1}$ when only $\nu_\mu \to \nu_\mu$ data are considered (left)
to $6 \times 10^{-2}$ when $\nu_\mu \to \nu_\mu$ and $\nu_\mu \to \nu_\tau$ data are combined (right).

Summarizing, the 50 GeV (20 GeV) setup can constrain $\theta_{34} \leq 12^\circ (14^\circ)$ and $\theta_{24}\leq 5.5^\circ (8^\circ)$ through the combination of $\nu_\mu \to \nu_\mu$ and $\nu_\mu \to \nu_\tau$ data; 
$\sin^2 2 \theta^{\rm (4fam)}_{13} \leq 6 \times 10^{-5} (1.5 \times 10^{-4})$, with a slight dependence on $\theta_{14}$, through the "golden channel". We can not improve the present constraint on $\theta_{14}$ (a near detector would be particularly useful for this purpose). Eventually, the combination of $\nu_\mu \to \nu_\mu$ and $\nu_\mu \to \nu_\tau$ data
is an extremely powerful tool to look for CP-violation beyond the standard three-family one.

\subsection{Non-standard Interactions in matter at a Nufact}
\label{sec:nsi}

It is anticipated that tiny oscillation signals 
will be explored with a neutrino factory experiment 
through its high statistics and precise information 
on the energy dependence.
They are not restricted only to the standard oscillation signals,
and we have a chance to discover non-standard signals
which are an evidence of physics beyond the standard model.
A non-standard neutrino interaction (NSI) emerges
in many classes of models at the high energy scale,
and it can be described as an effective four-fermion 
interaction at the electroweak scale.
Our concern in this section is to investigate the sensitivities
to the effective NSI in the neutrino factory and to find out 
the optimal setup for the NSI search.

A neutrino oscillation experiment can be classified into 
three parts --- beam production (source), propagation,
and detection, and NSIs can enter in all three parts.
Since the sensitivities to NSIs at the source and the detection
processes strongly depend on the setup of near detectors,
the specification of the near detector setup is currently discussed
(near detectors in a neutrino factory was studied in Ref.~\cite{Tang:2009na}).
Here, we focus on the NSIs in the propagation process, which are
parametrized as the additional matter effect potential in
the neutrino propagation Hamiltonian:
\begin{align}
H
=
\frac{1}{2 E}
\left[
 U
\begin{pmatrix}
 0 & & 
 \\ 
 & \Delta m_{21}^{2} & 
 \\
 & & \Delta m_{31}^{2}
\end{pmatrix}
 U^{\dagger}
+
 a_{\rm CC}
  \begin{pmatrix}
   1 + \epsilon^{m}_{ee} 
   & \epsilon^{m}_{e\mu} 
   & \epsilon^{m}_{e\tau} 
   \\
   (\epsilon^{m}_{e\mu})^{*} 
   &
   \epsilon^{m}_{\mu\mu}
   &
   \epsilon^{m}_{\mu\tau}
   \\
   (\epsilon^{m}_{e\tau})^{*}
   &
   (\epsilon^{m}_{\mu\tau})^{*}
   &
   \epsilon^{m}_{\tau\tau}
  \end{pmatrix}
\right],
\end{align}
where $a_{\rm CC}$ is the standard matter effect potential which
is defined as $a_{\rm CC} = 2 \sqrt{2} G_{F} n_{e} E$ with 
the electron number density $n_{e}$.
The experimental bounds to $\epsilon^{m}_{\alpha \beta}$
have been recently re-discussed in Refs.~\cite{Biggio:2009kv,Biggio:2009nt},
and the constraint from the loop-induced charged lepton flavour 
violating processes are revised.
In Ref.~\cite{Biggio:2009kv}, 
the authors concluded that there were no model-independent 
bounds from the loop-induced processes, and  
the bound to $|\epsilon^{m}_{e\mu}|$ is considerably 
relaxed according to the revision.
The bounds from the current experimental data are summarized 
in Ref.~\cite{Biggio:2009nt}, which are
\begin{gather}
|\epsilon^{m}_{ee}| < 3.8,
\qquad 
|\epsilon^{m}_{\mu \mu}| < 6.4 \times 10^{-2},
\qquad
|\epsilon^{m}_{\tau \tau}| < 21,
\nonumber \\
|\epsilon^{m}_{e\mu}| < 3.3 \times 10^{-1},
\qquad
|\epsilon^{m}_{e \tau}| < 3.1,
\qquad
|\epsilon^{m}_{\mu \tau}| < 3.3 \times 10^{-1},
\label{eq:bound}
\end{gather}
at the 90 \% confidence level.
When one assume some {\it naturalness conditions}, 
$|\epsilon^{m}_{e\mu}|$ 
is much more strongly constrained~\cite{Biggio:2009kv}.
Neutrino oscillation experiments
are expected to have a good sensitivity to 
the less constrained NSIs which are associated especially 
with the tau flavour.

\subsubsection{The setup}

The current baseline setup {\sf IDS-NF 1.0}~\cite{IDSNF}
includes total three detectors 
--- two Magnetized Iron Neutrino Detectors (MIND), 
and one Emulsion Cloud Chamber (ECC) for tau neutrino detection. 
With these detectors, we can observe 
the golden channel, the disappearance channel, and the silver channel, 
and 
the combination of the channels helps to solve the parameter
degeneracies among NSIs as in the standard oscillation parameters.
In Refs.~\cite{Ribeiro:2007ud,Kopp:2008ds}, it was shown that 
the combination of the golden and the disappearance channel with 
the MINDs placed at $L= 4000$ km and $L=7500$ km
solved the correlation between $\epsilon^{m}_{e\tau}$ and 
$\epsilon^{m}_{\tau \tau}$, which was the only NSI combination 
severely correlated in the golden channel.
Thus,
if two MINDs are included in the experiment,
one can safely study a sensitivity to each NSI separately
without an ECC.

\subsubsection{Results}

The impact of an ECC in the NSI search was discussed in 
Ref.~\cite{Kopp:2008ds}, in which
the sensitivity to $\epsilon^{m}_{e\tau}$ was investigated 
by scanning the baseline of an ECC and fixing the MINDs 
at $L=4000$ km and $L=7500$ km.
The results showed that 
the ECC was relevant only in the case that
the baseline is taken to be about 4000 km.
The optimum muon energy was also studied.
When two MINDs are fixed at $L=4000$ km and $L=7500$ km, 
the sensitivities to $\epsilon^{m}_{e\tau}$, $\epsilon^{m}_{\mu \tau}$
and $\epsilon^{m}_{\tau \tau}$ are getting higher proportional to 
the muon energy up to about 20 GeV, 
and they are saturated in the energy range of $E_{\mu} > 20$ GeV.
If we include an ECC set at $L=4000$ km, 
the sensitivity to $\epsilon^{m}_{e\tau}$ is improved
proportionally to energy even in $E_{\mu} > 20$ GeV, 
thanks to a synergetic effect.
However, the improvement is not impressive 
(less than a factor of two at $E_{\mu} = 100$ GeV),
because of the small statistics of the silver channel events.
From the simulation studies, 
we can conclude that the ECC is not a key component 
to search for NSIs in the neutrino propagation process,
when we already have two MINDs with the correct baseline configuration. 
Since a certain high energy (and also a long baseline) is an important
ingredient to search for the NSI in propagation, 
the low energy alternative~\cite{Geer:2007kn,Bross:2007ts} of a neutrino 
factory, which has been recently developed, is not preferred for 
this purpose.

The baseline of MINDs was also optimized for the NSI search 
in Ref.~\cite{Kopp:2008ds}, which showed that
the optimal setup for the standard oscillation parameters,
that is $L=4000$ km  and $L=7500$ km, is also favourable
for the $\epsilon^{m}_{e\tau}$ search.
For the $\epsilon^{m}_{\mu \tau}$ and $\epsilon^{m}_{\tau \tau}$ search,
an even longer baseline (such as a core-crossing baseline) is preferable.
However, the advantage of taking a longer baseline 
is just about a factor of two.

The robustness of the optimization for the standard oscillation parameters
with respect to the contamination of NSIs
is also an important issue.
When the additional fit of the NSI parameters is carried out,
the sensitivities to the standard oscillation parameters
are degraded somewhat. However, the optimal setup is hardly
changed from $L= 4000$ km and $L=7500$ km~\cite{Kopp:2008ds}.

In the neutrino factory with two MINDs set at $L= 4000$ km and $L= 7500$
km and with $E_{\mu} = 25$ GeV (based on {\sf IDS-NF 1.0}),
the sensitivities to the propagation NSI 
$\epsilon^{m}_{\alpha \beta}$ at the 90 \% confidence level
are listed as follows~\cite{Kopp:2008ds}: 
\begin{gather}
|\epsilon^{m}_{ee}| < 1.4 \times 10^{-1},
\qquad 
|\epsilon^{m}_{\mu \mu}| < 1.9 \times 10^{-2},
\qquad
|\epsilon^{m}_{\tau \tau}| < 1.9 \times 10^{-2} ,
\nonumber \\
|\epsilon^{m}_{e\mu}| < 3.4 \times 10^{-3},
\qquad
|\epsilon^{m}_{e \tau}| < 4.7 \times 10^{-3},
\qquad
|\epsilon^{m}_{\mu \tau}| <  1.8 \times 10^{-2}.
\end{gather}
From the comparison with Eq.~\eqref{eq:bound},
we can expect that the neutrino factory 
will improve the current bounds by one to three orders of magnitude.
The epsilon parameters of the order of $10^{-2}$ to $10^{-3}$
naively correspond to physics at the 
TeV scale~\cite{Berezhiani:2001rs,Davidson:2003ha,Antusch:2008tz,Gavela:2008ra}.

\subsection{Non-unitarity mixing at a Nufact}
\label{sec:nonunit}

Non-unitarity of the leptonic mixing
matrix is a generic manifestation of new physics in the lepton
sector. The MUV scheme~\cite{Antusch:2006vwa} provides an effective field theory extension of
the SM and is minimal in the sense that only three light neutrinos are
considered and that new physics is only introduced in the neutrino
sector, describing the relevant effects on neutrino oscillations in the
various types of models where the SM is extended by heavy
singlet fermions~\cite{Abada:2007ux}. 

In this scheme, the charged and neutral-current interactions of the neutrinos
are modified. The non-unitary leptonic mixing matrix $N$, which appears in the charged-current, interaction, contains the only additional degrees of freedom. 
Thus, instead of the three mixing angles and
three \CP-phases of the unitary PMNS leptonic mixing matrix (with only one
affecting neutrino oscillations), the non-unitary mixing matrix $N$
contains 15 parameters, out of which six are \CP-violating phases
(including two Majorana phases, which do not affect neutrino
oscillations). The goal of this work is to study the potential of a Neutrino Factory~\cite{Geer:1997iz,DeRujula:1998hd} in constraining or determining 
the whole parameter space of the MUV scheme, focusing in the new CP-phases associated. 

\subsubsection{The setup}

Our set-up is the Neutrino Factory proposed in the
International Design Study (IDS) \cite{Bandyopadhyay:2007kx,IDSNF},
which consists of $\nu_e$ and $\nu_\mu$ beams from $5\times 10^{20}$
muon decays per year per baseline. We consider a setting where the
experiment is assumed to run for five years in each polarity. The
parent muons are assumed to have an energy of 25~GeV. The beams are
detected at two far sites, the first located at 4000~km with a 50~kton
Magnetised Iron Neutrino Detector (MIND) \cite{Abe:2007bi} and a
10~kton Emulsion Cloud Chamber (ECC) for $\tau$ detection
\cite{Donini:2002rm,Autiero:2003fu}, and the second located close to
the magic baseline~\cite{Burguet-Castell:2001ez,Huber:2003ak} at
7500~km with an iron detector identical to the one at 4000~km.
In the simulations, we study the ``golden'' \cite{Cervera:2000kp}
$\nu_e \to \nu_\mu$ and $\nu_\mu$ disappearance channels in the MIND
detectors and the ``silver'' \cite{Donini:2002rm,Autiero:2003fu}
$\nu_e \to \nu_\tau$ and ``discovery'' \cite{Donini:2008wz} $\nu_\mu
\to \nu_\tau$ channels at the ECC detectors, both near and far. The detector efficiencies and backgrounds considered are described in~\cite{Antusch:2009pm}. 
We scan the complete MUV parameter
space, adding nine unitarity-violating parameters to the six standard
neutrino oscillation parameters. The scan is performed using the
MonteCUBES software \cite{Blennow:2009pk,mcubeshome}.

\subsubsection{Results}

First of all, let us parametrize the mixing matrix $N$ taking advantage from the fact that a general matrix can be
written as the product of a Hermitian matrix times a unitary
matrix. Writing the Hermitian matrix as $\mathbbm{1} + \eps$ (with
$\eps = \eps^\dagger$) and denoting the unitary matrix by $U$, we can
write \cite{FernandezMartinez:2007ms}
\begin{equation}
N = (\mathbbm{1} + \eps)\,U\;.
\label{param}
\end{equation}
where $\eps_{\alpha\beta} = |\eps_{\alpha\beta}| e^{i \phi_{\alpha\beta}}$ for $\alpha\neq\beta$. 
The present 90~\% CL bounds\footnote{Violation of unitarity can arise both in the production and in the detection processes. Coefficients parametrizing the former are usually labelled 
$\epsilon^s_{\alpha\beta}$,  the latter as $\epsilon^d_{\alpha\beta}$ . In the MUV scheme, $\epsilon^s = \epsilon^d = \epsilon$.}
 are
$|\eps_{\mu e}| < 3.5 \times 10^{-5}$, $|\eps_{\tau e}| < 8.0 \times
10^{-3}$, $|\eps_{\tau \mu}| < 5.1 \times
10^{-3}$~\cite{Antusch:2006vwa} and $|\eps_{ee}| < 2.0 \times 10^{-3}$,
$|\eps_{\mu\mu}| < 8.0 \times 10^{-4}$, $|\eps_{\tau\tau}| < 2.7 \times
10^{-3}$~\cite{Antusch:2008tz}.

Before to present the more relevant results of our simulations, we would like just to stress the importance of studying the ``zero-distance effect'' with near detectors. 
If the flavour basis is not orthogonal, it translates to a baseline-independent term in the oscillation probabilities:
\begin{equation} P_{\alpha
  \beta}(L=0) = 4 |\eps_{\alpha \beta}|^2 + \mathcal{O}(\eps^3)\,,\;\;\;\; \alpha\neq \beta
\end{equation}

This term is best probed at short distances, since the flux
is larger and it cannot be hidden by the standard oscillations. Near detectors are, thus, excellent for probing the zero-distance effect, in
particular $\tau$ detectors are of importance, since the present
bounds on $\eps_{\mu e}$ and $\eps_{\mu \mu}$ are rather strong. We
will therefore study the impact of near $\tau$ detectors of different
sizes located at 1~km from the beam source. In particular, we will
present all the results for near detector sizes of 100~ton, 1~kton,
and 10~kton, as well as the results without any near $\tau$ detector. 

One of the
most remarkable features resulting from our simulations is that the results do not contain
significant correlations between any of the unitarity-violating
parameters, nor are the unitarity-violating parameters significantly
correlated with the standard neutrino oscillation parameters. The only
exception are some mild correlations between $\theta_{13}$, $\delta$
and the modulus and phase of $\eps_{\tau e}$ in the absence of
near $\tau$ detectors which, however, do not lead to new degeneracies
between these parameters or spoil the determination of $\theta_{13}$
and $\delta$ at the Neutrino Factory. Furthermore, the addition of a
near $\tau$ detector of only 100~ton is enough to almost completely
erase these correlations.  This implies that the Neutrino Factory
setup considered here has enough sensitivity to distinguish the
effects induced by unitarity-violation from changes in the standard
parameters. Second, the sensitivities of the Neutrino Factory to the
diagonal parameters of the $\eps$ matrix, as well as to $\eps_{\mu
  e}$, do not improve with respect to the bounds derived from
electroweak decays, which are too stringent to allow for observable
effects at the Neutrino Factory.  
In addition, we find that a near $\tau$ detector with a mass as small as
100~ton would dominate the sensitivity to $\eps_{\tau e}$ through the
measurement of the zero-distance effect, providing sensitivities down
to $\mathcal O(10^{-3})$.


\begin{figure}
 \begin{center}
  \includegraphics[width=0.49\textwidth]{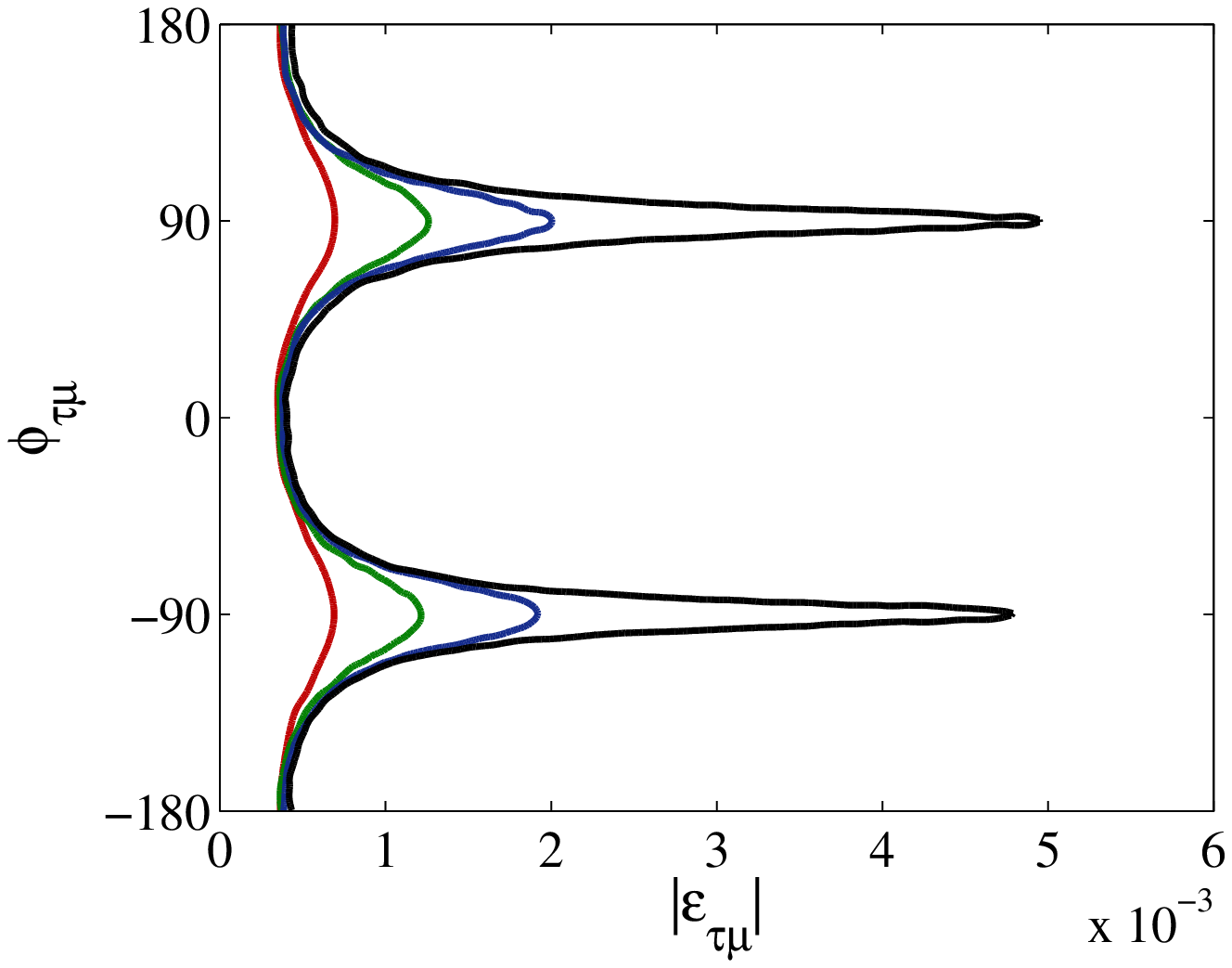}
   \includegraphics[width=0.49\textwidth]{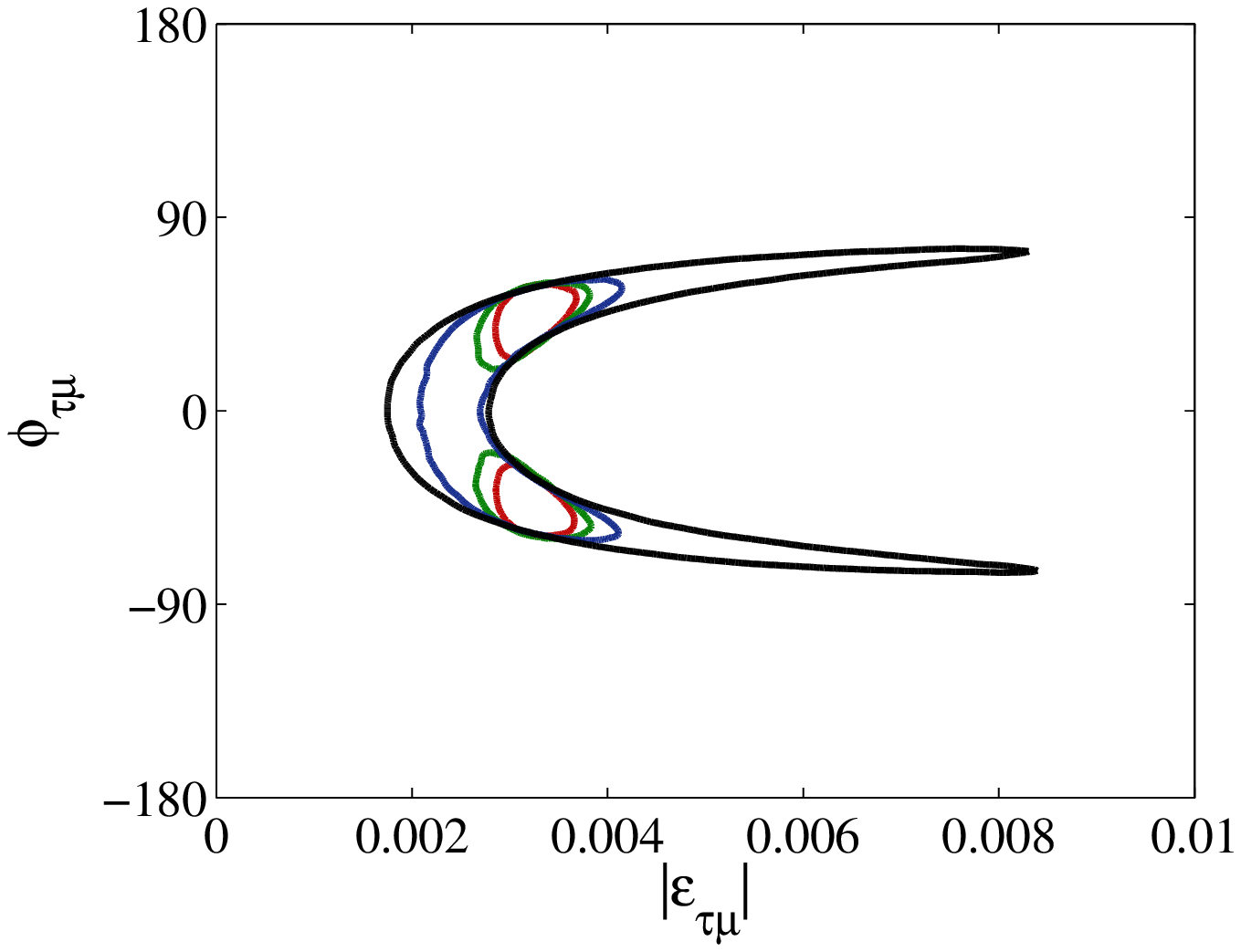}
 \end{center}
 \caption{The 90~\% confidence level sensitivity of the IDS Neutrino
   Factory to the unitarity-violating parameters $\eps_{\tau\mu}$
   (left), and sensitivity assuming that it takes the
   value $\eps_{\tau\mu} = 3.2\times 10^{-3} \exp(i\pi/4)$ (right). The different curves correspond
   to different sizes of the near $\tau$ detector, from left to right,
   10~kton, 1~kton, 100~ton, no near detector. Taken from Ref.~\cite{Kopp:2008ds}.}
 \label{fig:tm-sens}
\end{figure}
Now, we will thus concentrate on the sensitivities to $\eps_{\tau\mu}$, even though the other
unitarity-violating parameters and standard oscillation parameters are
allowed to vary in the simulations. In the left panel of \Fig~\ref{fig:tm-sens}, we show the sensitivity
to the $\eps_{\tau\mu}$ parameter for the four different sizes
considered for the near ECC.  The input values for all the
non-unitarity parameters and $\theta_{13}$ were set to zero to derive
these curves. We have checked that the results do not depend strongly
on this assumption.  The most remarkable feature of this figure is the
extreme sensitivity to the real part of $\eps_{\tau\mu}$ which is
present already without any near detector. This sensitivity mainly
originates from the matter effect on the disappearance channel, where
the leading non-unitarity correction to the oscillation probability is
given by
\begin{equation}
 \hat P_{\mu\mu} =  P_{\mu\mu}^{\rm SM}
 - 2\Real(\eps_{\mu\tau})AL\sin\left(\frac{\Delta m_{31}^2L}{2E}\right) +...\,,
\end{equation}
where $A = \sqrt{2} G_F n_e$. The terms we have omitted in the above equation, as well as expanded expressions for the rest of the probabilities, can be found in 
Ref.~\cite{Antusch:2009pm}.  Notice that the $\nu_\mu\rightarrow\nu_\tau$ channel also
depends linearly on $\eps_{\tau \mu}$ (see \cite{Antusch:2009pm}) and that the dependence is
\CP-violating. On the other hand, the mass and efficiency of the ECC
detector are much smaller compared to those of the MIND detectors for
the $\nu_\mu$ disappearance channel and therefore the sensitivity is
dominated by the latter.  As can be seen in the figure, a near $\tau$
detector will determine the modulus of $\eps_{\mu\tau}$ through the
zero-distance effect. This would translate into a vertical band in the
left panel of \Fig~\ref{fig:tm-sens} and thus the increase of the mass
of the near detector improves the measurement of the imaginary
part. However, given the linear dependence due to the matter effects
on propagation, the bound on the real part from the disappearance
channel remains stronger. We can also see that the bound on the
modulus does not require a very large near detector, the bound on the
imaginary part is essentially only improved by approximately 30~\% in
moving from a 1~kton to a 10~kton ECC detector.

Another important question is how well the Neutrino Factory would be
able to measure the unitarity-violating parameters if they are
non-zero. For this reason, in the right panel of \Fig~\ref{fig:tm-sens}, we show the
sensitivity to $\eps_{\tau\mu}$ assuming that $|\eps_{\tau\mu}| =
3.2\times 10^{-3}$ as well as $\phi_{\tau\mu} = 45^\circ$  which is disfavoured at only $1\sigma$ by
current bounds. 
Again, we can see that the sensitivity without the near detector is
only to the real part of $\eps_{\tau\mu}$. In this setting, there is a
degeneracy extending essentially as $|\eps_{\tau\mu}| \propto
1/\cos(\phi_{\tau\mu})$, along which the real part of $\eps_{\tau\mu}$
is constant and the imaginary part is changing. 
The
introduction of near detectors results in an effective measurement of
$|\eps_{\tau\mu}|$, \ie, a vertical band in the plot, which intersects
the far detector measurement giving rise to two degenerate solutions,
one for positive and one for negative phase value. Again, the
actual size of the near detector is not crucial and no significant
gain is seen beyond 1~kton.

These figures also show the strong complementarity between the near
and far detectors when it comes to measuring the phase of the
unitarity-violating parameter, and thus also a non-standard source of
\CP-violation. Neither the near nor the far detectors alone can
establish a \CP-violating phase by themselves. However, combining the
two results excludes \CP-conservation at 90~\% confidence level.

Taking into account the results just presented above, we conclude that a Neutrino Factory would provide powerful tool for
probing unitarity-violation in the leptonic mixing matrix. For the
parameters to which it is most sensitive, the sensitivity is an order
of magnitude better than the current experimental bounds. On the other hand, the
interplay between the near and far detectors would allow to test new
sources of \CP-violation in the lepton sector.

\subsection{Critical assessment on long baseline tau neutrino detection at Nufact}
\label{sec:taus}


\subsubsection{Standard oscillation physics}

The prime focus of a neutrino factory is to provide precision
measurements or tight constraints on the three-flavor oscillation
parameters.  Many studies done in the context of the
ISS~\cite{Huber:2006wb,Bandyopadhyay:2007kx} show that the potential
to achieve this is excellent if it is ensured that parameter
correlations and degeneracies can be resolved.  Any single rate
measurement at some fixed baseline $L$ and neutrino energy $E$ is
sensitive only to a \emph{combination} of parameters. To measure all
parameters separately, the following possibilities exist to resolve
the correlations and degeneracies:
\begin{itemize}
\item {\bf Use measurements at different energies.}  This is difficult
  at a neutrino factory due to the limited width of the neutrino
  spectrum and the limited energy resolution of the MIND detector.  It
  has been shown that the energy resolution of MIND is not enough for
  a single detector located at intermediate baseline to solve all of
  the degeneracies.
\item {\bf Perform measurements at two different baselines $L_1$ and
    $L_2$.}  This is an extremely powerful possibility, which is the
  reason, it is included in the current IDS-NF baseline setup. In
  particular, a measurement at the magic baseline~\cite{Huber:2003ak}
  turns out to be very important. A detailed optimization study for
  $L_1$ and $L_2$ has been performed in~\cite{Kopp:2008ds}, with the
  result that the combination $L_1 = 4\,000$~km, $L_2 = 7\,500$~km is
  optimal to study standard oscillation physics as well as
  non-standard neutrino interactions.
\item {\bf Study different oscillation channels.} With MIND detectors,
  the Golden ($\nu_e \rightarrow \nu_\mu$) and Disappearance ($\nu_\mu
  \rightarrow \nu_\mu$) channels are available, while an inclusion of
  a $\nu_\tau$ detector could in addition provide a window on the
  Silver ($\nu_e \rightarrow \nu_\tau$) and Discovery ($\nu_\mu
  \rightarrow \nu_\tau$) channels. Ref.~\cite{Kopp:2008ds}, however,
  shows that the combination of one MIND detector and one ECC at the
  intermediate baseline is not as good as the combination of two MINDs
  at two baselines, mainly because of the very low statistics at the
  $\tau$-detector for $\theta_{13} \leq 2^\circ$.  On the other hand,
  adding one ECC to the setup with two MINDs does not provide more
  than a marginal gain in sensitivity, independently of the neutrino
  energy and baseline. The reason is that the analytical expressions
  for the oscillation probabilities in the Golden and Silver channels
  are very similar (they differ only in the signs of certain terms and
  in the exchange $\sin\theta_{23} \leftrightarrow \cos\theta_{23}$ in
  several others), so that the Silver channel could help only to
  resolve degeneracies. This, however, is already done by the
  combination of the two Golden channel detectors. We, also, have
  checked numerically that also the inclusion of the Discovery channel
  does not improve the sensitivity of the neutrino factory to standard
  three-flavor oscillations.

  In Ref.~\cite{Meloni:2008bd}, the silver channel was studied to
  solve the octant degeneracy and as a tool to study deviations from
  maximality of the atmospheric angle $\theta_{23}$.  A comprehensive
  study of alternatives to the silver channel for these tasks is
  lacking, see Ref.~\cite{Donini:2005db}.  A likely outcome of such a
  study will be that alternatives are better than the silver channel.
  However, in the absence of such a study, we cannot draw a firm
  conclusion.
\end{itemize}

As for the standard three-family oscillations, we thus believe that an
ECC detector able to look for $\nu_e \to \nu_\tau$ and $\nu_\mu \to
\nu_\tau$ channels will not improve significantly the performances of
the baseline neutrino factory setup with two MINDs , due to the strong
statistical limitations of the present detector design and to the
relatively limited number of parameters to be measured.

\subsubsection{Non-standard oscillation physics}

There are several interesting cases of new physics that can be studied
through neutrino oscillation experiments.  We will address here the
potential of a detector capable of $\tau$-identification in searching
for Non-Standard Interactions (NSI) or additional singlet fermions
with some admixture with the three-family left-handed neutrinos,
so-called "sterile neutrinos".

Non-standard interactions are effective four-fermion interactions,
which arise if neutrinos couple to new, heavy particles. This is
similar to the Fermi theory of nuclear beta decay emerging as the
low-energy fingerprint of the Standard Model weak interactions.  NSI
can affect the neutrino production and detection mechanism if they are
of the charged current type, and the neutrino propagation if they are
of the neutral current type. In the first case, the NSI can be
parametrized as a small admixture of the ``wrong flavor''
$\ket{\nu_\beta}$ to a neutrino produced or detected in association
with a charged lepton of flavor $\alpha$:
\begin{align}
  \ket{\nu^s_\alpha} &= \ket{\nu_\alpha}
      + \sum_{\beta=e,\mu,\tau} \eps^s_{\alpha\beta} \ket{\nu_\beta},
      \qquad&\text{e.g. }& \pi^+ \xrightarrow{\eps^s_{\mu e}}  \mu^+ \nu_e \\[-0.2cm]
  \bra{\nu^d_\alpha} &= \bra{\nu_\alpha}
      + \sum_{\beta=e,\mu,\tau} \eps^d_{\beta\alpha} \bra{\nu_\beta}
      \qquad&\text{e.g. }& \nu_\tau N \xrightarrow{\eps^d_{\tau e}}  e^- X \,.
\end{align}
The second case corresponds to a non-standard contribution to the MSW potential:
\begin{align}
  \tilde{V}_{\rm MSW} = \sqrt{2} G_F N_e
  \begin{pmatrix}
    1 + \eps^m_{ee}       & \eps^m_{e\mu}       & \eps^m_{e\tau}  \\
        \eps^{m*}_{e\mu}  & \eps^m_{\mu\mu}     & \eps^m_{\mu\tau} \\
        \eps^{m*}_{e\tau} & \eps^{m*}_{\mu\tau} & \eps^m_{\tau\tau}
  \end{pmatrix} \,.
\end{align}
In the above expressions, the parameters $\eps^{s,d,m}_{\alpha\beta}$ give
the strength of the NSI relative to standard weak interactions. A generic
estimate is
\begin{align}
  |\eps^{s,d,m}_{\alpha\beta}| &\sim \frac{M_{\rm W}^2}{M_{\rm NSI}^2} \,,
  \label{eq:eps-estimate}
\end{align}
where $M_{\rm NSI}$ is the new physics scale, at which the effective
NSI operators are generated. Even though the present model independent
bounds on the $\eps^{s,d,m}_{\alpha\beta}$ are not very strong
($\mathcal{O}(0.1 - 1)$). However, these bounds are not likely to be
saturated in specific models~\cite{Gavela:2008ra, Antusch:2008tz}; at
least if one follows the usual guidelines of model building: no
fine-tuning, as few new particles as possible, new physics preferably
at or above the TeV scale, {\it etc}. Indeed, if the estimate
\eqref{eq:eps-estimate} is taken at face values, with $M_{\rm NSI}
\sim 1$~TeV, we expect $\eps^{s,d,m}_{\alpha\beta} < 0.01$. It is
important to keep in mind that, in any specific model, the
phenomenological parameters $\eps^{s,d,m}_{\alpha\beta}$ will in
general not be independent.

Phenomenological models in which $N$ new singlet fermions are mixed
with the three left-handed ones imply a straightforward generalization
of the PMNS matrix to a $(3+N) \times (3+N)$ unitary mixing matrix,
that for the case of $N=1$ is:
\begin{equation}
U_{\rm PMNS} =
\left (
 \begin{array}{cccc}
U_{e1} & U_{e2} & U_{e3} & U_{e4} \\
U_{\mu 1} & U_{\mu 2} & U_{\mu 3} & U_{\mu 4} \\
U_{\tau 1} & U_{\tau 2} & U_{\tau 3} & U_{\tau 4} \\
U_{s1} & U_{s2} & U_{s3} & U_{s4} 
\end{array}
\right )
\end{equation}
Some of these elements are strongly constrained by non-observation at
reactors and at the MiniBooNE experiment.  On the other hand, models
in which the mixing angles $\theta_{i4}$ between a new singlet fermion
$\nu_s$ and the three active ones are all very small cannot be
excluded. Notice that, both for NSI and sterile neutrino models, new
CP-violating phases are present in addition to the standard
three-family oscillation phase $\delta$.

\begin{enumerate}
\item {\bf NSI in production and detection}

NSI in production and detection imply non-unitarity of the PMNS
matrix.  Therefore, if some of the new parameters
$\epsilon^{s,d}_{\alpha\beta}$ are non-vanishing, it is not enough to
study the two channels available at the MIND detector (the $\nu_e \to
\nu_\mu$ golden channel and the $\nu_\mu \to \nu_\mu$ disappearance
channel) to measure all of the new parameters of the model. To study
non-unitarity of the leptonic mixing matrix, there are two options:
\begin{itemize}
\item Measure all the oscillation probabilities $P(\nu_\mu \rightarrow
  \nu_e)$, $P(\nu_\mu \rightarrow \nu_\mu)$, $P(\nu_\mu \rightarrow
  \nu_\tau)$ (or $P(\nu_e \rightarrow \nu_e)$, $P(\nu_e \rightarrow
  \nu_\mu)$, $P(\nu_e \rightarrow \nu_\tau)$, and check if they sum
  up to unity. A problem of this approach is that $\nu_e$ detection is
  very difficult in a MIND detector, so either there will be large
  uncertainties or a secondary detector with a different technology
  (for example, liquid argon) should be added to the two MINDs setup.
  Moreover, the systematical errors in the different oscillation
  channels will be different, which also limits the achievable
  sensitivity.
\item Use neutral current events. This is also
  difficult~\cite{Barger:2004db}, and, at present, only a sensitivity
  at the ten per cent level can be achieved. This might improve if the
  neutral current cross sections were known better and if more
  sophisticated event selection criteria could be developed.
\end{itemize}

Most of the new parameters could be measured using a dedicated near
detector. The detector design should be optimized so as to measure as
much oscillation channels as possible, and with very good
$\tau$-identification capability.  Therefore, this detector cannot be
a scaled version of MIND. At present, no detailed study of such a
detector has been performed, see Refs.~\cite{Gavela:2008ra,
  Antusch:2008tz} for the potential of an ECC near to a Neutrino
Factory source and the recent Ref.~\cite{Tang:2009na}.

\item {\bf NSI in propagation}

NSI in propagation do not imply a non-unitary PMNS matrix. In this
case it is therefore possible to obtain information on all of the new
parameters $\epsilon^m_{\alpha\beta}$ using the two channels available
at the MIND detector.

A detailed study of NSI in propagation at a neutrino factory has been
presented in Ref.~\cite{Kopp:2008ds} (see fig.~\ref{fig:nf-summary},
taken from that paper). The results obtained show that the IDS-NFS
baseline neutrino factory with two MIND detectors at $L \sim 4000$ km
and $L \sim 7500$ km is sensitive to $\eps^m_{\alpha\beta} \sim 0.01
- 0.1$, independent of whether a $\nu_\tau$ detector is present.
There might be a physics case for this detector if the process
$\nu_\tau + N \rightarrow \tau + X$ proceeds in an unexpected way
(e.g.\ an anomalous energy dependence), if $\tau$ leptons are produced
in a non-standard way (e.g.\ $\eps^d_{e\tau} \neq 0$ or
$\eps^d_{\mu\tau} \neq 0$), or if the muons stored in a neutrino
factory have a small branching to $\nu_\tau$, e.g.\ due to
$\eps^s_{\mu\tau} \neq 0$ or $\eps^s_{e\tau} \neq 0$.  In the first
case, a $\nu_\tau$ detector at around the first oscillation maximum
would be required because the $\nu_\tau$ flux first has to be
generated by oscillation from $\nu_\mu$; in the second case, a
$\nu_\tau$ \emph{near} detector would be optimal due to the higher
flux at the near site.

\begin{figure}[t!]
  \begin{center}
    \includegraphics[width=\textwidth]{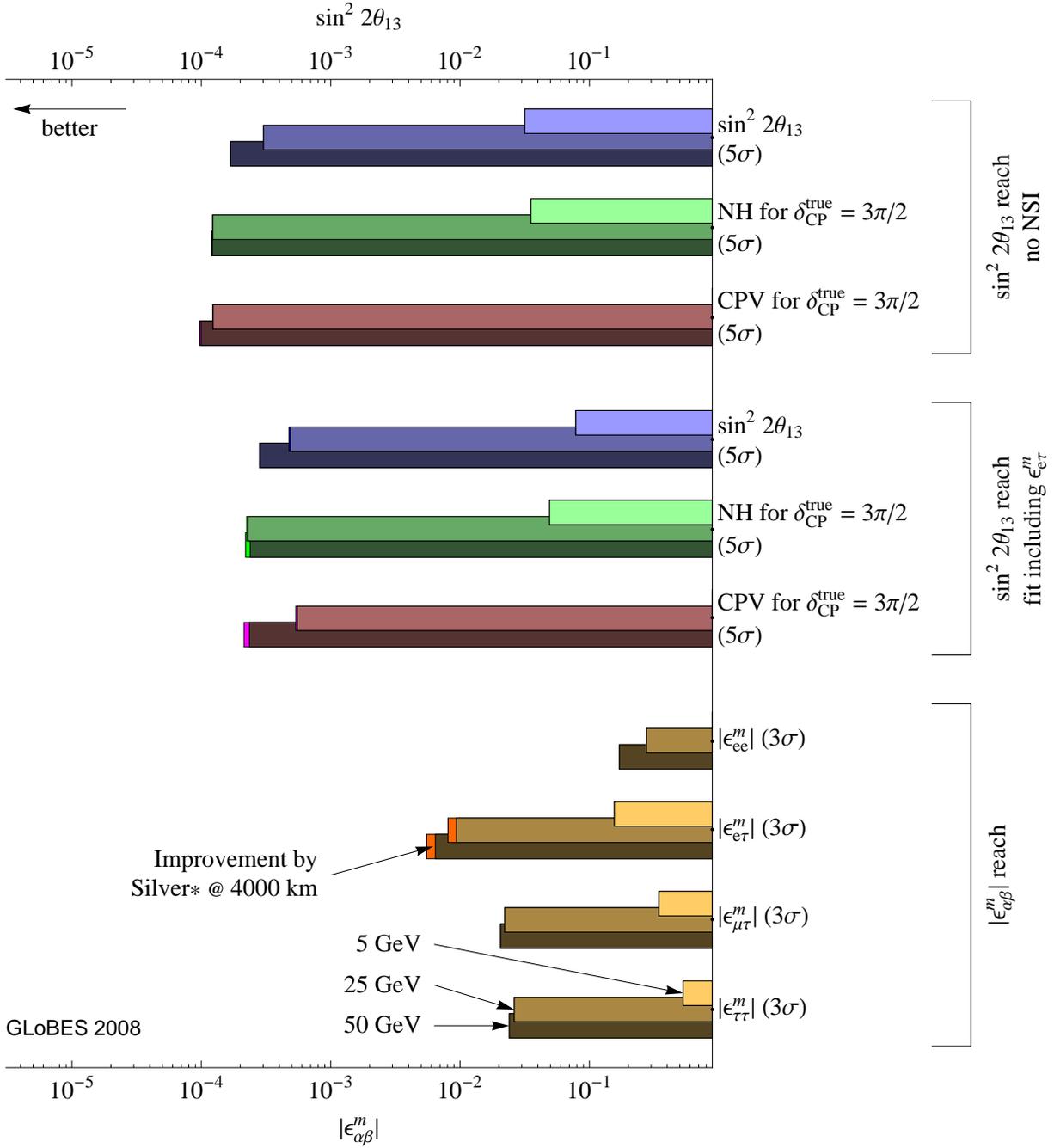}
  \end{center}
  \vspace{-0.5cm}
  \caption{Summary of the neutrino factory performance with and
    without the presence of non-standard interactions. The IDS-NF
    setup with two MIND detectors at $L_1 = 4\,000$~km, $L_2 =
    7\,500$~km was used, and the ``true'' parameter values $\sin^2
    2\theta_{13} =0.001$ and $\delta_{\rm CP}=3 \pi/2$ were assumed.
    The plot shows that sensitivities are poor at $E_\mu = 5$~GeV
    (light bars), but increase dramatically at $E_\mu = 25$~GeV
    (medium light bars). The benefit from increasing $E_\mu$ further
    to 50~GeV (dark bars) is only marginal, as is the benefit from
    including a silver channel detector.  Figure taken from
    \cite{Kopp:2008ds}; see that paper for details.}
  \label{fig:nf-summary}
\end{figure}

From this analysis, we conclude that an ECC detector to look for
$\tau$'s produced through $\nu_e \to \nu_\tau$ does not improve the
expected IDS-NF baseline setup sensitivity to NSI in propagation. A
thorough study of the impact of $\nu_\mu \to \nu_\tau$ data is
lacking, though. We do not expect, however, these data to have a
striking impact on the sensitivity, due to unitarity of the PMNS
matrix in models in which only NSI in matter are considered.

\item {\bf Sterile neutrinos}

Even though sterile neutrinos do no longer receive as much attention
nowadays as before the publication of the MiniBooNE results, they are
still a viable possibility, motivated by the fact that neutral
singlets $\nu_s$ appear in many models of new physics. If they are
light, the neutrinos produced in a neutrino factory may have a small
admixture of $\nu_s$, while heavy $\nu_s$ (such as right-handed
Majorana neutrinos in type-I see-saw models) would manifest themselves
in the form of a non-unitary mixing matrix of the light neutrinos.

In the case of one light $\nu_s$, a recent study~\cite{Donini:2008wz}
shows that the $\nu_\mu \to \nu_\tau$ appearance channel (mostly
disregarded up to now; see, however, Ref.~\cite{Bueno:2000fg}),
measured with a magnetized ECC, is extremely important when combined
to $\nu_\mu \to \nu_\mu$ to measure some of the parameters of the
model, and in particular some of the new CP-violating phases.  On the
other hand, the silver channel $\nu_e \to \nu_\tau$ is only of limited
impact when added to the golden channel $\nu_e \to \nu_\mu$, although
it is useful to solve some of the many degeneracies in the parameter
space.

A criticism to the use of magnetized ECC to study the $\nu_\mu \to
\nu_\tau$ channel is that the scanning load could be too high for this
analysis to be realistic. However, it has been found that the scanning
load for an emulsion detector at $L> 1\,000\,\mathrm{km}$ is not huge:
O(500) events per kton per year with a $2 \times 10^{20}$ flux are
expected , for perfect efficiency. Adding a similar number of
background events, this scanning load is compatible with extrapolation
for present capabilities

Notice that for standard three-family oscillation and in models with
NSI in propagation, due to the unitarity of the PMNS matrix, a good
knowledge of the golden and the disappearance channel (both studied at
MIND) should be enough to explore the whole parameter space.  This,
 however, is not the case in models in which the $3 \times 3$ PMNS
matrix is not unitary.  In sterile neutrino models, for example, since
we are not able to study the $\nu_\mu \to \nu_s$ appearance
channel(s), the information that can be extracted from the $\nu_\mu$
disappearance channel and the $\nu_\mu \to \nu_\tau$ channel are not
identical.  The same would happen in extensions of the standard model
in which NSI are considered both in propagation and production, such
as to violate unitarity of the PMNS matrix.

From the analysis of Ref.~\cite{Donini:2008wz} we conclude that the
combination of the IDS-NF baseline setup (with two MINDs) with one or
two magnetized ECC increase significantly the potential of the
Neutrino Factory to measure all the parameter space of the
(3+1)-neutrino model and, in particular, to increase its CP-violation
discovery potential. However, it has been shown that the present
design of the magnetized ECC is not optimized and that a dedicated
study of the detector to look for new physics is mandatory (see next
section).

The optimal location for a long baseline $\tau$-detector to study
sterile neutrinos is not clear, yet.  Whereas a detector whose purpose
is the study of the silver channel in the framework of the
three-family model or NSI in propagation is optimally located around
the intermediate IDS-NF baseline (see ISS Final Report and
Ref.~\cite{Kopp:2008ds}), it seems that to study (3+1) sterile
neutrinos to put the magnetized ECC detector at the Magic Baseline
could be more convenient.  This is particularly true for searches of
CP-violating signals. At the Magic Baseline, indeed, the standard
three-family CP-violating effect vanishes, and therefore if
CP-violation is observed this is clearly pointing out the existence of
physics beyond the standard model (either new particles, such as the
sterile neutrinos, or new effective operators, such as in NSI).
Notice that the $\nu_\mu \to \nu_\tau$ statistics at the Magic
Baseline is still large (of O(500) events for 1 kton MECC with perfect
efficiency and $2 \times 10^{20}$ useful muons per year).

\end{enumerate}

\subsubsection{Technological options for $\tau$-detectors}

\begin{figure}[t!]
\begin{center}
\hspace{-0.5cm}
\begin{tabular}{cc}
\includegraphics[width=7.5cm]{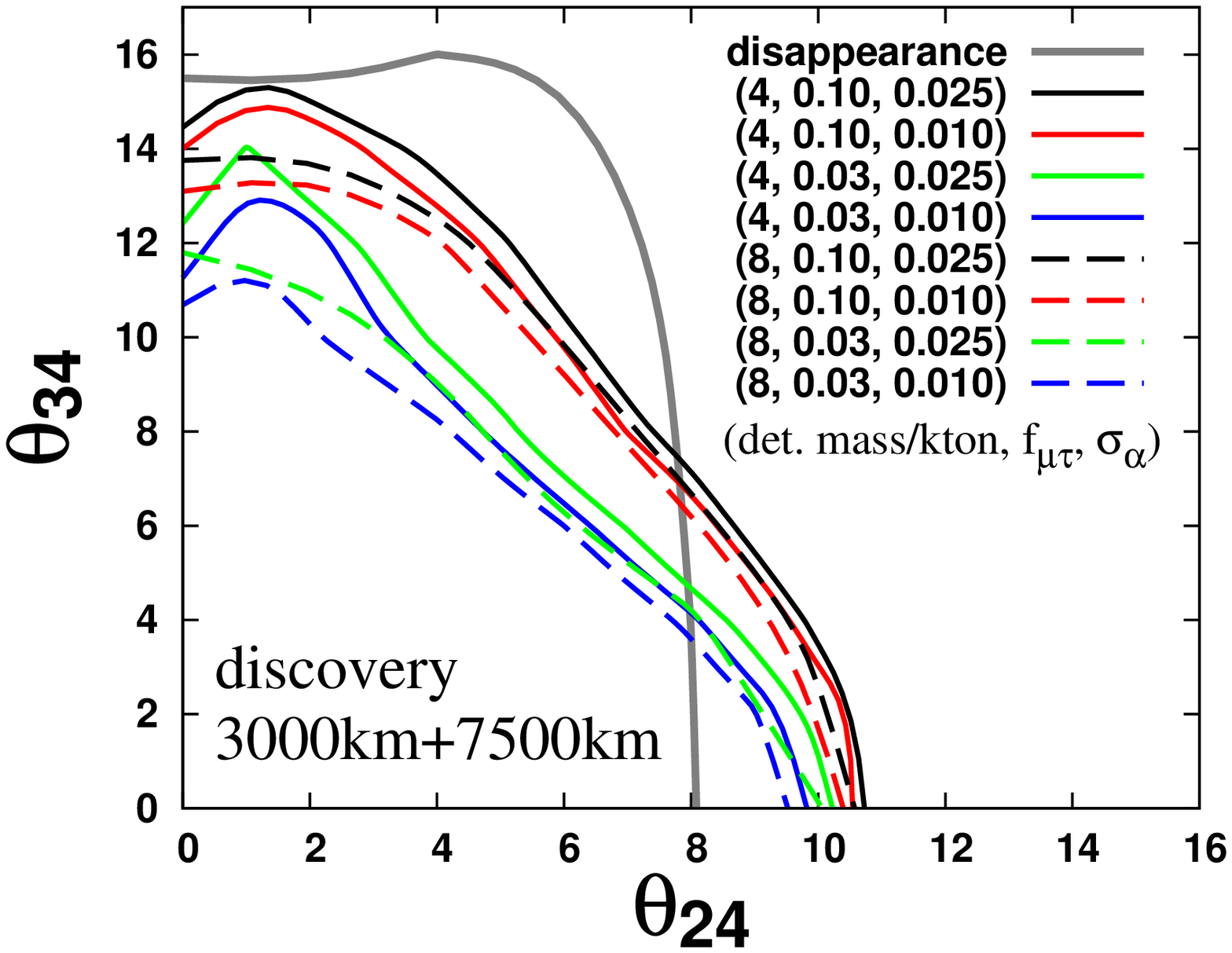} & 
\includegraphics[width=7.5cm]{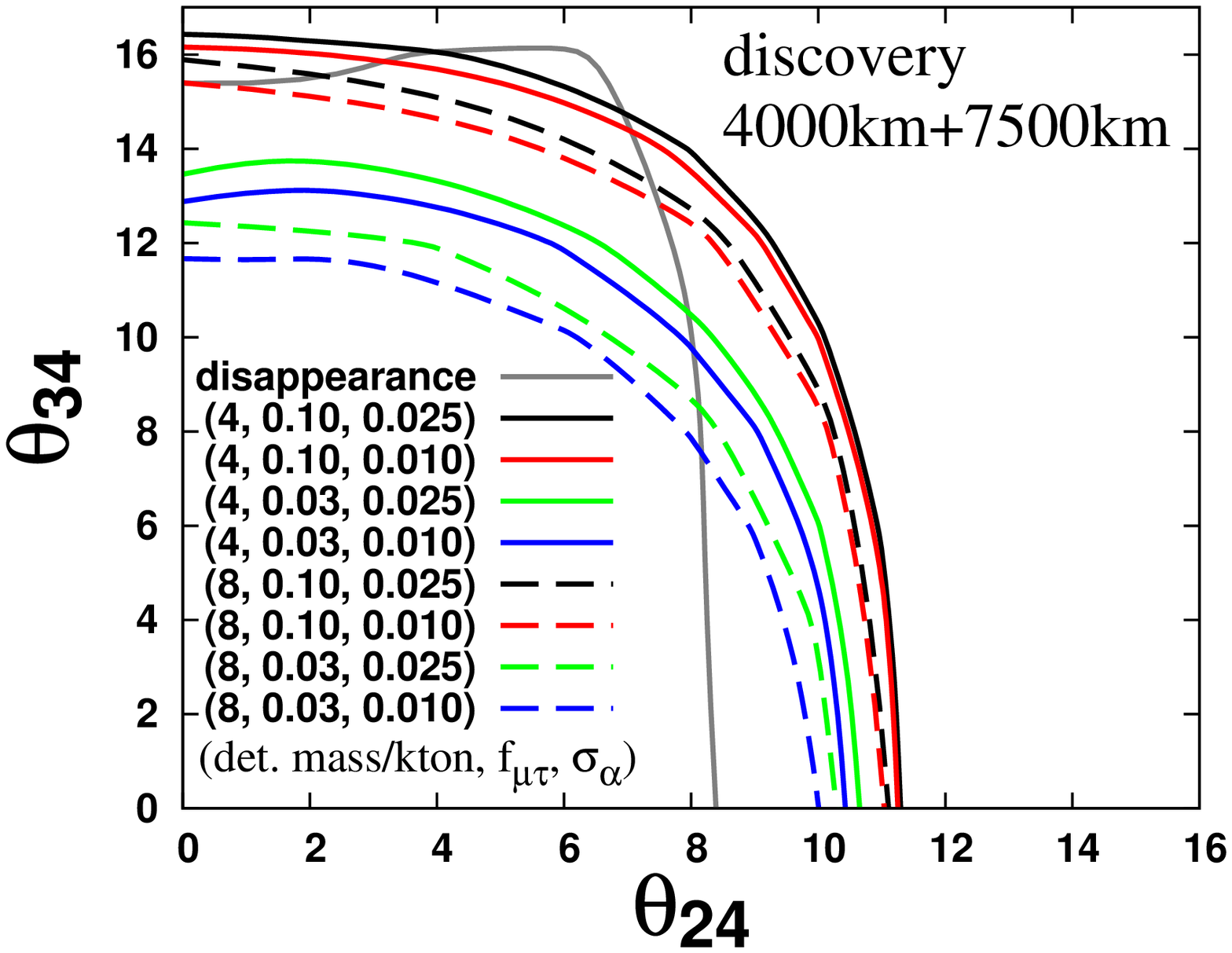}
\end{tabular}  
\caption{\label{fig:systematics}%
  Left (right) panel: Dependence of the excluded region in the
  ($\theta_{24}$, $\theta_{34}$)-plane on the systematic errors
  $f_{\mu\tau}\equiv f_j$ and $\sigma_\alpha$ for the discovery
  channel ($\nu_\mu \to \nu_\tau$) as well as the MECC detector mass
  at the 50 GeV (20 GeV) neutrino factory, where the excluded regions
  are by the discovery channel only.  The solid (dashed) lines assume
  4 kton (8 kton) for the tau detector mass.  The solid gray line,
  which stands for the excluded region by the $\nu_\mu$ disappearance
  channel, is also shown.  Taken from Ref.~\cite{Donini:2008wz}.}
\end{center}
\end{figure}

The technology for tau-detectors has not been fixed yet. The liquid
argon technology should be studied further (something compatible with
the time scale of a Neutrino Factory).  Furthermore, the impact of
systematics errors in the magnetized emulsion technique (MECC) is
shown to be very important, see Fig.~\ref{fig:systematics}.
 
In the figure, the sensitivity to two parameters of a model with three
active and one sterile neutrino (the "3+1" model) using the $\nu_\mu \to \nu_\tau$ 
channel is shown. The dashed gray line refers to the
sensitivity to those parameters achievable using two 50 kton MIND
detectors: one at an intermediate baseline, $L = 3000-4000$ km, and the
second at the Magic Baseline. In the two panels, we show the
sensitivity for a 50 GeV muon Neutrino Factory (left) and a 20 GeV
muon Neutrino Factory (right). It is clear from the left panel that a
huge increase in the sensitivity of the $\nu_\mu \to \nu_\tau$ channel
is achieved if the uncorrelated systematic errors are reduced from
10\%(black solid line) to 3\% (green solid line). This improvement is
actually much more important than an increase in the MECC detector
mass from 4 kton (green solid line) to 8 kton (green dashed line).

This systematic error is taking into account in a non-detailed way
systematics induced by normalization of the flux and cross-sections.
Both are expected to be better known after the first OPERA phase.
Moreover, $\nu_\tau N$ cross-sections must be studied with a near
detector, as it happens for the $\nu_\mu N$ one. This means that these
sources of systematics can be strongly reduced.  A study of the
possible improvement of the sensitivity with a better design of the
$\tau$-detector in the framework of NSI extensions of the standard
model is lacking.
 
\subsubsection{Conclusions \& recommendations}

In this note, we have discussed the potential of an ECC, added
to the IDS-NF baseline setup (with two MIND detectors located at $L
\sim 4000$ km and at the Magic Baseline), in three models: the
standard three-family oscillation scenario; an extension of the SM
with Non-Standard Interactions in matter; and an extension of the SM
with one extra light singlet fermion (the so-called 3+1 sterile
neutrino model).  In the first two cases, the $\nu_\tau$ detector does
not improve the potential of the IDS-NF baseline setup to measure the
oscillation parameters or to uncover new physics effects in neutrino
oscillations. The reason is that, due to the large mixing in the
$\mu$-$\tau$ sector, most effects that are present for
$\tau$-neutrinos, will have a similar impact also for $\mu$-neutrinos.
In these models, the $\tau$-detector could only serve as a tool for
resolving parameter degeneracies. This, however, could be also
achieved combining the Golden and Disappearance channels and data from
two different baselines $L = 4\,000$~km and $L = 7\,500$~km.  We must
remind that a comparison of the potential of the IDS-NF baseline setup
and the same setup with an additional $\tau$-detector to measure the
$\theta_{23}$-octant in the standard three-family oscillation model is
missing, though (see Ref.~\cite{Meloni:2008bd}).

In the case of the (3+1)-sterile neutrino model, studied in
Ref.~\cite{Donini:2008wz}, the availability of the $\nu_\mu \to
\nu_\tau$ data using a magnetized ECC has been shown to be extremely
important to measure the whole parameter space of the model and, in
particular, to study CP-violating phases different from the standard
three-family oscillation one, $\delta$.

There may also be a physics case for $\nu_\tau$ detection if new
physics should manifest itself in the $\nu_\tau$ detection process, or
if non-standard couplings of $\nu_\mu$, $\nu_e$ to $\tau$ leptons or
of $\nu_\tau$ to muons and electrons should exist.  However,
non-standard contributions to the $\nu_\tau$ detection process would
require a $\nu_\tau$ detector at a long baseline (e.g.\ $4\,000$~km),
while non-standard $\tau$ and $\nu_\tau$ production can be most
efficiently observed in a $\nu_\tau$ near detector.

The outcome of this short review is that it is very difficult, at the
present stage, to draw a final conclusion on the increase in the
Neutrino Factory physics potential to discover new physics if a
$\tau$-detector is added to the IDS-NF baseline setup. It is also far
from clear which detector technology would be optimal: a good
knowledge of the ECC technology will be available only after some
years of OPERA data taking; it is not clear if a magnetized ECC,
important to increase the ECC statistics, is feasible; the liquid
Argon technology has not been studied in detail.  Eventually, the
technology to be used if a near $\tau$-detector should be built could
be completely different from what proposed up to now: due to the high
neutrino flux at the near site if exposed to a Neutrino Factory beam,
more powerful techniques than what suggested for a large detector
could be used, since a smaller detector mass could be sufficient.

In view of these arguments, we suggest that the ECC $\tau$-detector is
\emph{not} to be included in the IDS-NF \emph{baseline} setup due to
the absence of a compelling physics case and, given the present very
preliminary status of the detector design.  This does not exclude the
option that a $\nu_\tau$ detector (not necessarily based on the ECC
technology) is added to the neutrino factory at a \emph{later stage} of the
project if unexpected results from the LHC or from the neutrino
factory itself should create a physics case for it.

However, we think it is mandatory to further pursue the study of the
potential of such a detector, especially in view of the fact that we
do not know what new physics may be out there. Having access to more
flavors can only increase the discovery potential of the Neutrino
Factory. Notice, eventually, that if $\theta_{13}$ results to be large
(see solar, atmospheric and MINOS results), part of the statistical
problems of the $\tau$-channels become less relevant. At the same
time, the main motivation for a Neutrino Factory would become the
search for new physics beyond the Standard Model, and therefore the
option of an increased flavor sensitivity becomes extremely
interesting.

\subsection{Relevance of Near Detectors at Nufact}
\label{sec:near}

As far as the importance and requirements for near detectors at a neutrino factory
are concerned, the questions raised by the {\em International Design Study for the Neutrino Factory} (IDS-NF) include:
What is the potential of near detectors to cancel systematical errors?
When do we need a near detector for standard oscillation physics?
What (minimal) characteristics do we require, such as technology, number, and sites?
What properties do near detectors need for new physics searches?
From these questions, we can read off already two obstacles: First of all, one has to address
which kind of systematics near detectors should reduce, and second, one has to address 
what kind of new physics near detectors may help for. In the following, we split the discussion
into standard oscillation physics and new physics searches.

\subsubsection{Near detectors for standard oscillation physics}

\begin{figure}[t]
\includegraphics[width=\textwidth]{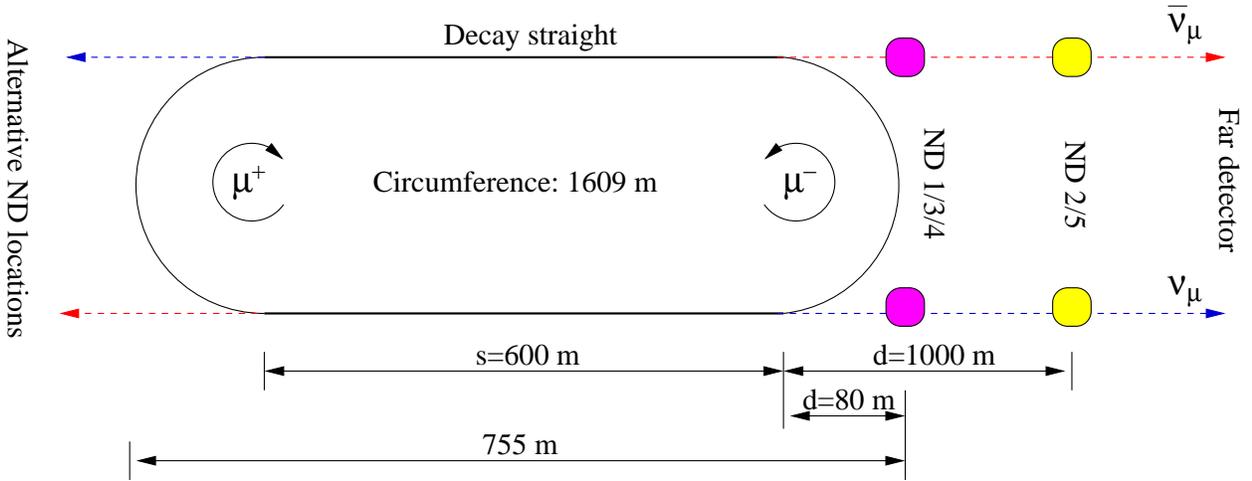}
\caption{\label{fig:ring}Geometry of the muon storage ring and possible near detector (ND) locations (not to scale). The baseline $L$ is the distance between production point and near detector, \ie, $d \le L \le d+s$.
Figure taken from \Ref~\cite{Tang:2009na}.}
\end{figure}

As it is illustrated in \figu{ring}, (at least) two near detectors are required for a neutrino factory if the $\mu^-$ and $\mu^+$ circulate in different directions in the ring. For the same reason, charge identification is, in principle, not required, since there are no wrong sign muons produced by oscillations so close to the source. However, for background measurements (such as from charm decays), a magnetic field may be necessary. As it is demonstrated in \Ref~\cite{Tang:2009na}, the size, location, and geometry of the near detectors hardly matter for standard oscillation physics even in extreme cases of possible near detectors. Because of the high statistics in all energy bins of the near detectors, the physics potential is generally limited by the statistics in the far detector(s). However, note that rare interactions used for flux monitoring, such as inverse muon decays or elastic scattering, may require large enough detectors. A possible near detector design for a neutrino factory is, for instance, discussed in \Ref~\cite{Abe:2007bi}. 

\begin{figure}[t]
\begin{center}
\includegraphics[width=\textwidth]{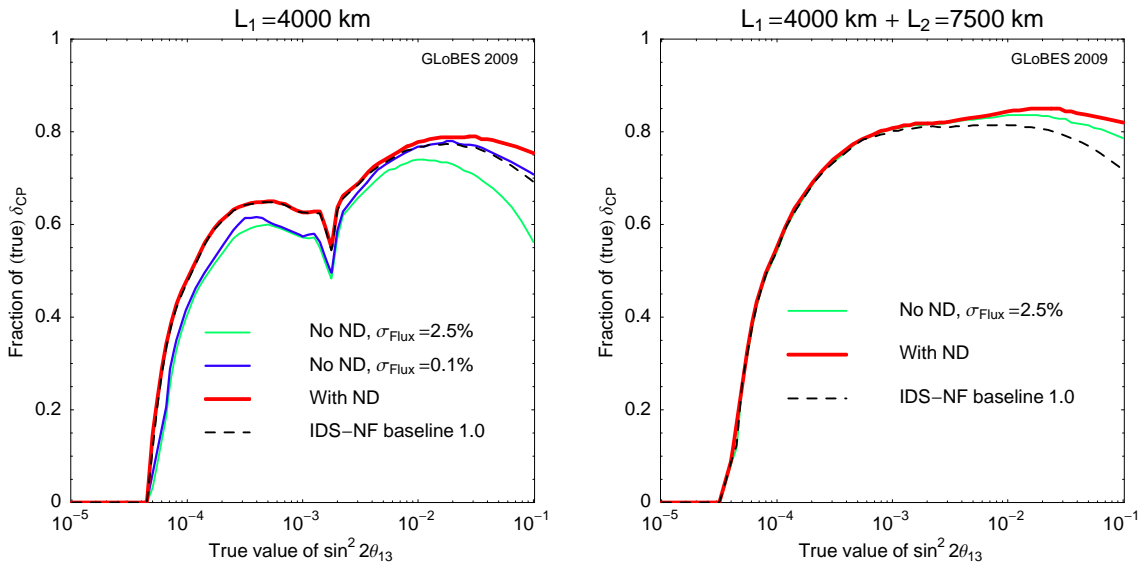}
\end{center}
\caption{\label{fig:ndcomp} CP violation discovery reach as a function of true $\stheta$ and the fraction of (true) $\deltacp$ for one far detector (left) and two far detectors (right); $3\sigma$ CL. Figure taken from \Ref~\cite{Tang:2009na}.}
\end{figure}

As far as the systematics treatment is concerned, the current IDS-NF baseline setup relies on uncorrelated (among all oscillation channels, detectors, and neutrinos-antineutrinos) signal and background normalization errors treating the near detector(s) implicitly, whereas realistic systematics implies particular correlations. For example, the cross section errors are correlated among all channels and detectors measuring the same $\nu_\mu$ or $\bar\nu_\mu$ (inclusive) charged current cross sections, but there may be a shape error, \ie, the errors may uncorrelated among the energy bins. On the other hand, flux errors are correlated among all detectors and channels from the same decays in the same storage ring straight, and they are correlated among different energy bins. In \Ref~\cite{Tang:2009na}, the systematical errors have been tested which are, in principle, reducible by the use of near detectors. Note that there may be other types of systematics, such as fiducial volume errors, which have not yet been discussed. The refined systematics treatment is illustrated for the CP violation discovery reach in \figu{ndcomp}, in comparison to the IDS-NF current parameterization (dashed curves). In the left panel, only one baseline is used.  In this case, the near detectors turn out to be very important. In the right panel, the combination of two baselines is shown. The result using the new systematics treatment is already better than the previous one even without near detector, because the cross sections are fully correlated between the two detectors. This means that possible cross section errors cancel. The near detectors improve the result even further.

\subsubsection{Near detectors for new physics searches}

In order to address the requirements for near detectors for new physics searches, one has to specify which
type of new physics. Here we show a number of examples to produce a list of detector requirements which should be as complete as possible. If the new physics originates from heavy mediators, which are integrated out, the new physics can be parameterized in the effective operator picture. The lowest possible effective operators affecting the production, propagation, or detection of neutrinos are dimension six operators, suppressed by $v^2/\Lambda^2$ by the new physics scale $\Lambda$ compared to the SM Higgs VeV $v$. At tree level, they can be mediated by heavy neutral fermions, leading to a non-unitary mixing matrix after the re-diagonalization and re-normalization of the kinetic terms of the neutrinos (see, \eg,  \Ref~\cite{Antusch:2009pm} for a short summary), or by scalar or vector bosons, leading to so-called non-standard interactions (NSI; see, \eg, \Ref~\cite{Kopp:2007ne} for the terminology). A very different type of new physics is the oscillation into light sterile neutrinos, because it may lead to oscillation signatures. We consider near detectors for these three applications. As the common requirement to near detectors compared to standard oscillation physics, the detector mass is very important, because the new physics sensitivity is in many cases limited by the statistics in the near detectors.

The most interesting non-standard interactions for near detectors may be $\epsilon^s_{e \tau}$ and $\epsilon^s_{\mu \tau}$, which replace an initial $\nu_e$ or $\nu_\mu$ in the beam by a $\nu_\tau$. Since these NSI lead to a zero-distance effect $\propto | \epsilon_{\alpha \tau}|^2$. As illustrated in \Ref~\cite{Tang:2009na}, the sensitivity to these parameters can be significantly improved in the presence of $\nu_\tau$ near detectors.  If the two effects need to be distinguished, charge identification is required. Assuming that this new physics effect originates from $d=6$ NSI  and that the bounds from charge lepton flavor violation will be strong enough, the bounds from the source effects can be translated into bounds on the matter effect $\epsilon_{\mu \tau}^m$~\cite{Gavela:2008ra} (see also \Ref~\cite{Ohlsson:2009vk} for a particular model). In particular models, also other types of NSI at the source may be interesting, such as $\epsilon_{e \mu}^s$ from a Higgs triplet as type-II see-saw mediator~\cite{Malinsky:2008qn}. This application relies on excellent charge identification properties of the near detector(s). The above mentioned non-unitarity leads to a particular, fundamental correlation among source, matter, and detector effects.  In this case, near detectors improve the measurements in a similar way~\cite{Antusch:2009pm}. Both the NSI and non-unitarity searches have in common that the near detector location only affects statistics, whereas there is no relevant oscillation effect close to the source.

\begin{figure}[t!]
\begin{center}
\includegraphics[width=9cm]{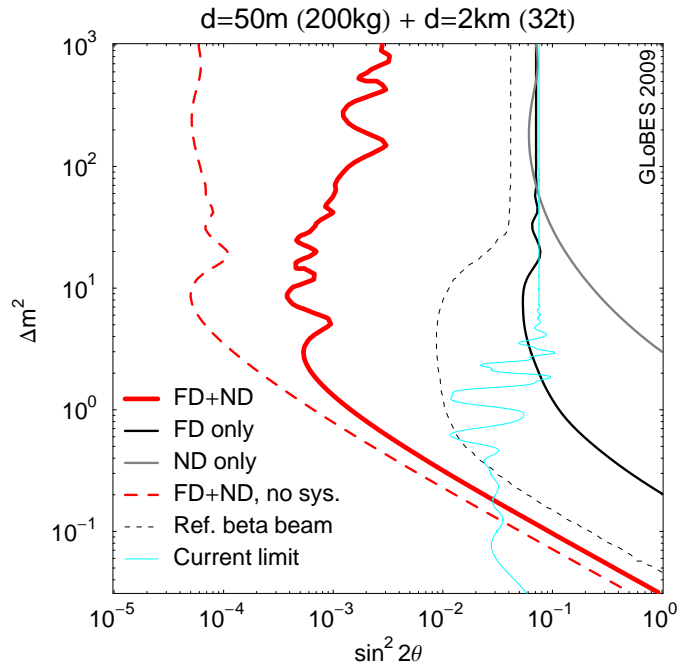}
\end{center}
\caption{\label{fig:fnsys} 
Exclusion limit in the $\sin^2 2 \theta$-$\Delta m^2$ plane for $\nu_e$ disappearance into sterile neutrinos from the two detector setup discussed in the main text (thick solid curve, 90\% CL, 2 d.o.f.). Figure taken from \Ref~\cite{Giunti:2009en}.
}
\end{figure}

The search for sterile neutrinos is qualitatively different: The location of the near detectors affects the sensitive $\Delta m^2$ range. An interesting example is the disappearance of $\nu_e$ at short baselines, discussed in \Ref~\cite{Giunti:2009en}. In this case, the oscillations have to be averaged over the decay straight, which leads to significant effects for near detectors very close to the source (or large $\Delta m^2$). In addition, the unknown cross section errors play the same role as the unknown fluxes in the two-detector reactor experiments, such as Double Chooz or Daya Bay. Therefore, a similar approach has been proposed in \Ref~\cite{Giunti:2009en}: two sets of near detectors at different (short) baselines, which are sensitive to different $\Delta m^2$-ranges, are important for the systematics cancellation of the cross section errors. The result is illustrated in \figu{fnsys} for $d=50 \, \mathrm{km}$ (ND) and $d=2 \, \mathrm{km}$ (FD), where the thick curve shows the combined sensitivity. If only ND or FD is used (medium thin solid curves), the sensitivity is limited by the knowledge of the cross sections. Obviously, the current bound can be exceeded by two orders of magnitude.
 
\subsubsection{Summary}

The requirements/characteristics of near detectors at a neutrino factory can be summarized as follows:
For standard oscillation physics, the exact location, size, and geometry of the detectors hardly matters because of the large statistics. Data for the $\nu_\mu$ and $\bar\nu_\mu$ cross sections are already sufficient, because only these two are needed in the far detectors (for a high energy neutrino factory). If the $\mu^+$ and $\mu^-$ circulate in different directions in the storage ring(s), at least two near detectors are needed. A magnetic field may be necessary for background measurements.

For nearly all new physics applications, the sizes (masses) of the near detectors are important. Since the size of the detector cannot be arbitrarily increased beyond the opening angle of the beam, the detectors should be as long as possible in order to capture a large portion of the on-axis flux. For new physics searches, all flavors should be measured, and charge identification is mandatory for many applications. For some purposes, such as the short baseline electron neutrino disappearance, more than one set of near detectors may be required. In addition, for any light sterile neutrino oscillation search, the baselines of the near detectors are very important. Energy resolution, on the other hand, is of secondary importance, since the energy resolution is, close to the source, limited by the extension of the decay straight.

In conclusion, near detectors at a neutrino factory are important for both standard oscillation physics and new physics searches. From the physics point of view, however, the requirements, such as size and location, may be driven by new physics searches.

\section{Update on betabeam performance}
\label{sec:bbeam}

\subsection{Baseline scenario optimization}
\label{sec:enrique}

The $\beta$-beam concept was first introduced in Ref.~\cite{Zucchelli:2002sa}.  
It involves producing a large number of $\beta$-unstable ions, accelerating them to some reference
energy, and allowing them to decay in the
 straight section of a storage ring, resulting in a very intense and pure $\nu_e$ or $\bar \nu_e$ beam. 
``Golden'' sub-leading transitions, $\nu_e \to \nu_\mu$ and $\bar \nu_e \to \bar \nu_\mu$, can then be measured through muon observation 
in a distant detector. In the original proposal $^{18}$Ne ($^{6}$He) ions are accelerated to $\gamma \sim 100$ and stored so that $\nu_e$ ($\bar{\nu}_e$) beams are produced and detected at a Mton class water \v Cerenkov detector located at $L=130$km at the Frejus site. Numerous modifications of this basic setup have been studied, most of them being different combinations of two basic ingredients: the possibility of accelerating the ions to higher $\gamma$ factors \cite{Burguet-Castell:2003vv,Burguet-Castell:2005pa}, thus increasing the flux and the statistics at the detector, and the possibility of considering the decay of different ions to produce the neutrino beam. In particular $^8$B and $^8$Li
have been proposed as alternatives to $^{18}$Ne and $^{6}$He respectively \cite{Rubbia:2006pi,Donini:2006dx,Rubbia:2006zv}

Here we investigate the physics potential of the $\gamma = 100$ $\beta$-Beam option, accessible at the CERN SPS, comparing the different possible choices of ions and baselines. We also take into account the atmospheric background expected at the detector and study the suppression factor of this background required through the bunching of the beam to attain the best sensitivities in the different setups.


%
\begin{table}
\begin{center}
\begin{tabular}{|c|c|c|c|c|} \hline \hline
   Element  & $A/Z$ & $T_{1/2}$ (s) & $Q_\beta$ eff (MeV) & Decay Fraction \\ 
\hline
  $^{18}$Ne &   1.8 &     1.67      &        3.41         &      92.1\%    \\
            &       &               &        2.37         &       7.7\%    \\
            &       &               &        1.71         &       0.2\%    \\ 
  $^{8}$B   &   1.6 &     0.77      &       13.92         &       100\%    \\
\hline
 $^{6}$He   &   3.0 &     0.81      &        3.51         &       100\%    \\ 
 $^{8}$Li   &   2.7 &     0.83      &       12.96         &       100\%    \\ 
\hline
\hline
\end{tabular}
\caption{\label{tab:ions}  $A/Z$, half-life and end-point energies for three $\beta^+$-emitters ($^{18}$Ne  and $^8$B)
and two $\beta^-$-emitters ($^6$He and $^8$Li). All different $\beta$-decay channels for $^{18}$Ne are presented.}
\end{center}
\end{table}
In Tab.~\ref{tab:ions} we show the relevant parameters for the $\beta$ decay of four ions: $^{18}$Ne and $^6$He, $^8$Li and $^8$B. As can be seen, the main difference between the two sets of ions consists in their decay energy: $Q_\beta \sim 3.5$ for $^{18}$Ne and $^6$He and $Q_\beta \sim 12.5$ for $^8$Li and $^8$B. Thus, the neutrino beams produced by the decay of the latter set of ions are around $3.5$ times more energetic than the ones produced by the former when accelerated to the same $\gamma$ factor. This
also means that, for the oscillation to be on peak, a baseline $3.5$ times longer is required to achieve the same $L/E$ value. As a consequence,
a suppression of the neutrino flux at the detector of one order of magnitude, since the flux decreases with $L^{-2}$. For this reason, $\beta$-Beams based on $^8$Li and $^8$B decays usually suffer from low statistics. This was one of the reasons why ions such as $^{18}$Ne and $^6$He with rather low decay energies were chosen in the original proposal.
Despite their statistical limitations, however, $^8$Li and $^8$B offer the interesting opportunity to probe higher neutrino energies (with respect to $^6$He and $^{18}$Ne beams)
using the maximum $\gamma$ factor achievable at the existing facilities. The higher energies accessible translate in stronger matter effects and generally higher sensitivity to the neutrino mass hierarchy through them.

A key factor in the determination of the best ion candidates is the achievable number of decays per year for each ion. This factor is at present extremely uncertain. For 
$^{18}$Ne and $^6$He ``standard'' fluxes of $1.1 \times 10^{18}$ and $2.9 \times 10^{18}$ useful decays per year are usually assumed. These fluxes would grant the $\gamma =100$ $\beta$-Beam proposal enough sensitivity to compete with similar Super-Beam facilities. Preliminary studies show that this requirement should be achievable for $^6$He ions (the estimations actually yield a flux somewhat larger, of $3.18 \times 10^{18}$ useful decays per year). In the case of $^{18}$Ne, on the other hand, the production of an intense flux is much more challenging and the present estimates fall two orders of magnitude short of the mark, yielding a flux of $4.64 \times 10^{16}$ useful decays per year. We will present results for both assumptions of the fluxes for $^{18}$Ne and $^6$He, the nominal one and that achievable with present studies. A similar analysis of the achievable fluxes for $^8$Li and $^8$B is only in its very early stages. Assuming that the two ions can be produced with the rates described in \cite{Rubbia:2006pi} a preliminary estimate of its fluxes after the acceleration up to $\gamma=100$ through the SPS yielded $3.49 \times 10^{18}$ and $7.56 \times 10^{18}$ useful decays per year for $^8$B and $^8$Li, respectively. We will evaluate the sensitivities achievable with neutrino beams based on the decays of $^8$Li and $^8$B assuming these production fluxes. However, the extraction of this intense $^8$B fluxes could be challenging, since B is known to be very reactive and difficult to release.

The choice of the baseline should match the neutrino energy, since neutrino flavour change oscillates with $L/E$. Too short a baseline will not allow oscillations to develop and a baseline much larger than the one matching the first oscillation peak will reduce unnecessarily the statistics at the detector due to the beam divergence as $L^{-2}$.
There are two sites capable of housing the Mton class water Cerenkov detector proposed to observe the neutrino beam produced at CERN and that match the baseline requirements. 
The shorter baseline is $130$ Km and matches the CERN to Frejus distance. It is suited to observe the first oscillation peak of neutrinos from $^{18}$Ne and $^6$He decays accelerated to $\gamma =100$, and it was the baseline suggested in the original $\beta$-Beam proposal. A longer baseline of $650$ Km would correspond to the CERN to Canfranc distance and would roughly match the first oscillation peak of neutrinos from $^8$Li and $^8$B decays accelerated to $\gamma =100$. Oscillations of neutrinos from $^{18}$Ne and $^6$He could also be observed at this baseline at the second peak for $\gamma=100$ or at the first peak for $\gamma=350$ (achievable with a refurbished SPS \cite{Burguet-Castell:2003vv}).

\begin{figure}[t!]
\vspace{-0.5cm}
\begin{center}
\begin{tabular}{cc}
\hspace{-0.55cm} \epsfxsize7.5cm\epsffile{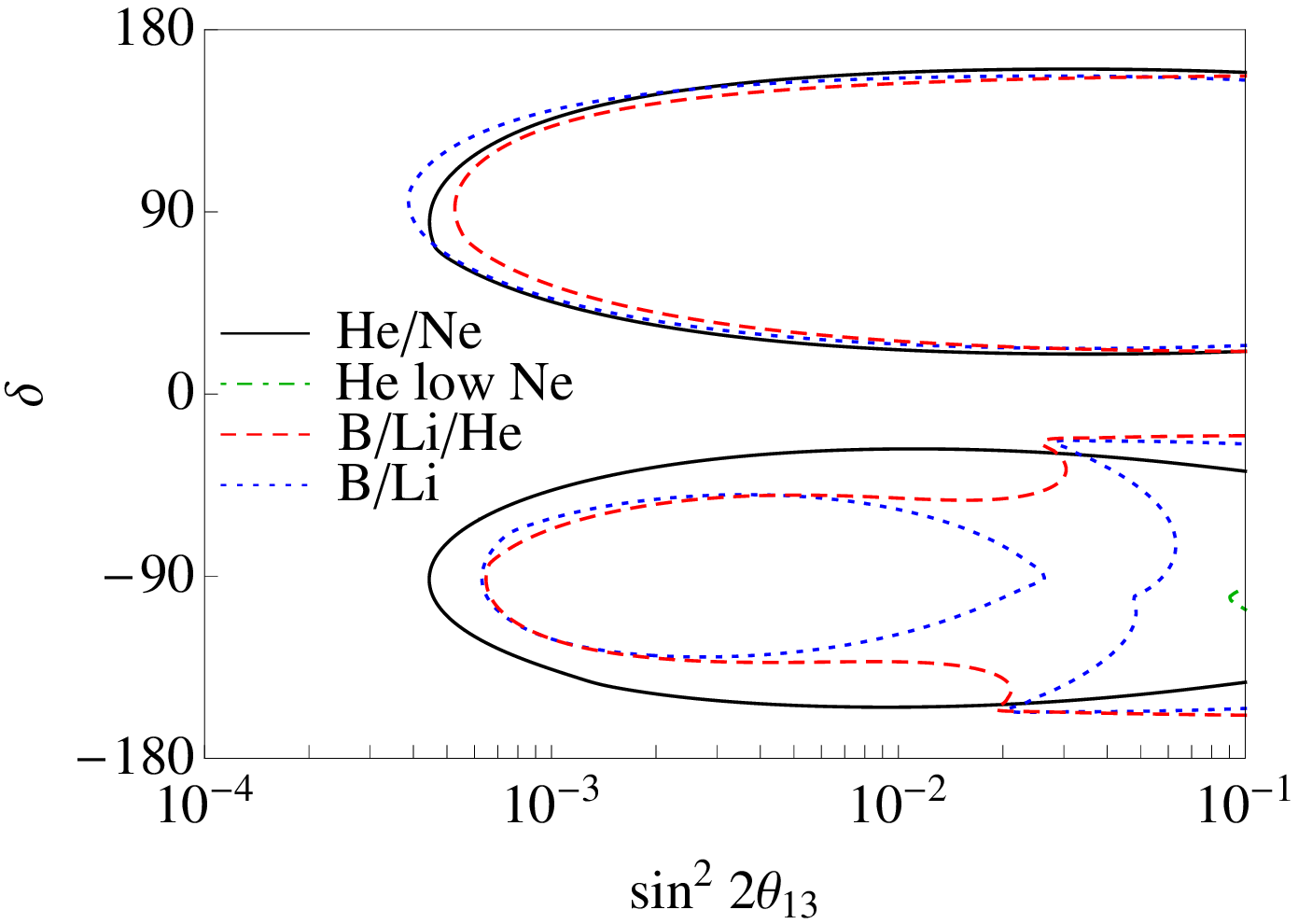} & 
                 \epsfxsize7.5cm\epsffile{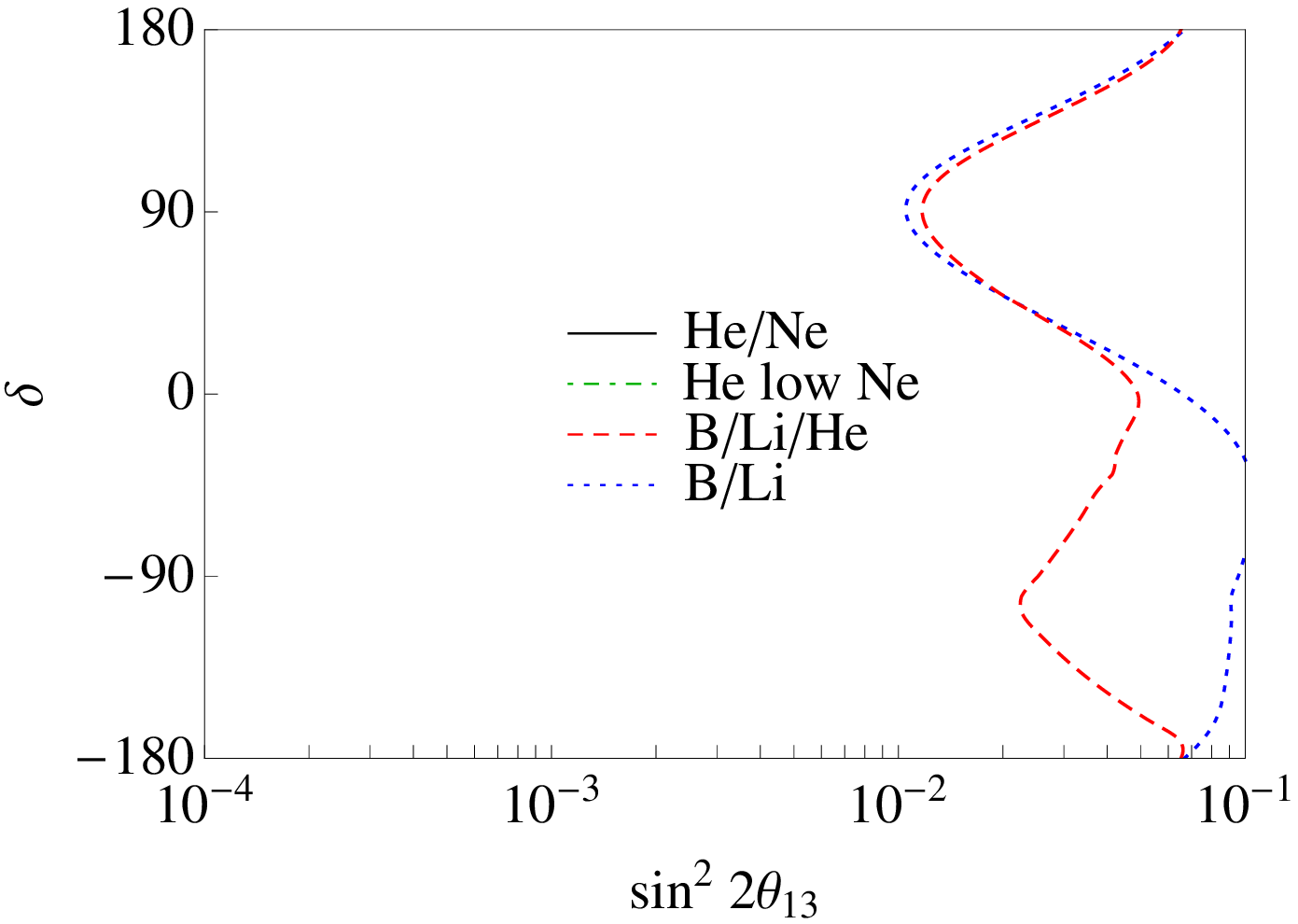}
\end{tabular}
\caption{\label{fig:setup}
Comparison of the 3 $\sigma$ discovery potential of different $\beta$-Beam setups to leptonic CP-violation (left panel) and the mass hierarchy (right panel).
Solid, dotdashed, dotted and dashed lines correspond to the setups, 1, 2, 3 and 4 described in the text, respectively. Taken from Ref.~\cite{FernandezMartinez:2009hb}.}
\end{center}
\end{figure}

With the four ion candidates and two baselines available we have explored the sensitivities to the unknown parameters of four possible $\beta$-Beam setups:
\begin{itemize}
\item 1. The ``standard'' setup with neutrinos from $^{18}$Ne and $^6$He decays accelerated to $\gamma=100$. The baseline is $L=130$ Km and the fluxes are given by $1.1 \times 10^{18}$ and $2.9 \times 10^{18}$ useful decays per year for $^{18}$Ne and $^6$He, respectively. We assumed a 5 year run with each ion. This setup is represented by the black solid line in Fig~\ref{fig:setup}.

\item 2. The same as setup 1 but with the estimated fluxes of $4.64 \times 10^{16}$ and $3.18 \times 10^{18}$ useful decays per yer
for $^{18}$Ne and $^6$He, respectively. This is depicted by the green dotdashed line in Fig~\ref{fig:setup}.

\item 3. A setup based in $^8$Li and $^8$B decays accelerated to $\gamma=100$. The baseline is $L=650$ Km and the fluxes $3.49 \times 10^{18}$ and $7.56 \times 10^{18}$ useful decays per year for $^8$B and $^8$Li, respectively. We assumed a 5 year run with each ion. This setup is represented by the blue dotted line in Fig~\ref{fig:setup}.

\item 4. A setup based in $^8$Li, $^8$B and $^6$He decays accelerated to $\gamma=100$. The baseline is $L=650$ Km; the same fluxes as for setups 2 and 3 were assumed. We assumed a 5, 3 and 2 year run with $^8$B, $^8$Li and $^6$He, respectively. This setup is represented by the red dashed line in Fig~\ref{fig:setup}.
\end{itemize}

The expected efficiencies and beam-induced backgrounds of the detector when exposed to the considered beams has been added as migration matrices extracted from Ref.~\cite{Burguet-Castell:2005pa}. In all the simulations the following best fit values and $1 \sigma$ errors for the known oscillation parameters were assumed: $\Delta m^2_{21} = (8.0 \pm 0.3)\times 10^{-5}$ eV$^2$, $\Delta m^2_{31} = (2.6 \pm 0.1)\times 10^{-3}$ eV$^2$, $\theta_{12} = 33.0^\circ \pm 1.3^\circ$ and $\theta_{23} = 45.0^\circ \pm 4.5^\circ$. These parameters were marginalized over to present the final curves. The background caused by atmospheric neutrinos in the detector was neglected in this section but will be discussed in detail in the next section. The evaluation of the performance of the detector made use of the GLoBES software \cite{Huber:2004ka,Huber:2007ji}. 

The expected CP discovery potential (defined as the values of $\theta_{13}$ and $\delta$ that would allow to discard at $3 \sigma$ the CP-conserving values $\delta=0$ and $\delta=\pi$) of the four setups considered is presented in the left panel of Fig~\ref{fig:setup}. The discovery potential down to the smallest values of $\theta_{13}$ corresponds to setup 1, since the shorter baseline guarantees higher statistics. However, taking into account the very low fluxes estimated for $^{18}$Ne in setup 2, the sensitivity to CP-violation is lost. Setup 3, based on $^8$B and $^8$Li decays has a smaller statistics and worse sensitivity for small values of $\theta_{13}$. Moreover, the stronger matter effects present at these higher energies and baseline can mimic true CP-violation and lead to degeneracies that translate into the loss of sensitivity for negative values of $\delta$ around $\sin^2 2 \theta_{13} = 3 \times 10^{-2}$. In order to alleviate this degeneracy problem we introduced setup 4: the combination of information at the first oscillation peak from $^8$B and $^8$Li beams with that from a $^6$He beam at the second oscillation peak is enough to solve the sign degeneracies and fill in the gap in sensitivity for negative values of $\delta$ \cite{Donini:2006dx}. 

The term in the oscillation probability that provides the sensitivity to CP-violation is suppressed by $\sin 2 \theta_{13}$ and $\Delta m^2_{21} L/E$ and has to compete with a $\delta$-independent term suppressed by $\sin^2 2 \theta_{13}$. The sensitivity to CP-violation thus decreases for the largest values of $\theta_{13}$, where the $\delta$-independent term dominates. However, in setups 3 and 4 the value of $L/E$ is larger for the lower energy bins than in setup 1 and, thus, their sensitivity to CP-violation outperforms that of setup 1 for large values of $\theta_{13}$ even if their statistics is lower. It should also be remarked that the expected statistics in the detector strongly depends of the assumed value of the cross section which, at very low energies below $1$ GeV, is plagued by nuclear effects uncertainties and different computations can differ up to a factor two, leading to very different estimations of the sensitivity for the lower energy setups 1 and 2. The cross section assumed here is particularly optimistic at these low energies, which emphasizes the statistics difference between setup 1 and setups 3 and 4, allowing it to reach smaller values of $\theta_{13}$.   

The expected discovery potential to a normal mass hierarchy, defined as the values of $\theta_{13}$ and $\delta$ that would allow to discard an inverted hierarchy at $3 \sigma$, of the four setups considered is presented in the right panel of Fig~\ref{fig:setup}. Notice that only setups 3 and 4 show some sensitivity to mass hierarchy. Setups 1 and 2 have too small matter effects due to the low energy of the beam and the shorter baseline to be able to measure the mass hierarchy by themselves. Nevertheless, some sensitivity to the mass hierarchy could be gained when combining their information with atmospheric neutrino oscillations measured at the detector \cite{Campagne:2006yx}. The better sensitivity of setup 4 with respect to setup 3 for negative values of $\delta$ is again due to the complementarity of the information on the oscillation probability of $^6$He at the second oscillation peak that allows to break degeneracies when combined with $^8$Li and $^8$B at the first peak \cite{Donini:2006dx}. 

We conclude that the setups 3 and 4 based on $^8$Li and $^8$B decays outperform the standard setup 1 based on $^{18}$Ne and $^6$He decays for values of $\sin^2 2 \theta_{13} > 2 \times 10^{-2}$, providing a better sensitivity to CP-violation and to the mass hierarchy. For small values of $\theta_{13}$, setup 1, with its better statistics, is the better option. However, if it turns out that the present estimation of the achievable $^{18}$Ne fluxes cannot be improved, all the sensitivity to CP-violation is lost (setup 2) and alternative setups based on $^8$Li and $^8$B decays would remain the only option. 

\subsection{The atmospheric neutrino background}
\label{sec:atmo}

One of the main sources of background that can spoil the $\beta$-Beam sensitivity is the background from atmospheric neutrinos. This background can be reduced by imposing angular cuts in the direction of the beam. The typical scattering angle of the lepton produced in CC interactions with respect to the incident neutrino is $\sim \sqrt{1/E}$. We have evaluated the expected number of muon neutrinos arriving within a solid angle $\sim \sqrt{1/E}$ in the beam direction for the different energy beams and studied its impact on the sensitivities. Even after imposing this directional cut the background level dominates the expected signal and an additional cut must be imposed to reduce it to acceptable levels. This can be achieved by accumulating the signal in small bunches so as to use timing information to reduce the constant atmospheric background. Previous analysis showed that for the standard setup the decaying ions must be accumulated in very small bunches occupying just a $\sim 10^{-3}$ fraction of the storage ring so as to achieve a $10^{-3}$ suppression factor of the background. Since the atmospheric neutrino background decreases with the energy as $\sim E^{-2}$, for the setups considered here based on $^8$Li and $^8$B decays with $\sim 3.5$ times higher energy, an order of magnitude less atmospheric background can be expected at the detector. We studied whether this allows to relax the stringent $\sim 10^{-3}$ bunching required on the signal. Notice that the achievable ion flux is strongly affected by this strong requirement and a relaxation of this value could allow an increase in statistics.

\begin{figure}[t!]
\vspace{-0.5cm}
\begin{center}
\begin{tabular}{cc}
\hspace{-0.55cm} \epsfxsize7.5cm\epsffile{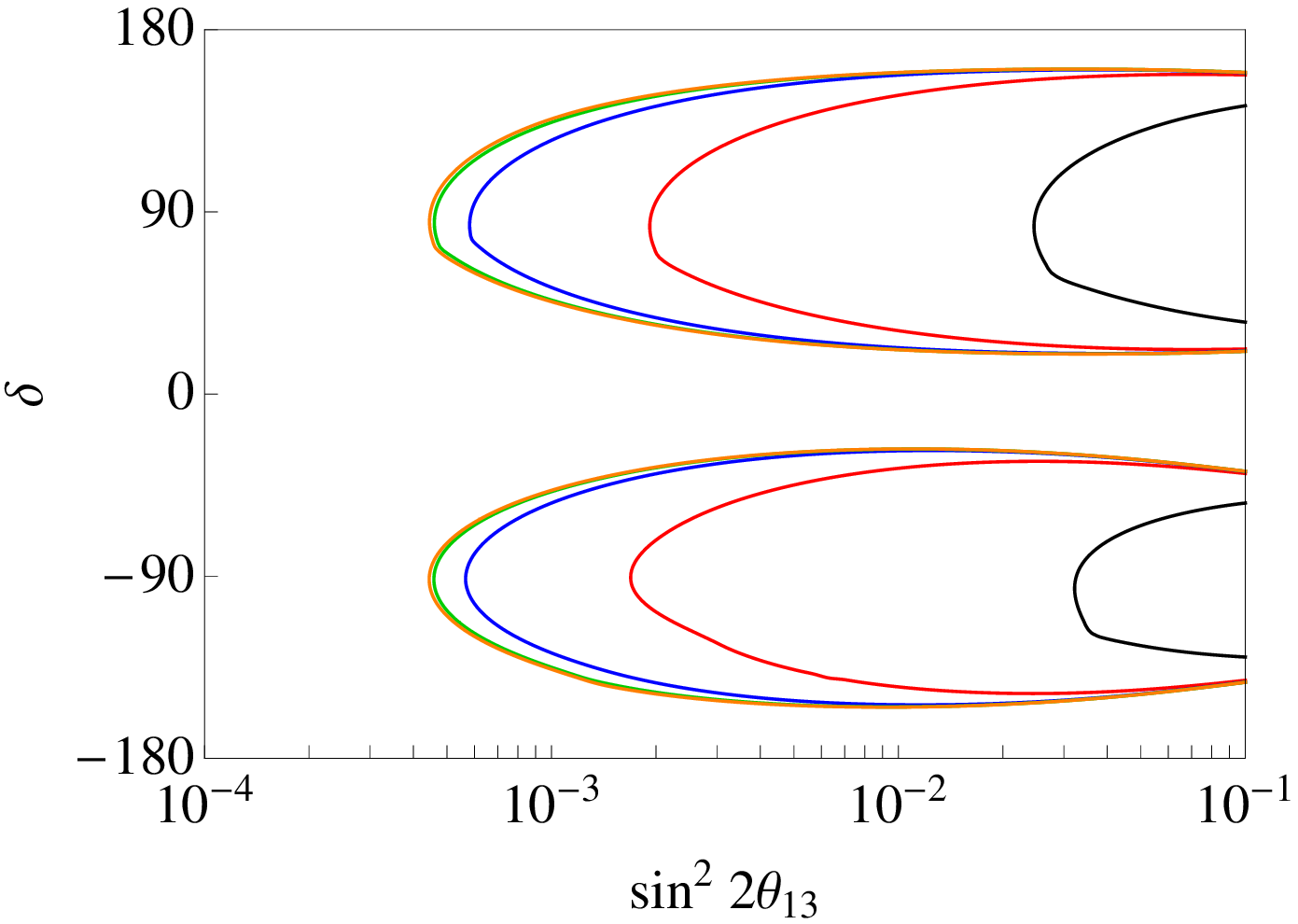} & 
                 \epsfxsize7.5cm\epsffile{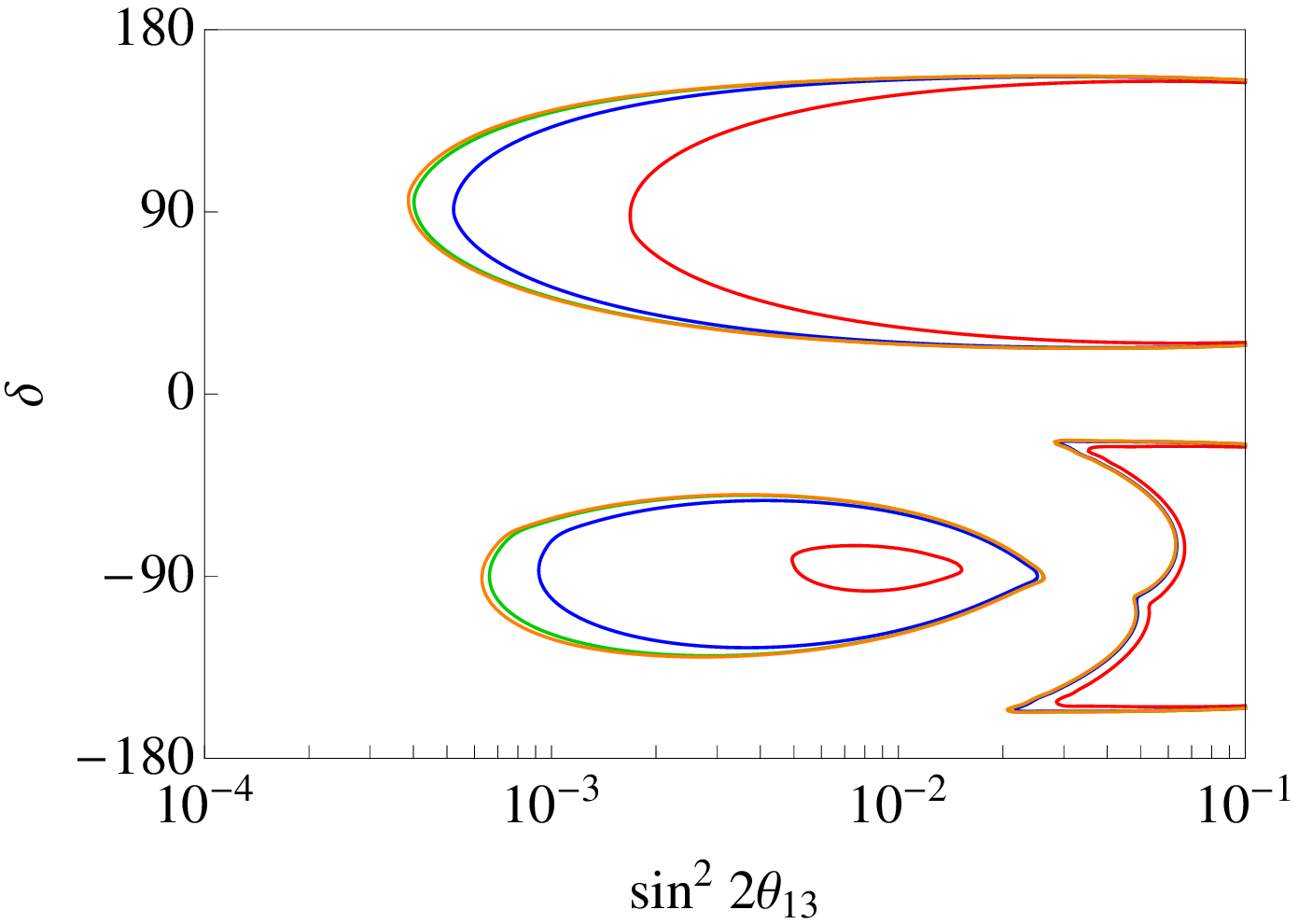} \\
\hspace{-0.55cm} \epsfxsize7.5cm\epsffile{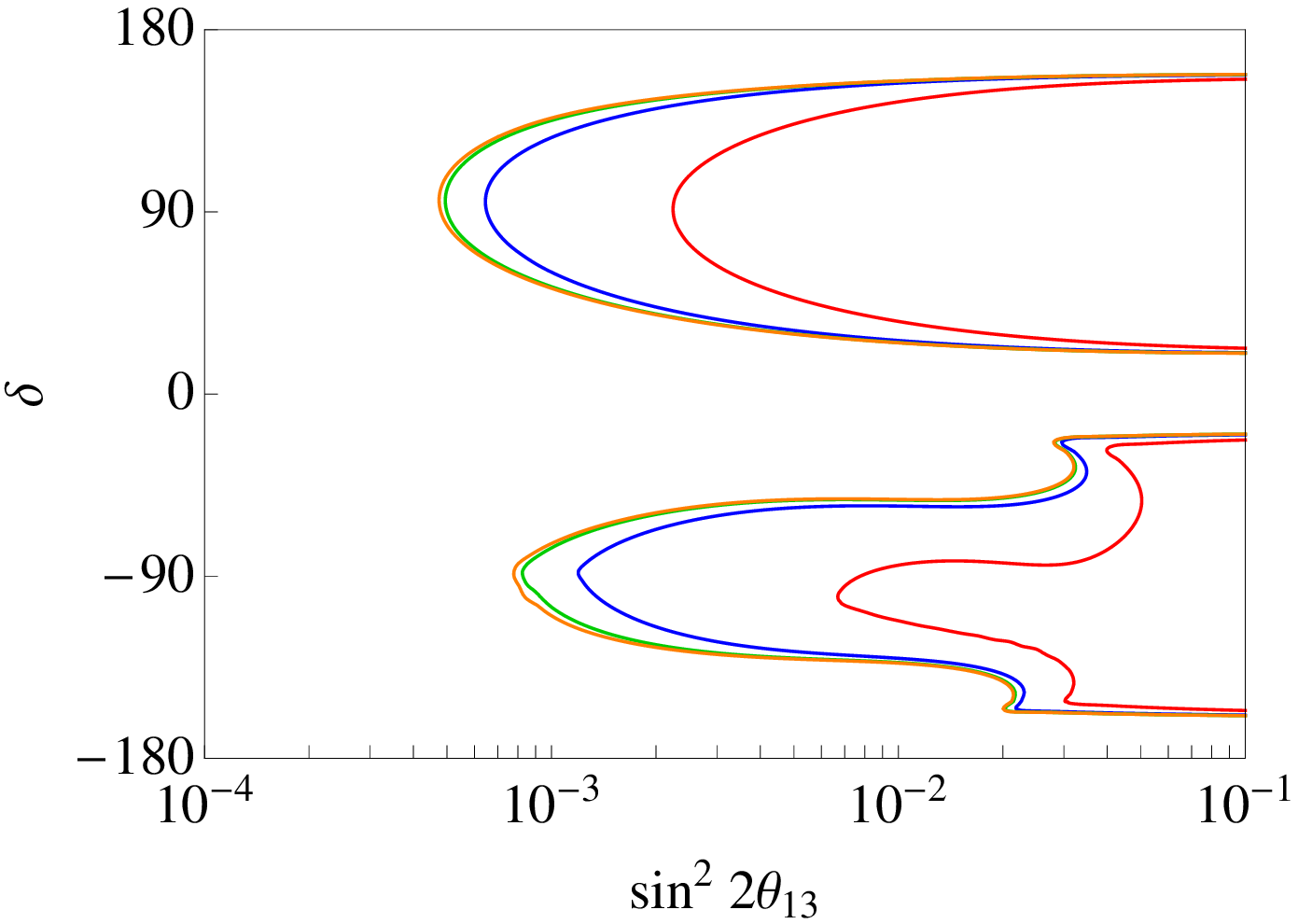} & 
                 \epsfxsize7.5cm\epsffile{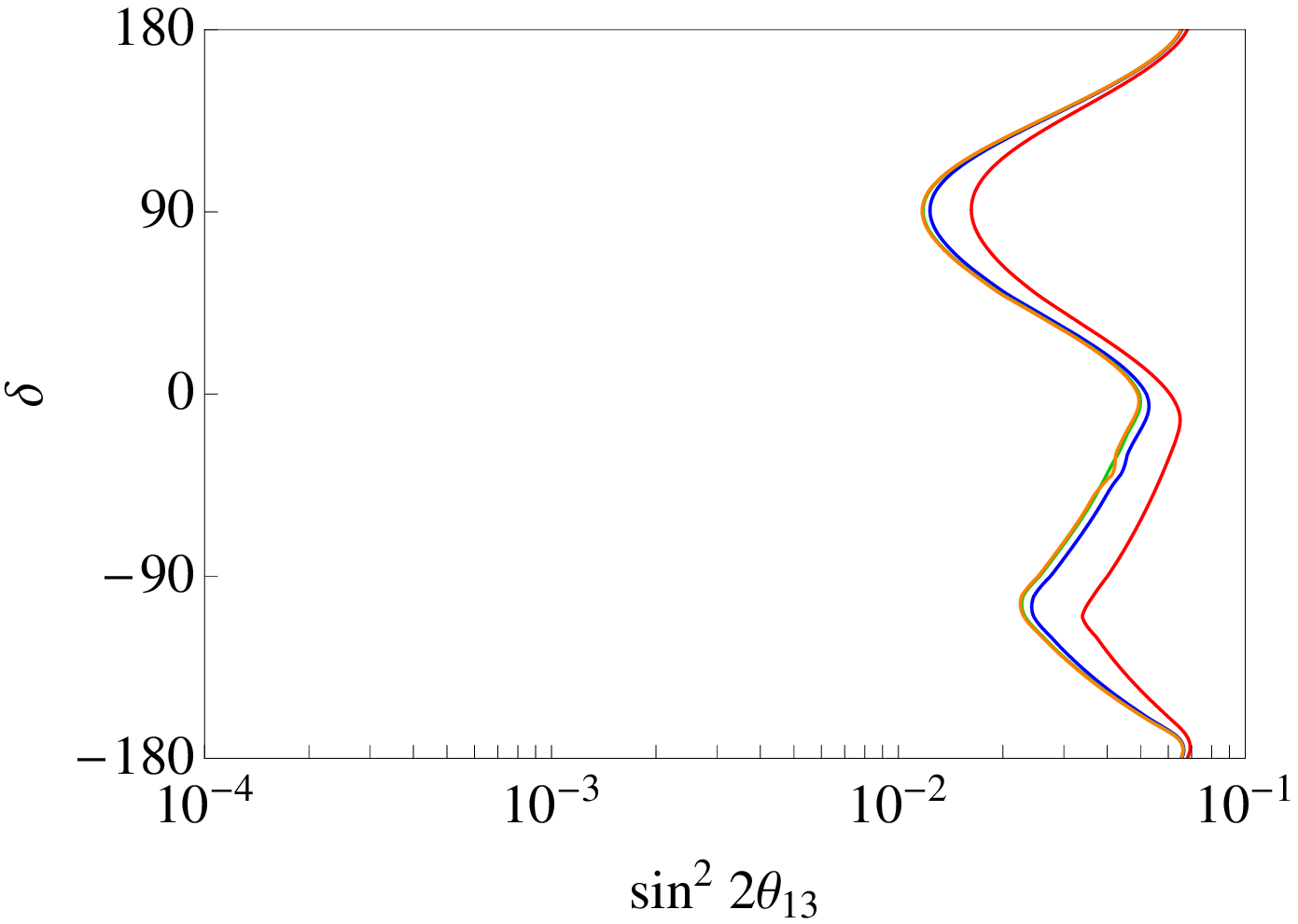} 
\end{tabular}
\caption{\label{fig:atmo}
Dependence on the background suppression factor of the the 3 $\sigma$ discovery potential to leptonic CP-violation for setup 1 (upper left), setup 3 (upper right) and setup 4 (bottom left). 
The bottom right panel represents the sensitivity to the mass hierarchy for setup 4. 
The lines correspond to no atmospheric background (orange) and to suppression factors of $10^{-4}$ (green), $10^{-3}$ (blue) and $10^{-2}$ (red).  
Taken from Ref.~\cite{FernandezMartinez:2009hb}.}
\end{center}
\end{figure}

In Fig.~\ref{fig:atmo} we show the dependence on the background suppression factor of the 3$\sigma$ discovery potential to leptonic CP-violation (left panels) and the mass hierarchy (right panels). The lines correspond to no atmospheric background (orange) and to suppression factors of  $10^{-4}$ (green), $10^{-3}$ (blue) and $10^{-2}$ (red). The upper panels
correspond to setup 1 (left) and 3 (right), respectively. We can see from these two plots that there is almost no difference between the sensitivity with no atmospheric background (orange line) and with a $10^{-4}$ background (green line). The atmospheric background thus becomes negligible if a $10^{-4}$ suppression is achieved. The sensitivities for the $10^{-3}$ suppression case (blue lines) are only slightly worse than those for $10^{-4}$. Thus, $10^{-3}$ seems the goal that should be achieved in order to exploit the full potential of the  $\beta$-Beam. However, it can be seen that, whilts setup 1 have a CP discovery potential that is independent on the sign of the CP-violating phase, setup 3 (that is more sensible to matter effects) suffers from degeneracies for negative $\delta$. This is cured by the addition of a component of $^6$He beam in setup 4 that, combined with the two high-$Q$ beams solve the sign degeneracy and restore symmetry between positive and negative $\delta$ values (lower panel, right). The sensitivity to the mass hierarchy in setup 4 can be seen in lower panel (right)
as a function of the suppression factor. Notice that, for all setups, the CP discovery potential is strongly degraded for a suppression factor of $10^{-2}$.

We conclude that in all the setups studied a background suppression of $\sim 10^{-3}$ is required to achieve the full sensitivity of the $\beta$-Beam. The fact that the higher energy beams provided by $^8$Li and $^8$B decays require the same background suppression than those based on $^{18}$Ne and $^6$He is the non-trivial consequence of many factors.

The suppression of the signal for the high-$Q$ beam is not fully compensated by the higher cross section at higher energies (more than a factor $3.5$ larger since the cross section grows faster than linearly at those energies) and by the higher flux assumed here for $^8$B and $^8$Li, around three times larger. 
Eventually, the efficiency of the detector gets degraded at high energies, since water \v Cerenkov detectors are not optimal beyond the quasi-elastic regime of the cross section. In particular, following Ref.~\cite{Burguet-Castell:2005pa}, we find the efficiencies to be around more than a factor two smaller for the higher $Q$ ions than for $^{18}$Ne and $^6$He. This means that the signal/background fraction remains similar in all scenarios and a suppression factor $\sim 10^{-3}$ is always required to achieve the full $\beta$-Beam potential.

\subsection{Greenfield scenario optimization: two baseline betabeam}
\label{sec:twobaseline}

It is well-known that for $L\sim 7000 $ km the density encountered by the (anti)neutrinos allows for a resonant enhancement in the probability when $E_\nu \sim 6$~GeV if the mass hierarchy is normal (inverted). Therefore, \br and \li turn out to be the best candidates to determine the mass hierarchy, as their end point energies are much higher than those of the ``standard'' ions, \he and $^{18}$Ne.

However, another detector at a short baseline is needed in order to probe CP-violation, as $\delta$ cannot be measured at the magic baseline. For neutrinos coming from $^6$He and $^{18}$Ne decays at $\gamma = 350$, the mean neutrino energy of $E_0 \gamma \sim 1.2$~GeV translates into an on-peak baseline of $L=618$~km, while for \br and \li this baseline would be $L\sim 1500-2000$ km. As the flux is proportional to $L^{-2}$, \he and \neon turn out to be the best candidates to probe CP-violation.


The original design of the storage ring proposed in~\cite{Zucchelli:2002sa} must be modified when the boost factor is increased up to $\gamma=350$. If we use LHC dipolar magnets to bend the ions, and keeping the straight sections untouched, the useful fraction of ion decays for this ring (also called ``livetime'') would be\footnote{$l_{\rm racetrack} = \frac{L_{\rm straight}}{2 L_{\rm straight} + 2 \pi R}$} {\em l} = 0.28, the total length being $L_r = 8974$ m. The tilt angle needed to aim at the near detector is $\vartheta = 3^\circ$: this means that the maximum depth of the far end of the ring is $d = 197$ m. However, the tilt angle to aim at $L=7000$ km is $\vartheta = 34.5^\circ$. In this case, the ring would reach a depth of $d = 2132$ m, something well beyond any realistic possibility. 

Note that with the refurbished SPS (SPS+) the \br and \li ions could be accelerated up to $\gamma = 650$ and $\gamma=390$, respectively. Due to the resonance, this $10\%$ increase in the $\gamma$ factor for \li produces an increase of a $40\%$ in the number of antineutrino events at the detector. Therefore, we can reduce the livetime {\em l} = $0.6 \times 0.28 \sim 0.17$ by reducing the straight section of the ring to $L=998$ m, and the physics reach of the setup will remain practically unaffected. This ring would be almost 1 Km less deep than the one described above, $d = 1282$ m. 


We make the following choices for our detectors: (1) Since CP measurements are better at lower energies, it is preferable to have a detector with low threshold and good energy resolution. We opt for a water \chr detector with 500 kton fiducial mass (as in Refs.~\cite{Burguet-Castell:2005pa,Burguet-Castell:2003vv}). This detector could be hosted at Canfranc, for example, at a distance of 650 km from the $\beta$-Beam at CERN; (2) Higher energy neutrinos from highly boosted \br and \li ions will be sent to the far detector. We prefer thus to choose a 50 kton magnetized iron detector for the magic baseline: the ICAL@INO detector in India \cite{Athar:2006yb}, located at a distance of 7152 km, which will soon go under construction. 

To simulate the response of the water \chr detector we use the migration matrices given in \cite{Burguet-Castell:2003vv}, while for the iron detector, we follow the efficiencies and backgrounds derived in \cite{Abe:2007bi} for the Neutrino Factory (NF) fluxes. Notice that this is a very conservative assumption, since for a \bb charge identification is not mandatory, unlike in the NF. Therefore, the charge ID capability of the detector could be used to further reduce the background. Finally, the performance of the iron detector is affected by large uncertainties at energies around $1-5$ GeV. However, we need this detector only to observe the resonance in the probability which takes place around $6-7$ GeV, so these uncertainties will practically have no effect at all in the performance of our setup. 

Preliminary studies on the ion production rates show that these could be enhanced in the near future. Therefore, as $\beta$-Beams are facilities under study for construction in the next two decades, we will assume that $10^{19}$ ions per year can be stored into the ring, for all ion species\cite{Donini:2008zz}, with a livetime $l = 0.28$ for the ring aiming at the 650 Km baseline and $l = 0.6 \times 0.28$ for the ring aiming at the magic baseline.  For $^6$He and $^{18}$Ne beams, this means that we are considering $\sim 3 \times 10^{18}$ useful decays per year aiming at the 650 Km detector. This choice corresponds to the nominal flux for $^6$He ions and roughly twice the nominal flux for $^{18}$Ne ions\footnote{Notice that, as it was 
stressed in Sect.~\ref{sec:enrique}, at present a flux of only $4.6 \times 10^{16}$ useful $^{18}$Ne ion decays are achievable.}.
For the four ions setup proposed in Ref.~\cite{Choubey:2009ks}, a runtime with each ion species of 2.5 years is considered. 

\begin{figure}[t]\begin{tabular}{cc}
\includegraphics[width=0.40\textwidth,angle=0]{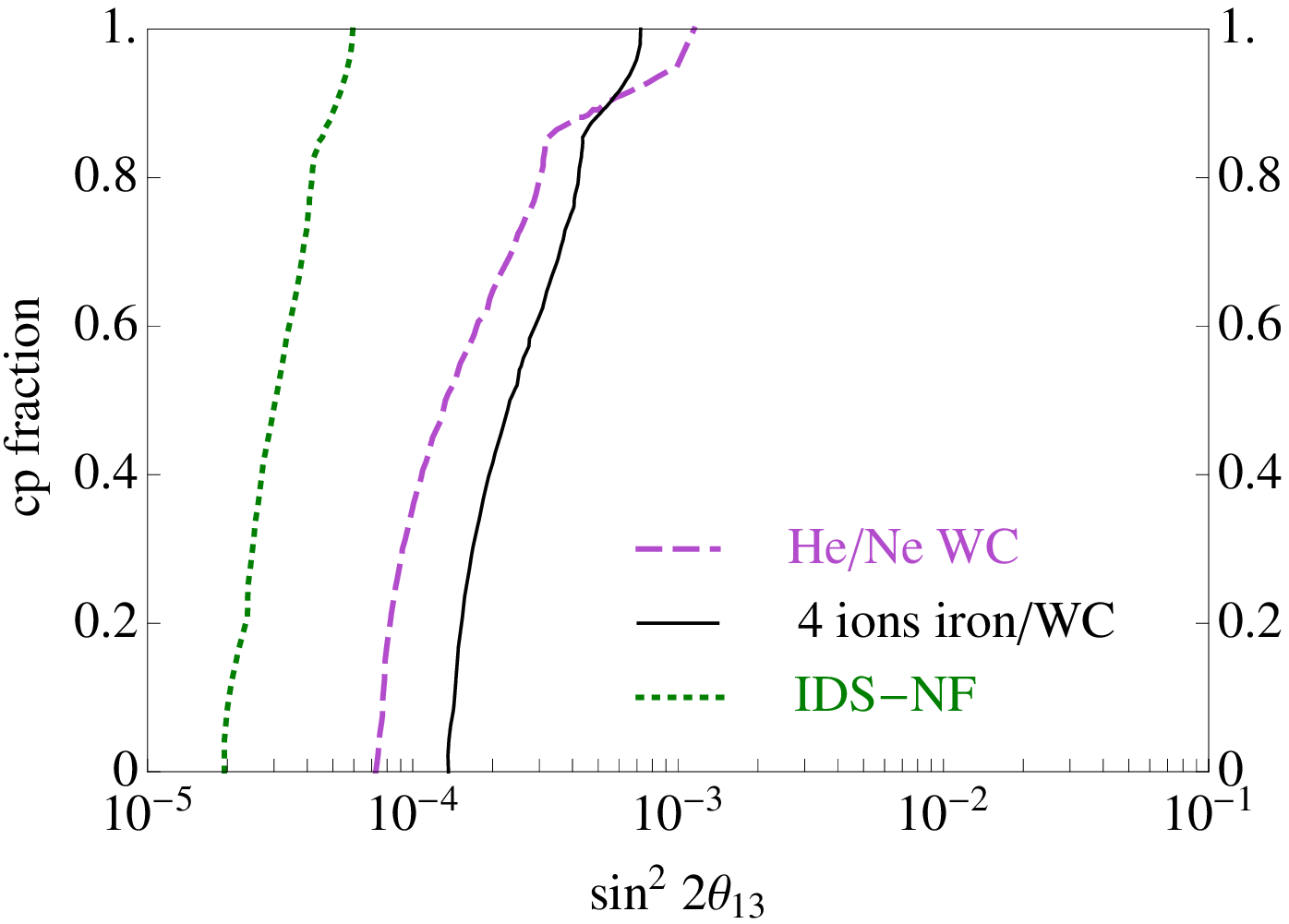}&
\includegraphics[width=0.40\textwidth,angle=0]{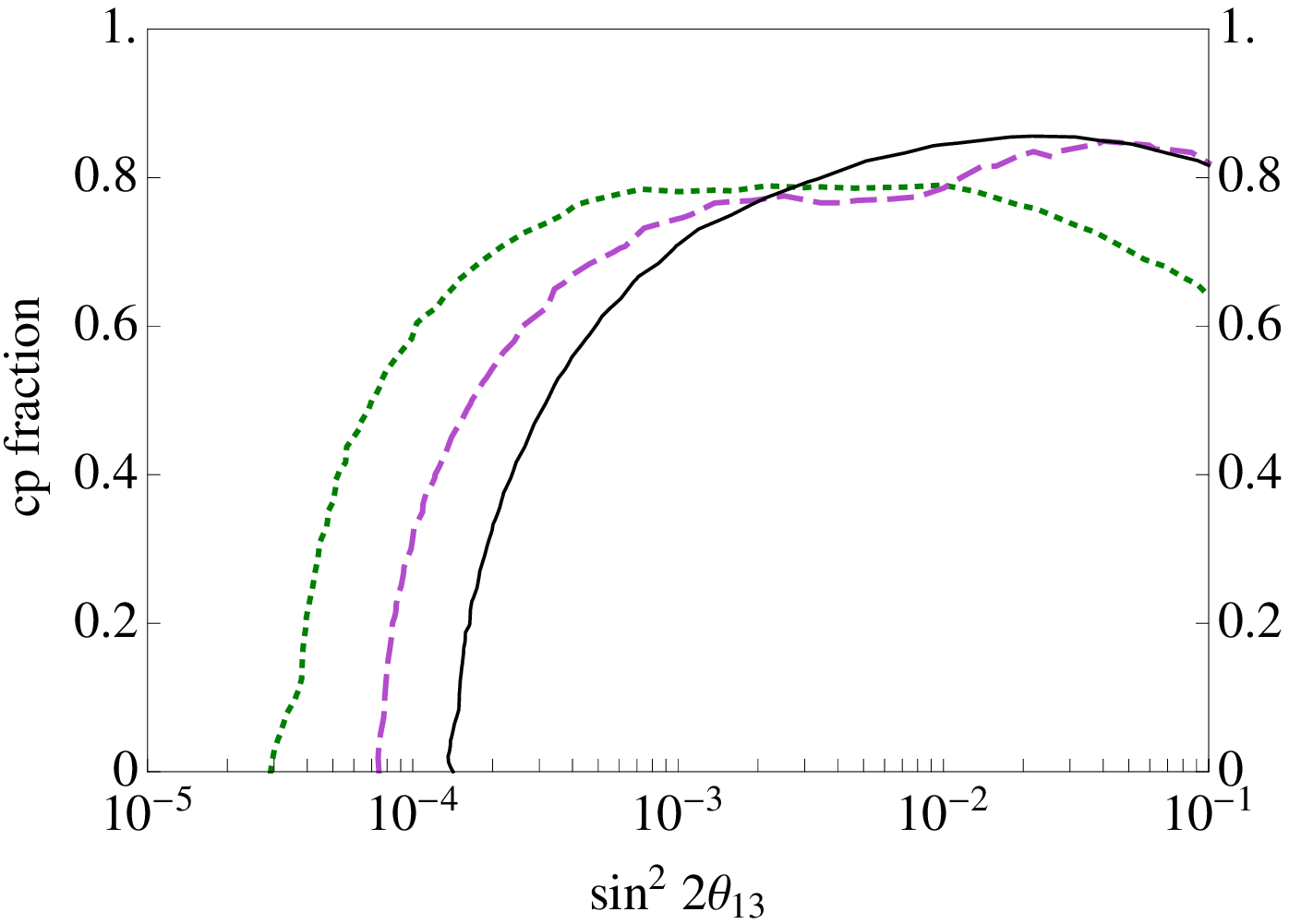}\\
\includegraphics[width=0.40\textwidth,angle=0]{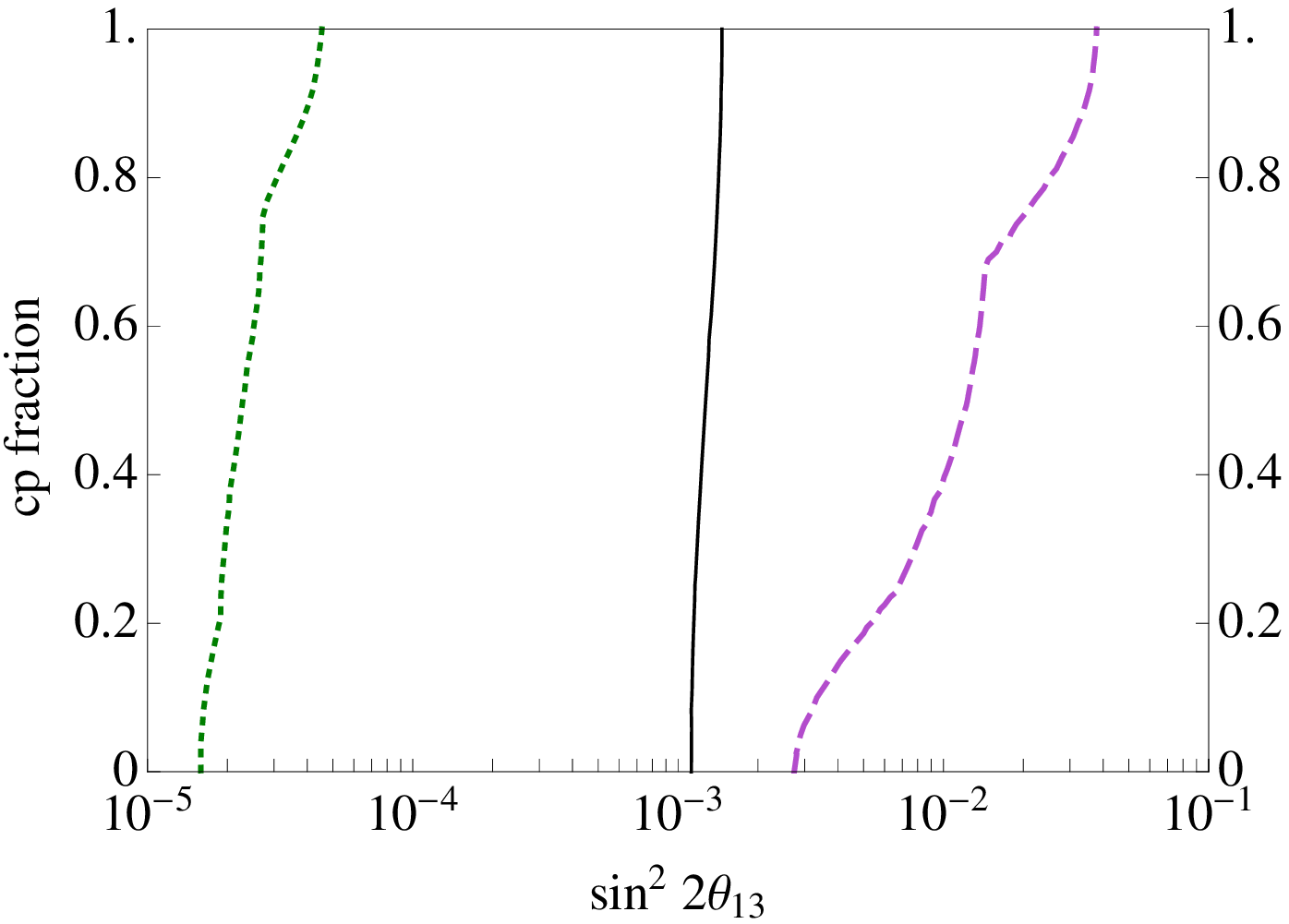}&
\includegraphics[width=0.40\textwidth,angle=0]{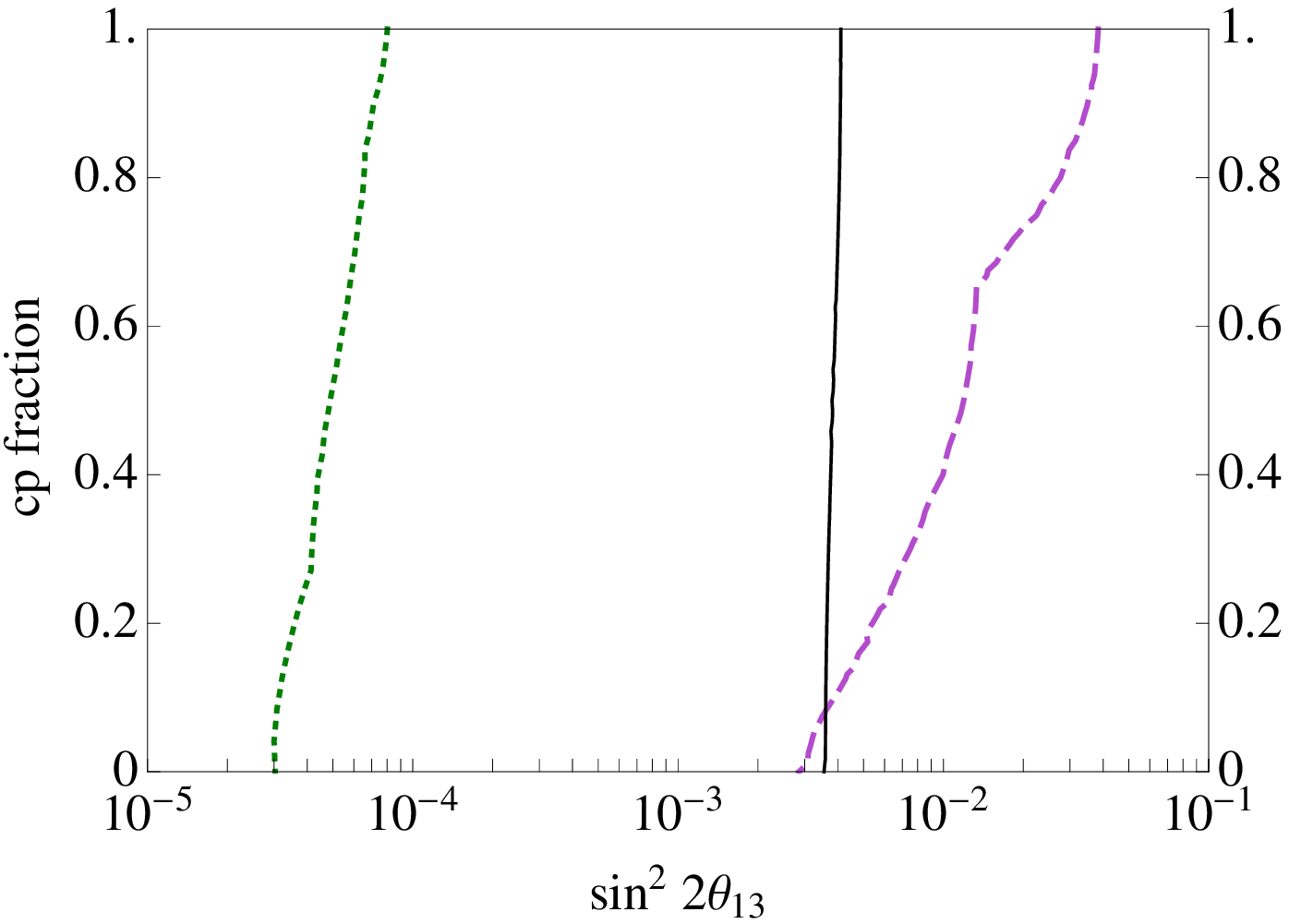}\\
\end{tabular}

\caption{\label{fig:senscpf}
Comparison of our proposed set-up (black solid lines) with the IDS Neutrino Factory baseline design (green dotted lines) and 
the high $\gamma$ \bb set-up from \cite{Burguet-Castell:2005pa,Burguet-Castell:2003vv}. The upper left hand panel shows $\stch$ discovery reach, the upper right hand 
panels shows the CP violation reach, and the lower panels show the mass hierarchy discovery reach for normal (left panel) and inverted (right pannel) hierarchy. All sensitivities are presented as a function of the fraction of the values of $\delta$ for which they can be discovered. Taken from Ref.~\cite{Choubey:2009ks}.}
\end{figure}

We present in Fig.~\ref{fig:senscpf} the comparison of the performance of our setup with the IDS Neutrino Factory design \cite{Bandyopadhyay:2007kx} (with 25 GeV muons stored in
two racetrack rings aiming at two 50 Kton magnetized iron detectors of the MIND-type located at $L = 4000$ Km and $L = 7500$, respectively, with $5 \times 10^{20}$ useful muon decays per year aiming at each detector) and the one-baseline \bb set-up proposed in~\cite{Burguet-Castell:2005pa,Burguet-Castell:2003vv}, where neutrino beams produced by \neon and \he decays accelerated to $\gamma =350$ are detected in a 500 kton water \chr detector located at 650 km. In both setups, five years of data taking with neutrinos and anti-neutrinos are
considered.

We have considered 2.5\% and 5\% systematic errors on the signal and on the beam-induced background, respectively. They have been included as ``pulls'' in the statistical $\chi^2$ analysis. The following $1 \sigma$ errors for the oscillation parameters were also considered: 
$\delta \theta_{12} = 1\%$, $\delta \theta_{23} = 5\%$, $\delta \Delta m^2_{21} = 1 \%$ and $\Delta m^2_{31} = 2\%$. 
Eventually, an error $\delta A = 5\%$ has been considered for the Earth density given by the PREM model \cite{Dziewonski:1981xy,prem}. 
Marginalization over these parameters has been performed for all observables. The Globes 3.0 \cite{Huber:2004ka,Huber:2007ji} software was used to perform the numerical analysis.


We quantify the physics reach of the experiments in terms of three different performance indicators: 
\begin{enumerate}
\item The $\stch$ discovery reach: This is the minimum true value of $\stch$ for which the experiment can rule out at 
the 1 d.o.f. $3\sigma$ the value $\stch=0$ in the fit, after marginalizing over all the other parameters. 
\item The CP-violation 
reach: This is the range of $\delta$ as a function of $\stch$ which can rule out no CP-violation ($\delta=0$ and $180^\circ$) at the 1 d.o.f. $3\sigma$, 
after marginalizing over all the other parameters. 
\item The $sgn(\ma)$ reach in $\stch$: This is defined as the limiting value of $\stch$ for which the wrong hierarchy can be 
eliminated at $3\sigma$. 
\end{enumerate}


From Fig.~\ref{fig:senscpf} it is clear that the facility with sensitivity to the different observables down to smallest values of $\stch$ is the Neutrino Factory. This can be understood from the much larger fluxes assumed for the IDS setup: $5\times10^{20}$ useful muon decays per year and per baseline to be compared to the $3 \times 10^{18}$ assumed for the $\beta$-Beams. On the other hand, the high energy of the Neutrino Factory beams implies a very small value of $L/E_\nu$. This translates into a stronger suppression of the CP-violating term of the oscillation probability with respect to the one suppressed by two powers of $\theta_{13}$ for large values of this parameter. Therefore, the CP discovery potential of $\beta$-Beams outperforms that of the Neutrino Factory in Fig.~\ref{fig:senscpf} when $\sin^2 2 \theta_{13}~>~10^{-3}$.  Since this large value of $\sin^2 2 \theta_{13}$ also guarantees a discovery of the mass hierarchy and $\stch$ regardless of the value of $\delta$, this makes $\beta$-Beams the best option when $\sin^2 2 \theta_{13}~>~10^{-3}$. Furthermore, even if the statistics at the near detector is reduced by half in our proposal compared to the one in Ref.~\cite{Burguet-Castell:2005pa,Burguet-Castell:2003vv} (we are running only 2.5 years with $^6$He and $^{18}$Ne beams, instead of five years), the CP-discovery potential for $\sin^2 2 \theta_{13}~>~10^{-3}$ is better in the two-baseline set-up due to the lifting of the degeneracies that can mimic CP-conservation when combining the information from the two detectors.

\subsection{Measuring absolute neutrino mass with Beta Beams}
\label{sec:mossbauer}

\begin{figure}
\begin{center}
\resizebox{0.4\textwidth}{!}{%
\includegraphics{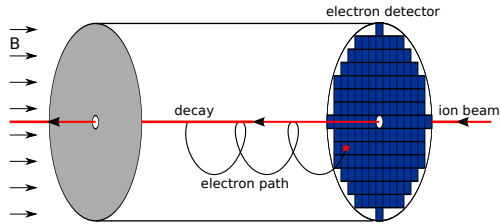}
}
\end{center}
\caption{Measuring absolute neutrino mass with ion beams.  The ion
  beam enters an evacuated cavity whose back wall holds an electron
  detector. Each ejected electron follows a helical
  trajectory. Electrons moving in the backward direction in the
  laboratory frame are counted by the detector. Taken from Ref.~\cite{Lindroos:2009mx}.}
\label{Fi:diagram}
\end{figure}

In Ref.~\cite{Lindroos:2009mx} a method has be proposed to measure
the neutrino mass kinematically using beams of ions which undergo beta
decay. The idea is to tune the ion beam momentum so that in most
decays, the electron is forward moving with respect to the beam, and
only in decays near the endpoint is the electron moving backwards, see
Fig.~\ref{Fi:diagram}. Then, by counting the backward moving electrons
one can observe the effect of neutrino mass on the beta spectrum close
to the endpoint. 
There are stringent requirements of the proposed setup in order to
exceed the sensitivity of 0.2~eV of the latest generation of Tritium
and Rhenium decay experiments. A crucial question is whether it will
be possible to accelerate enough ions within reasonable time such that
of order $10^{18} - 10^{20}$ decays can be observed. This issue is
also related to the identification of a suitable ion with a low enough
$Q$-value (in order to maximize the effect of the neutrino mass), and
a small enough half life (in order to have a high enough decay rate).

As an example, for a very low $Q$-value of 2~keV, one needs $4\times
10^{16}$ decays to obtain the KATRIN sensitivity of 0.2~eV, while
$10^{19}$ decays will allow for a $m^{\rm eff}_\nu$ measurement at
0.04~eV. On the other hand, if no suitable ion with such a low
$Q$-value can be identified the requirements on the total number of
decays increases drastically: for $Q = 4 \, (8)$~keV the 0.2~eV
sensitivity is reached for $5\times 10^{17}$ ($8\times 10^{18}$) decays,
respectively.  The sensitivity goal of $m^{\rm eff}_\nu<0.04$~eV,
which will separate the normal and inverted neutrino mass hierarchy
regions, requires in excess of $10^{19}$ counts across the run of the
experiment together with a $Q$-value of 2~keV.
Further requirements are control over ion momentum with a precision
better than $\delta p/p < 10^{-5}$, and separation of forward and
backward going electrons with very good precision.

\subsection{Progress in monochromatic betabeams}
\label{sec:mono}

The next generation of long baseline neutrino oscillation
experiments will aim at determining the unknown mixing angle $\theta_{13}$,
the type of neutrino mass hierarchy and CP-violation. We discuss the
separation of these properties by means of the energy dependence of
the oscillation probability and we consider a hybrid setup which
combines the electron capture and the $\beta^+$ decay from the same
radioactive ion with the same boost. We study 
the CP discovery potential for different boosts and baselines. 
We conclude that the combination of the two decay channels, with different neutrino energies in a single experiment, achieves remarkable results.
 
\subsubsection{Energy Dependence}

The magnitude of the T-violating and CP-violating interference in neutrino oscillation
probabilities is directly proportional to $\sin\theta_{13}$~\cite{Krastev:1988yu,Arafune:1997hd,Bernabeu:1999gr,Dick:1999ed,Bernabeu:1999ct}. 
CP-violation can be observed either by an Asymmetry between
neutrinos and antineutrinos and/or by Energy Dependence in the neutrino
channel. In the last case, the CP phase $\delta$ plays the role of a phase shift
in the interference pattern between the atmospheric and solar amplitudes
for the appearance oscillation probability. This result is a consequence
~\cite{Bernabeu:1999ct} of the assumptions of CPT-invariance and No Absorptive part in the
oscillation amplitude: the Hermitian character of the Hamiltonian
responsible of the time evolution says that the CP-odd$=$T-odd
probability $P (\nu_e \rightarrow \nu_{\mu}) - P (\bar{\nu_e} \rightarrow \bar{\nu_{\mu}})$  
is an odd function of time, i.e., an odd function of the baseline $L$. In vacuum neutrino 
oscillations for relativistic neutrinos, the oscillation phase depends
on the ratio $L/E$, and then the CP-odd term becomes an odd function of the
energy $E$ for fixed $L$. With the same reasoning, the CP-even terms are
even functions of the energy $E$ in the oscillation probability. In 
this way, Energy Dependence in the appearance oscillation probability
is able to disentangle CP-even and CP-odd terms. One can check these
properties in the explicit expression for the suppressed appearance probability for neutrinos in vacuum oscillations.  

This result suggests the idea of disentangling $\delta$ from $|U_{e3}|$ without a need
of comparing neutrino and antineutrino events, which have different
beam systematics and different cross sections in the detector: either monochromatic
neutrino beams with different boosts or a combination of channels
with different neutrino energies in the same boost are able of
separating the CP-violating phase.

Due to neutrino propagation through the Earth, matter effects
can "fake" CP-violation in the sense that the presence of matter affects
neutrino and antineutrino oscillations in a different way.
It is not easy to disentangle matter effects from CP-violation since
there is the so-called "mass hierarchy degeneracy", which swaps the
effect of matter for neutrino and antineutrino oscillation according to
the sign of $\Delta m^2_{31}$. The energy dependence of the $\nu_e \rightarrow \nu_{\mu}$
oscillation probability, in presence of matter effects, can be studied \cite{Cervera:2000kp,Akhmedov:2004ny}  observing that the energy dependence induced by the presence of 
matter  is different in the three terms $T_{atm}$, $T_{sol}$ and in the interference $T_{int}$ and, in fact, 
different from the energy dependence associated with the CP-even versus the CP-odd separation. 
On the other hand, the mass hierarchy degeneracy in vacuum
is now removed  because $T_{atm}$ and $T_{int}$
are now changing under the change of sign of $\Delta m^2_{31}$, although $T_{sol}$ remains the same. 
All in all, we observe the virtues of studying the neutrino appearance probability
as a function of the neutrino energy.

From the current discovery phase of $\theta_{13}$, next generation
experiments will hence aim at precision measurements of the $\nu_e \rightarrow \nu_{\mu}$
oscillation probability. This will require large underground detectors 
coupled to more intense and pure neutrino beams. These aspects are being
studied within the LAGUNA and EURONu design studies. The knowledge
on the possible values of $\theta_{13}$ is a necessary input to best optimize
the search for CP-violation in the leptonic sector.

 A sister approach to the beta-beam is to use the
neutrinos sourced from ions that decay mainly through electron
capture (EC)~\cite{Sato:2005ma,Bernabeu:2005jh}. We discuss here a hybrid 
\cite{Bernabeu:2009np} of the EC and $\beta^{+}$ approaches. By selecting a 
nuclide with $Q_{\rm EC}\sim 4$~MeV, we can make use of neutrinos from
an electron capture spike and $\beta^{+}$ continuous spectrum
simultaneously. Assuming a detector with low energy threshold, the use
of such ions allows one to exploit the information from the first and
second oscillation maxima with a single beam. The use
of the hybrid approach makes it possible to use a
monochromatic beam at higher energies and a beta-beam at lower
energies. The need for good neutrino energy resolution at the higher
energies will therefore be less crucial than for high-$\gamma$
beta-beam scenarios.
 
\subsubsection{Combined EC plus $\beta^+$ Beam}

In this section we discuss the results of the idea \cite{Bernabeu:2009np} of the 
beta-beam and electron capture hybrid approach, as applied to the decay of Yterbium ($^{156}_{70}$Yb).
We simulate appearance experiments divided into four setups with the following characteristics:
\begin{enumerate}
\item 50~kton detector (LAr or TASD) with $2 \times 10^{18}$~ions/yr
\begin{itemize}
\item \textbf{Setup I}: CERN-Frejus (130~km) and $\gamma = 166$
\item \textbf{Setup II}: CERN-Canfranc or Gran Sasso (650~km) and $\gamma = 166$
\item \textbf{Setup III}: CERN-Canfranc or Gran Sasso (650~km) and $\gamma = 369$
\item \textbf{Setup IV}: CERN-Boulby (1050~km) and $\gamma = 369$ 
\end{itemize}
\item 0.5~Mton water-\v{C}erenkov detector with $2 \times 10^{18}$~ions/yr 
\begin{itemize}
\item \textbf{Setup III-WC}: CERN-Canfranc or Gran Sasso (650~km) and $\gamma = 369$
\item \textbf{Setup IV-WC}: CERN-Boulby (1050~km) and  $\gamma = 369$
\end{itemize}
\end{enumerate} 
Setups I and II correspond to present SPS energies for the boost,
whereas III(-WC) and IV(-WC) need an upgraded SPS with proton energy 1 TeV.

\begin{figure}
\begin{center}
\includegraphics[width=6cm]{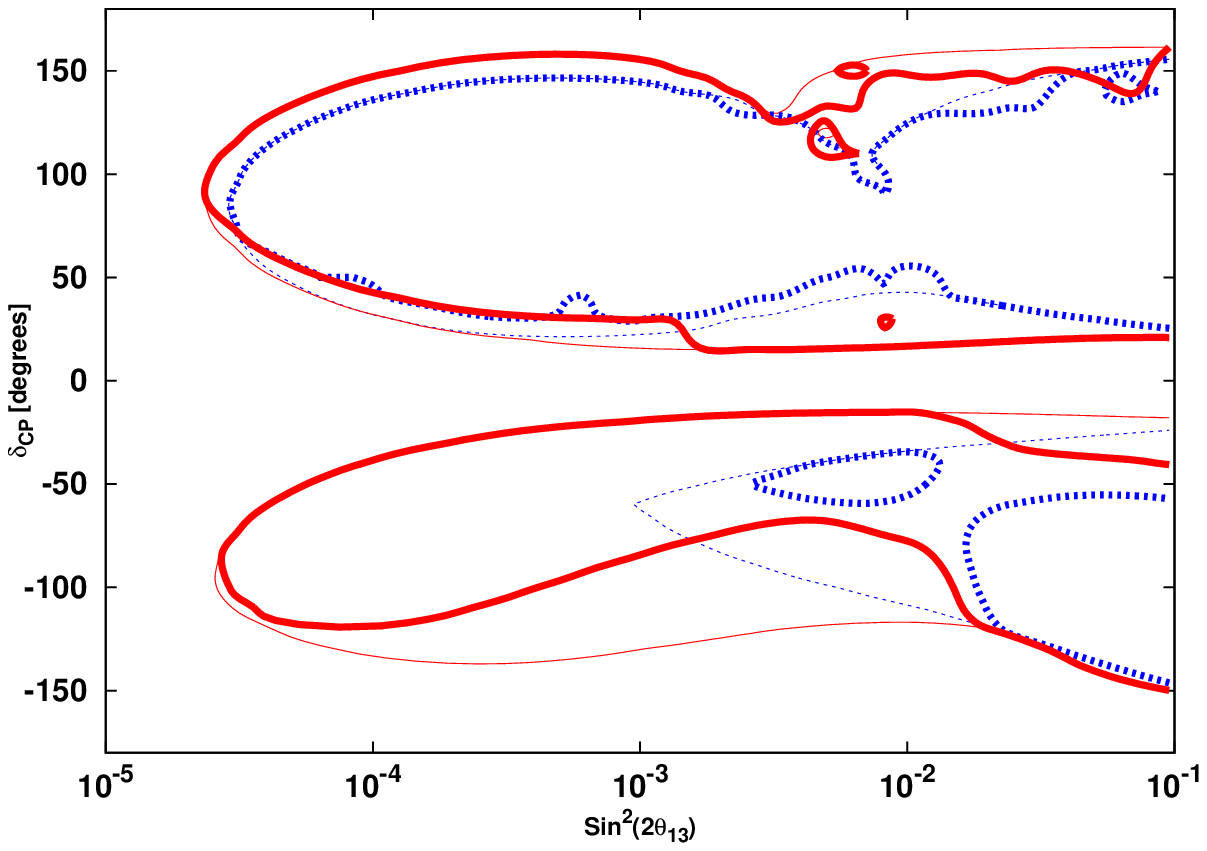} \hspace{1.5cm}
\includegraphics[width=6cm]{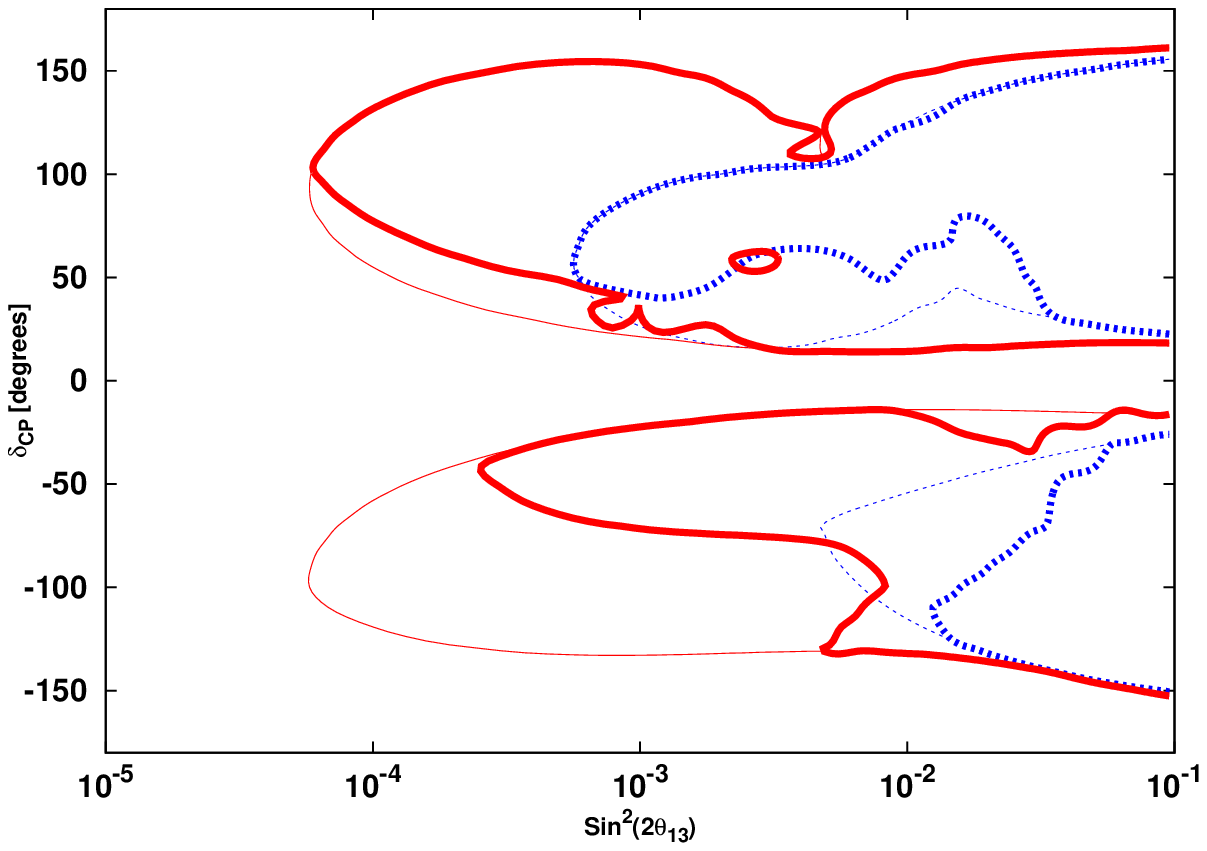}
\end{center}
\caption{{\small CP-violation discovery potential at 99\% CL for setup III-WC
  (left panel) and IV-WC (right panel). In each case, we present the
  results for the beta-beam only (blue dotted lines) and the
  combination with the electron capture result (red solid lines), both 
  without (thin lines) and with (thick lines) taking the hierarchy
  degeneracy into account. Taken from Ref.~\cite{Bernabeu:2009np}.}}
\label{Fi:CPWC}
\end{figure}

The best results are obtained for the setups with large Water Cerenkov
detector. The plots in Fig.~(\ref{Fi:CPWC}) represent the CP discovery potential for the setups with intermediate and larger baseline for high boost and large detector. 
    Note that the 650~km baseline has a significally better reach for CP violation at negative values of $\delta$ than the Boulby baseline.  
    
\begin{figure}[!ht]
\includegraphics[width=6cm]{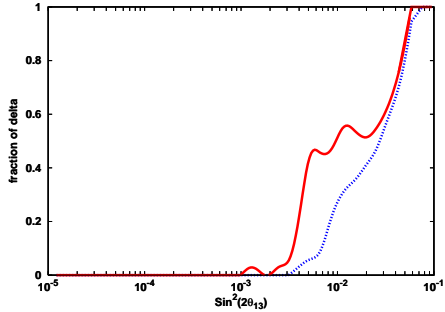}\hspace{2pc}
\begin{minipage}[b]{20pc}\caption{{\small Fraction of $\delta$ for which the neutrino mass hierarchy can be determined at 99\% CL for setup IV-WC. We present the results for the beta-beam only (blue dotted lines) and the combination with the electron capture result (red solid lines). Taken from Ref.~\cite{Bernabeu:2009np}.}}
\label{Fi:hierarchyWC}
\end{minipage}
\end{figure}

    The plot in Fig.~(\ref{Fi:hierarchyWC}) represent the fraction of the CP phase $\delta$ for which the neutrino mass hierarchy can be determined for setup IV-WC.
    The Boulby baseline, with its larger matter effect, is better for the
    determination of the mass hierarchy.

The principle of energy dependence to separate out the CP-even and CP-odd contributions to the neutrino oscillation probability works. The combined channels result in a good resolution on the intrinsic degeneracy, and on CP phase sensitivity. While the SPS upgrade is crucial for a better sensitivity to CP violation, it is so provided it is accompanied by an appropriate baseline of 650~km. The setups with the larger baseline provide a better determination of the hierarchy and still a good reach to CP violation for negative $\delta$, even if the mass ordering is unknown. The main conclusion is that the combined experiment achieves remarkable results. Up to now, past results on the production rate of these proton-rich ions are of the same order as for light ions.
It would be desirable to study the expectactions in the future radioactive
ion facility.

\section{Update of the SPL super-beam physics potential}
\label{sec:mezzetto}

The simulations developed so far for the SPL-Fr{\'e}jus Super Beam
foresee the use of a liquid mercury-jet target in order to
efficiently dissipate the heat produced by the 4 MW incoming proton
beam from the HP-SPL \cite{Campagne:2004wt}.
%
Due to the low energy of incoming protons (4 GeV) the emission angle
of secondary pions is large enough to force the use of a 
horn-embedded target in order to preserve a good collection
efficiency.
It should be noted that recent results from the MERIT collaboration
\cite{cit:MERIT} support the importance of a high magnetic field to
mitigate the explosion of the mercury jet. This can be achieved by
using superconducting solenoids for the capturing system. 
This solution is acceptable for the neutrino factory design but not
for a Super Beam due to the lack of charge discrimination. In the
current scenario the focusing system is composed of two concentric
magnetic horns.
%
Recent efforts \cite{Longhin,indicoWP2} have then been focused on the study of a solid target
option which would greatly simplify the problem of the integration of
the target and the focusing system and more importantly would avoid
the difficult issues related to the mercury jet handling in a magnetic
field free region.
The impact of using a solid target has been studied in terms of both
technical aspects (the power dissipation in the target) and physics
performance related aspects (pion and kaon yields, pion collection
efficiency with a long target, $\nu$ fluxes and sensitivity to
$\delta_{CP}$ and $\theta_{13}$). 

%
A graphite target was chosen since it is already an adopted technology
in current experiments.
As a first attempt the graphite ($\rho$=1.85 gcm$^{-3}$) target
was chosen to have the same radius of the previos mercury target (0.75
cm) and a length of 78 cm (instead of 30 cm for mercury) to roughly
preserve the prescription of having $\sim 2 \lambda_I$ of material.
%
%
The power released in the target has been estimated using
FLUKA2008.3\footnote{FLUKA 2002.4 was used in previous studies} and GEANT4.
At 4 GeV the deposited power is $\sim$ 250 kW
for the graphite target and 700 kW for the mercury one. 
The evolution of absolute particle yields for different particles
($K^\pm$, $K^0(\bar{K}^0)$, $\pi^\pm$, $n$) has been studied as a 
function of $E_k(p)$ from 2 to 10 GeV with FLUKA working at constant power. 
%
%
%
The pion yield for the mercury target is reasonably stable with energy
at $\sim$ $3 \cdot 10^{15}$ $\pi^+/s$ and $2.5 \cdot 10^{15}$
$\pi^-/s$.  The graphite target gives a rather flat rate of $\sim$
$2.5 \cdot 10^{15}$ $\pi^-/s$ while the larger $\pi^+$ flux decreases
from $\sim$ 4.5 to $\sim$ 3 $\cdot 10^{15}$ $\pi^+/s$ at 10 GeV.
The most striking difference between the two targets is the
neutron yield which is about a factor $\times$ 15 larger in the case
of mercury.  
A reduced neutron flux is highly
beneficial in terms of aluminum radiation damage.
%
%
At 5 GeV a structure occurs in the yields of $\pi^-$ and neutrons. 
At this energy the
matching of different inelastic hadron-nucleus production models
(Glauber-Gribov multiple scattering + GINC model below and
PEANUT model above) occurs. A similar structure used to be observed in
kaon spectra at 3.5 GeV in FLUKA2002.4 \cite{Campagne:2004wt}.
%
%
%
%
Neutrino fluxes have been computed with GEANT3
and the standard horn for kinetic energies of
2.2,3.5,4.5 and 8.0 GeV for both both positive and negative
focusing \cite{Longhin}. 

The obtained fluxes reflects the pion yields and thus
graphite fluxes result to be of the same order or even larger than the
ones obtained with graphite depending on energy.  On the other hand a
quite higher contamination of $\bar{\nu}$ in the neutrino beam and
particularly $\nu$ in the $\bar{\nu}$ beam is observed due to the fact
that with the standard horn many wrong charge pion emerging in the
downstream part of the target and at low angles are not effectively
defocused.
%
%
The $\sin^22\theta_{13}$ sensitivity curves (at $3\sigma$ C.L.) 
have been re-evaluated after the substitution of the standard mercury
target with the graphite one.
A worsening of the limit with graphite in the $\delta_{CP}<\pi$
region which is driven by $\bar{\nu}$ running ($\pi^-$ focusing) has
been observed.  The effect was found to be related to a sizable
contamination of $\nu_e^{cc}$ in the $\bar{\nu}$ beam from cascade
decays of defocused $\pi^+\to \mu^+\to e^+ \nu_e \bar{\nu}_\mu$. The
same behavior is not as evident in the $\nu$ running driven region
($\delta_{CP}>\pi$) due to the combined effect of the reduced cross
section of $\bar{\nu}_e$ and the fact that $\pi^-$ are less abundantly
produced than $\pi^+$.  This consideration motivated a reoptimization
of the horn shape in view of using a graphite target taking into
account in particularly the need for a reduced contamination from
wrong--charge $\pi$.

The horn optimization has been performed after a full rewriting of the 
simulation from GEANT3 \cite{Campagne:2004wt} to GEANT4 
in order to easily change the geometrical parameters and have a quick feed-back.
%
%
Two horn geometries have been implemented in GEANT4: the standard
one reproducing the existing CERN prototype and a more general one
based on a parametric model inspired by the shape of the MiniBOONE horn.
In order to debug and validate the new GEANT4--based software, a
comparison has been done with the fluxes obtained with GEANT3 using
the standard horn geometry and the graphite target.  Good agreement
has been achieved.
The parametric model is flexible enough to reproduce also the standard
conical geometry with an appropriate choice of the parameters.  This
possibility has also been used to cross check the parametric model by
comparison with the standard horn geometry.

A subset of the nine available geometrical parameters were sampled uniformly.
The horn currents and the horn+reflector
structure for the moment were maintained as in the original
design. The resulting fluxes were analyzed and ranked according to the
requirement of having low enough ``wrong-CP'' neutrino contamination
and high flux for the signal component. More sophisticated selection
techniques (i.e. based on final sensitivity on physical parameters and
energy spectrum shape) and further tuning would be possible but has
not yet been fully pursued. 
One of the horn shapes selected with the outlined heuristic procedure
has been studied in more detail (will be denoted as ``test
horn'' in the following). The most evident modifications with respect
to the previous design are the presence of a forward ``end-cap'' in
the horn (effective in removing low--angle wrong--sign pions) and the
thickness of the reflector which is larger by $\sim$ 10 cm. The radius
of the inner conductor is as in the previous design (3.7 cm).
With the test horn the $\nu_\mu$ and $\nu_e$ energy spectra are
shifted to higher energies with an increase in statistics particularly
around 5-600 MeV. The wrong-CP component on the other hand is reduced
by more than a factor two. The beam composition for the standard and
test horn is detailed in Tab.\ref{tab:comp} for positive and negative
focusing.
\begin{table}
\begin{center}
\begin{tabular}{ccc}
\hline
 & + focusing & - focusing\\
\hline
$\nu_\mu$ (\%)& 88.9 $\to$ {\bf{95.6}} &26.1 $\to$ {\bf{11.2}}\\
$\bar{\nu}_\mu$ (\%) & 10.5 $\to$ {\bf{3.9}} &73.4 $\to$ {\bf{88.4}}\\
$\nu_e$ (\%) & 0.60 $\to$ {\bf{0.56}} & 0.17 $\to$ {\bf{0.09}}\\
$\bar{\nu}_e$ (\%) & 0.052 $\to$ {\bf{0.025}} &0.340 $\to$ {\bf{0.352}}\\
\hline
\end{tabular}
\end{center}
\caption{Standard horn $\to$ {\bf{test horn}}.}
\label{tab:comp}
\end{table}
%

Profiting of the relative horn ($r=0.5$ m) and tunnel ($L=40$ m, $r=2$
m) compactness 
the idea of using a battery of four horns in parallel has been
proposed. This arrangement would imply reduced stress on the targets via
lower frequency (12.5 Hz) or lower proton flux depending on the
injection strategy. This choice would bring the incoming beam power in
the regime which is currently considered as a viable upper limit for
solid targets operations ($\sim$ 1 MW). 
This scenario has been implemented and tested with the GEANT4 simulation.
Small flux loss even up to big lateral displacements ($r$) are
found. In the extreme case of putting the four horns at the tunnel
edge ($r=r_{TUNNEL}-r_{HORN}$) the flux of $\nu_\mu$ is reduced by
13\% at 4.5 GeV. The baseline configuration with horns as central as
possible ($r \sim r_{HORN}\sqrt{2}$) causes an almost negligible loss
of $\nu_\mu$.  The presence of a magnetic field in all the horns
simultaneously or in each horn separately does not change
significantly the predicted fluxes.

\begin{figure}[ht]
\includegraphics[width=7.5cm]{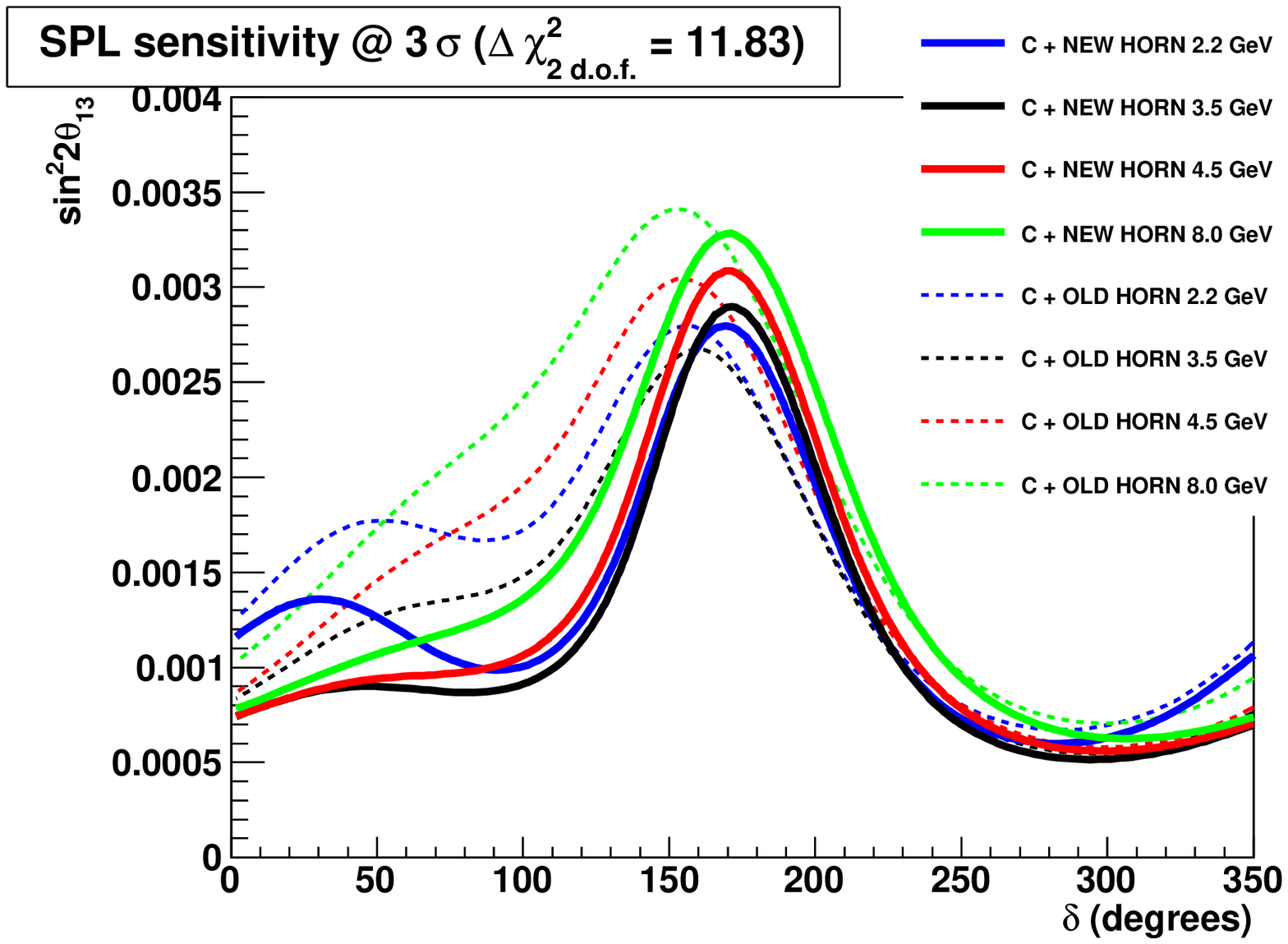}%
\includegraphics[width=7.5cm]{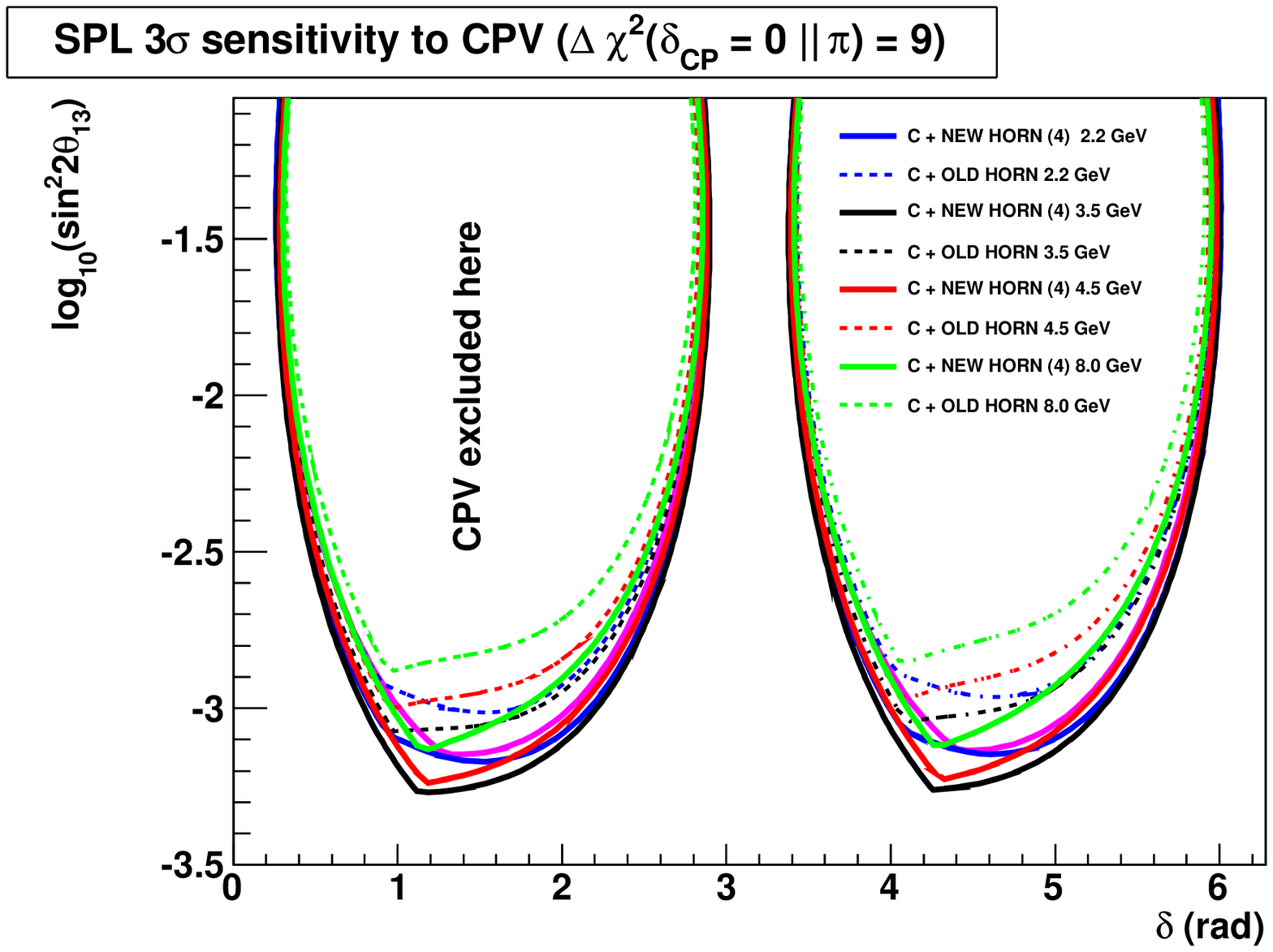}
\caption{$3\sigma$ sensitivity to $\sin^22\theta_{13}$. The standard and the test horn
both with a graphite target are compared. $3\sigma$ CP violation discovery. Taken from Ref.~\cite{Longhin}.
}
\label{fig:sensnew}
\end{figure}

Sensitivity limits on $\sin^22\theta_{13}$ 
calculated with GLoBES 3.0.14 are shown in Fig.~\ref{fig:sensnew}.
The performance of the MEMPHYS Water Cherenkov
detector\cite{MEMPHYS06} at the level of physics performance
(efficiencies, background rejection, etc.) is implemented in the AEDL
file \verb+SPL.glb+ which is distributed with GLoBES\cite{Campagne:2006yx}. A
mass of 0.44 Mton and a data taking of 8+2 years $\bar{\nu}$+$\nu$-running 
has been assumed.
The dashed curves refer to the standard horn
design in combination with the graphite target while the continuous ones
refer to the test horn. A significant improvement is observed in the
$\bar{\nu}$ running driven region as wanted. The graphite limits after the 
horn upgrade are in general even more performing than those obtained 
with the standard liquid mercury design.
It must be noted that this result is still to be considered as
preliminary since the NC-$\pi^0$ background has not yet been corrected
for the change in the neutrino energy spectrum. This correction anyway
will not alter the conclusions of this study since the bulk of the
background is coming from the intrinsic $\nu_e$+$\bar{\nu}_e$ beam
contamination which has been exactly taken into account. Increasing
the background by 30\% induces a worsening in the limit which is
$<$ 1$\cdot$10$^{-4}$ (mainly in the $\bar{\nu}$ driven $\delta$
region).
The CP violation discovery potential is shown in Fig.\ref{fig:sensnew}.
Parameter regions for which a $\Delta\chi^2>9$ is obtained when fitting under the
CP conserving hypotheses ($\delta_{CP}=0, \pi$) allow the CPV discovery at more
than 3$\sigma$. Also in this case a sizable improvement is obtained
(lowest $\sin^22\theta_{13}$ passes from $\sim 8\cdot10^{-4}$ to $\sim 5\cdot10^{-4}$).
It can be noticed than in general the 3.5 GeV and 4.5 GeV energies are
still the preferred ones also within the test focusing.

In summary the possibility to use a solid target looks very appealing for the 
SPL-Fr\'{e}jus Super Beam. Further steps which are foreseen include the
use of the HARP experiment ``thick target'' data to put the results on 
pion yields in graphite on a stronger experimental basis.

\newpage

\section{Updated comparison}

\begin{figure}[htbp]
\begin{center}
\includegraphics[width=12cm]{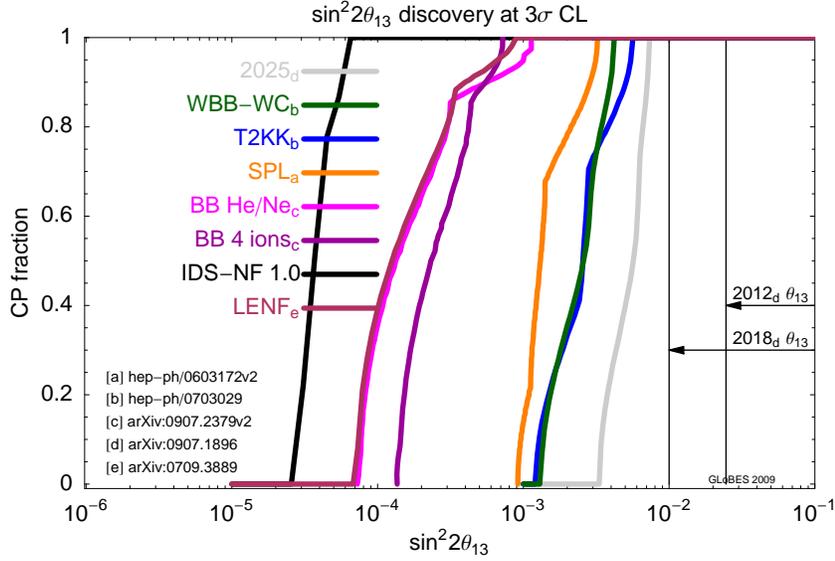}\\
\caption{Update of the comparison of the physics reach of different future facilities in  $\sin^2 2 \theta_{13}$. Prepared by P.~ÊHuber for this EURONU report using the GLoBES package \cite{Huber:2004ka,Huber:2007ji}. Curves are taken from [a] \cite{Campagne:2006yx}, [b] \cite{Barger:2007jq} , [c] \cite{Choubey:2009ks},  [d] \cite{Huber:2009cw} and [e] \cite{Bross:2007ts}.}
\label{fig:euronu20091}
\end{center}
\end{figure}

\begin{figure}[htbp]
\begin{center}
\includegraphics[width=12cm]{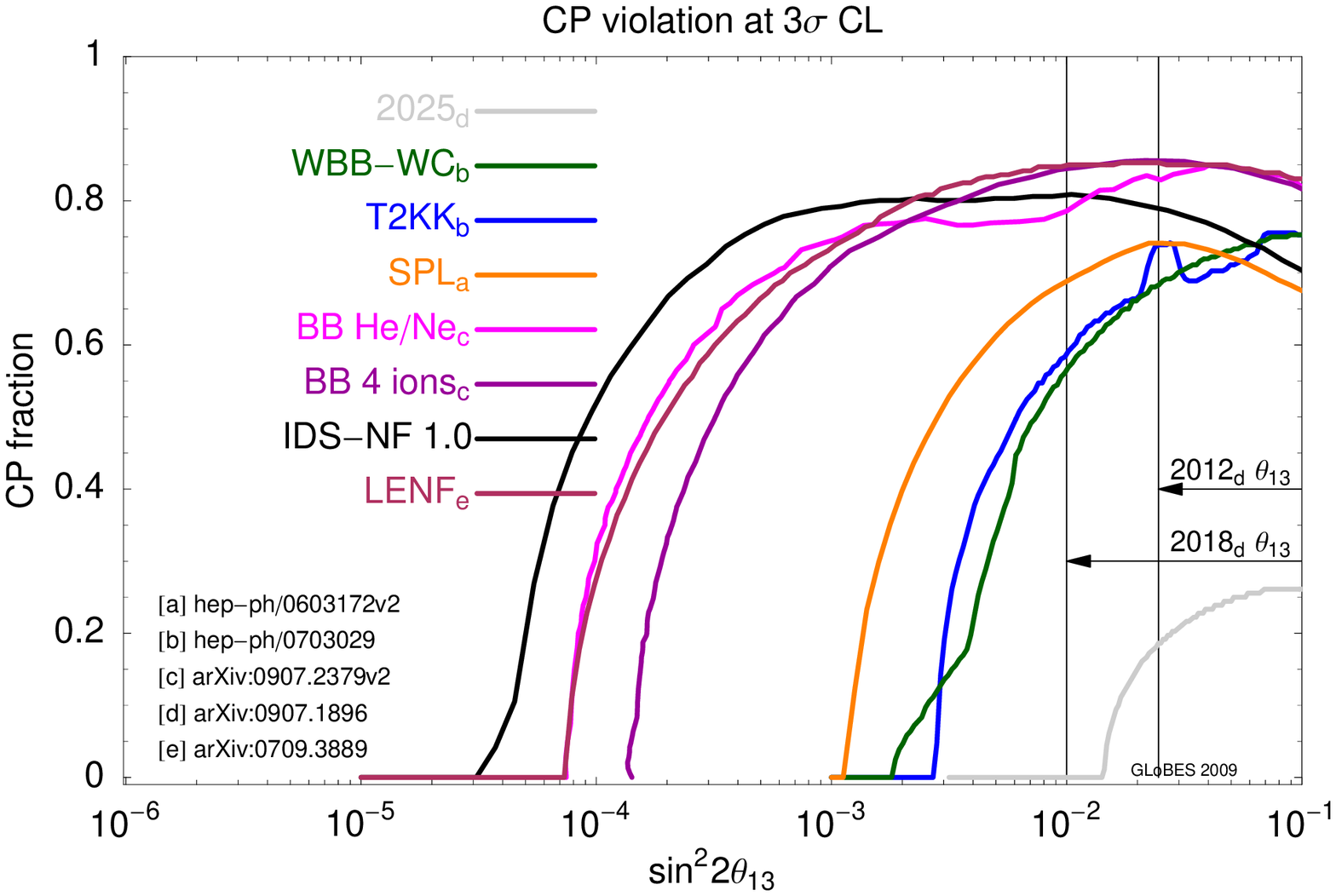}\\
\caption{Update of the comparison of the physics reach of different future facilities in  leptonic CP violation. Prepared by P.~ÊHuber for this EURONU report using the GLoBES package \cite{Huber:2004ka,Huber:2007ji}. Curves are taken from [a] \cite{Campagne:2006yx}, [b] \cite{Barger:2007jq} , [c] \cite{Choubey:2009ks},  [d] \cite{Huber:2009cw} and [e] \cite{Bross:2007ts}.}
\label{fig:euronu20092}
\end{center}
\end{figure}

\begin{figure}[htbp]
\begin{center}
\includegraphics[width=12cm]{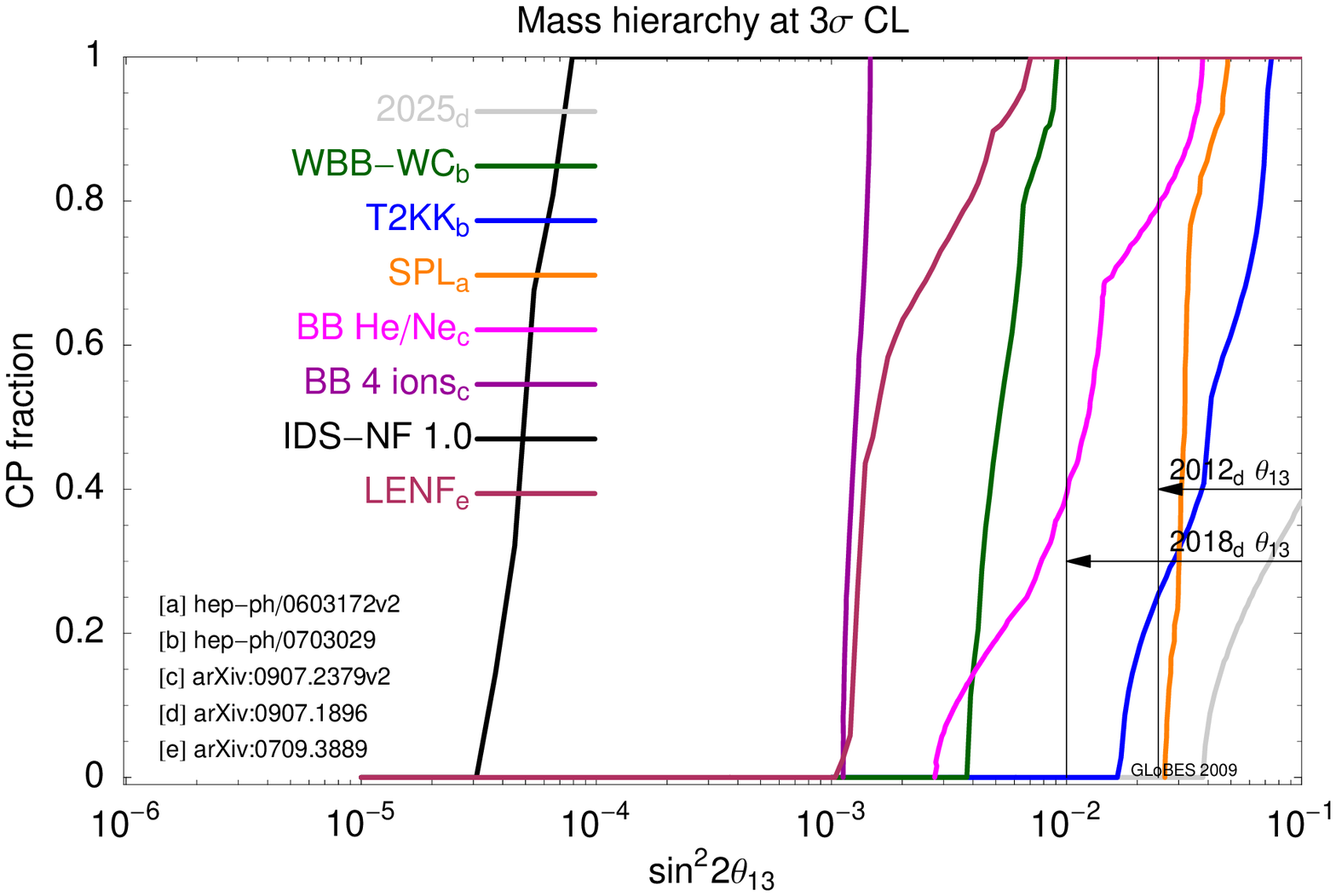}
\caption{Update of the comparison of the physics reach of different future facilities in  the neutrino mass hierarchy. Prepared by P.~ÊHuber for this EURONU report using the GLoBES package \cite{Huber:2004ka,Huber:2007ji}. Curves are taken from [a] \cite{Campagne:2006yx}, [b] \cite{Barger:2007jq} , [c] \cite{Choubey:2009ks},  [d] \cite{Huber:2009cw} and [e] \cite{Bross:2007ts}.}
\label{fig:euronu20093}
\end{center}
\end{figure}

In Figs.~\ref{fig:euronu20091}-\ref{fig:euronu20093} we present the comparison of the updated setups for the SPL superbeam, beta-beams and Nufact. The curves correspond to $3\sigma$CL  (1 dof) and 
the known parameters have been fixed to $\Delta m^2_{31}= 0.0024$eV$^2$, $\Delta m^2_{21}= 8 \times 10^{-5}$ eV$^2$ and $\theta_{23} =45^\circ$.

\vspace{1cm}

\newpage

\bibliographystyle{unsrt}
\bibliography{references}
\end{document}